%% file: bhabha2.tex
\documentclass[10pt]{article}
\usepackage{colordvi}
\usepackage{epsfig}
\usepackage{axodraw}
\usepackage{epsfig}
\usepackage{graphicx}
\usepackage{rotate}
\usepackage{latexsym}
\usepackage{amssymb}
\usepackage{amsmath}
\usepackage{multirow}
\newcommand\pubnumber{TTP08-26}
\newcommand\pubdate{\today}

\def\csum{Institut f\"ur Theoretische Teilchenphysik,\\
Universit\"at Karlsruhe, 76128 Karlsruhe, Germany}
\def\Title#1{\begin{center} {\Large\bf #1 } \end{center}}

\def\Author#1{\begin{center}{ \sc #1} \end{center}}
\def\Address#1{\begin{center}{ \it #1} \end{center}}

\newcommand\pubblock{\rightline{\begin{tabular}{l} \pubnumber\\
         \pubdate\\ \end{tabular}}}
\newenvironment{Abstract}{\begin{quotation}  }{\end{quotation}}

\def\Acknowledgments{\bigskip  \bigskip \begin{center}
          \large\bf Acknowledgments\end{center}}
\def\email#1{\footnote{#1}}
\makeatletter
\def\section{\@startsection{section}{0}{\z@}{5.5ex plus .5ex minus
 1.5ex}{2.3ex plus .2ex}{\large\bf}}
\def\subsection{\@startsection{subsection}{1}{\z@}{3.5ex plus .5ex minus
 1.5ex}{1.3ex plus .2ex}{\normalsize\bf}}
\def\subsubsection{\@startsection{subsubsection}{2}{\z@}{-3.5ex plus
-1ex minus  -.2ex}{2.3ex plus .2ex}{\normalsize\sl}}

\renewcommand{\@makecaption}[2]{%
   \vskip 10pt
   \setbox\@tempboxa\hbox{\small #1: #2}
   \ifdim \wd\@tempboxa >\hsize     
       \small #1: #2\par          
     \else                        
       \hbox to\hsize{\hfil\box\@tempboxa\hfil}
   \fi}
 \def\citenum#1{{\def\@cite##1##2{##1}\cite{#1}}}
\def\citea#1{\@cite{#1}{}}
\newcount\@tempcntc
\def\@citex[#1]#2{\if@filesw\immediate\write\@auxout{\string\citation{#2}}\fi
  \@tempcnta\z@\@tempcntb\m@ne\def\@citea{}\@cite{\@for\@citeb:=#2\do
    {\@ifundefined
       {b@\@citeb}{\@citeo\@tempcntb\m@ne\@citea\def\@citea{,}{\bf }\@warning
       {Citation `\@citeb' on page \thepage \space undefined}}%
    {\setbox\z@\hbox{\global\@tempcntc0\csname b@\@citeb\endcsname\relax}%
     \ifnum\@tempcntc=\z@ \@citeo\@tempcntb\m@ne
       \@citea\def\@citea{,}\hbox{\csname b@\@citeb\endcsname}%
     \else
      \advance\@tempcntb\@ne
      \ifnum\@tempcntb=\@tempcntc
      \else\advance\@tempcntb\m@ne\@citeo
      \@tempcnta\@tempcntc\@tempcntb\@tempcntc\fi\fi}}\@citeo}{#1}}
\def\@citeo{\ifnum\@tempcnta>\@tempcntb\else\@citea\def\@citea{,}%
  \ifnum\@tempcnta=\@tempcntb\the\@tempcnta\else
  {\advance\@tempcnta\@ne\ifnum\@tempcnta=\@tempcntb \else\def\@citea{--}\fi
    \advance\@tempcnta\m@ne\the\@tempcnta\@citea\the\@tempcntb}\fi\fi}
\makeatother
\input macro.tex
\begin{document}
\begin{titlepage}
\pubblock
\vfill
\Title{Two-loop QED hadronic corrections to Bhabha scattering}
\vfill
\Author{
Johann H. K\"uhn\email{jk@particle.uni-karlsruhe.de}
\hspace{0.1cm} {\rm and} \hspace{0.1cm}
Sandro Uccirati\email{uccirati@particle.uni-karlsruhe.de}
}
\Address{\csum}
\vfill
\vfill
\begin{Abstract}
\noindent 
Theoretical predictions for Bhabha scattering at the two-loop level require
the inclusion of hadronic vacuum polarization in the photon propagator.
We present predictions for the contributions from reducible amplitudes which
are proportional to the vacuum polarization $\pi(q^2)$  and from irreducible
ones where the vacuum polarization appears in a loop representing vertex or
box diagrams. The second case can be treated by using dispersion relations
with a weight function proportional to the $R$-ratio as measured in
electron-positron  annihilation into hadrons and kernels that can be
calculated perturbatively. 
We present simple analytical forms for the kernels and, using two convenient 
parametrizations for the function $R(s)$, numerical results for the 
quantities of interest. 
As a cross check we evaluate the corresponding corrections resulting from 
light and heavy lepton loops and we find perfect agreement with previous 
calculations. For the hadronic correction our result are in good agreement 
with a previous evaluation.
\end{Abstract}
\vfill
\begin{center}
\\[5mm]
\end{center}
\end{titlepage}
\def\thefootnote{\arabic{footnote}}
\setcounter{footnote}{0}
\section{Introduction}
Electron-positron colliders, with their potential for precise and specific
measurements of cross sections, have been and are being operated from
the very low energy region around the pion threshold up to more than 200 GeV
and may, in the future, reach up to energies of one or perhaps even several 
TeV. To determine the luminosity, one necessarily uses a reaction, whose cross
section can be well measured and furthermore calculated with sufficient
precision. Reactions which involve only leptons in the final state, like 
electron-positron annihilation into muon pairs or elastic electron-positron 
(Bhabha-)scattering are ideally suited for this purpose. In particular Bhabha
scattering, with its relatively large cross section, has always been
the standard luminosity monitor reaction. Precise theory predictions are, 
therefore, mandatory and, in view of recent interest in precise measurements 
with high counting rates, must be pushed to two-loop order.

Various ingredients are necessary in this connection. 
A major step has been made in~\cite{Penin:2005kf} where the photonic 
two-loop virtual corrections plus the corresponding soft real radiation 
has been evaluated for the case of interest $m_e^2/s\ll 1$, using earlier 
results for the completely massless case~\cite{Bern:2000ie} and exploiting 
the relation between the soft and collinear singularities for these two 
limiting cases. 
These results were confirmed in~\cite{Becher:2007cu}, where
in addition also the contributions from muon loops (again in the limit
$m_\mu^2/s \ll 1$ ) were calculated. These muon-loop contributions, in
the same high-energy limit were also evaluated in~\cite{Czakon:2006pa}, 
the corresponding results for arbitrary mass of the internal lepton were
presented in~\cite{Bonciani:2008ep}.
The electron loop corrections involving the exact $m_e^2/s$ dependence 
were computed in~\cite{Bonciani:2003te}, while further efforts towards 
the full electron mass depedence at two-loop level can be found 
in~\cite{Czakon:2004wm}.

All these contributions can be calculated strictly within Quantum
Electrodynamics (we do not consider electroweak corrections that are relevant
for high energies). However, in two loop approximation contributions from
virtual hadrons come into play. From general considerations it is obvious
that, generally speaking, their magnitude is comparable to or larger than those
from virtual muons. 
It is well known~\cite{Cabibbo:1961sz} that these virtual hadronic
contributions can be evaluated through dispersion relations, folding the
absorptive part of the hadronic vacuum polarization, the $R$-ratio measured in
electron-positron annihilation, with a kernel that can be calculated
perturbatively, in the present case in a one-loop calculation. 
This approach has been adopted in~\cite{Actis:2007fs}, where the kernel 
for the box diagram has been calculated for non-vanishing electron mass 
and the limit $m_e\to 0$ has been considered only subsequently. 
A more compact form for this kernel can be
obtained by using directly the well known results for the (direct plus
crossed) box with one photon and one massive vector boson, the $Z$-$\gamma$
box, contributing to Bhabha scattering at high energies. Since it is well known
that in this case the electron mass can be safely set to zero from the
beginning , the calculation becomes significantly simpler. Vertex corrections
with a hadronic insertion have been evaluated with this technique long time
ago and it is only this box contribution, that was not yet available since 
long. 

As stated above, results for the hadronic contributions have been 
presented in~\cite{Actis:2007fs}.
In view of the fact, that we are using a somewhat different approach
and furthermore, to provide an independent cross check the results of  our 
calculation will be presented in some detail. To allow for an easier
comparison with~\cite{Actis:2007fs}, we shall use the same 
parametrization~\cite{Burkhardt:1981jk} for the $R$-ratio. 
In addition we shall compare the results to the ones derived from a second 
parametrization~\cite{Hagiwara:2006jt} that includes more recent 
data and these latter ones should be considered as our definite predictions.

The paper  will be organized as follows: In chapter 2 we present the
general analysis and classify the various reducible and irreducible 
contributions.
In chapter 3 we give the details of the calculation, the explicit forms of
the kernels and identify the contributions from real radiation needed to
render the results infrared finite. In this connection it is convenient to
split the virtual (plus soft real) corrections into different building blocks
that will be described in more detail in this chapter.
The handling of the dispersion integrals with their poles is described in
chapter 4. Of specific interest is the high energy limit, with $s$, $t$ and
$u$ in an region where the $R$- ratio  has approached an approximately
constant value. 
In this limit and in analogy with the treatment on the form factor 
in~\cite{Kniehl:1988id} a particularly simple form can be derived where 
the information about hadron physics can be encoded into three ``moments'' 
of the $R$ ratio. 
Using this method and evaluating the moments for a lepton, e.g. the
muon or the tau-lepton, the results from~\cite{Becher:2007cu} are easily 
recovered.
In chapter 5 we present the numerical results for the two parametrizations. We
give the results for the building blocks, vacuum polarization, vertex and
boxes, and the complete corrections, split into the various contributions and
discuss their physical relevance.
The final Chapter 6 contains a brief summary and our conclusions.
\section{General analysis}
\begin{figure}[h]
$$
\scalebox{1}{
\begin{picture}(50,20)(0,-3)
 \Photon(0,0)(50,0){2}{10}          \Text(50,5)[cb]{$q^2$}
 \GCirc(25,0){10}{0.5}
 \Text(25,-30)[cb]{a}
\end{picture}
}
\qquad\qquad\qquad\qquad
\scalebox{1}{
\begin{picture}(40,20)(0,-3)
 \ArrowLine(20,0)(0,20)
 \ArrowLine(0,-20)(20,0)
 \Photon(20,0)(40,0){2}{5}          \Text(35,5)[cb]{$q^2$}
 \Photon(5,15)(5,-15){1.5}{8}
 \GCirc(5,0){5}{0.5}
 \Text(20,-30)[cb]{b}
\end{picture}
}
\qquad\qquad\qquad\qquad
\scalebox{1}{
\begin{picture}(60,20)(0,-3)
 \ArrowLine(15,15)(0,15)
 \ArrowLine(15,-15)(15,15)
 \ArrowLine(0,-15)(15,-15)
 \Photon(15,15)(45,15){2}{7}
 \Photon(15,-15)(45,-15){2}{7}
 \ArrowLine(60,15)(45,15)
 \ArrowLine(45,15)(45,-15)
 \ArrowLine(45,-15)(60,-15)
 \GCirc(30,15){5}{0.5}
 \Text(30,-30)[cb]{c}
\end{picture}
}
+
\;\cdots
$$
\\
\caption{The building blocks for the QED hadronic corrections}
\label{fig:buildingblock}
\end{figure}
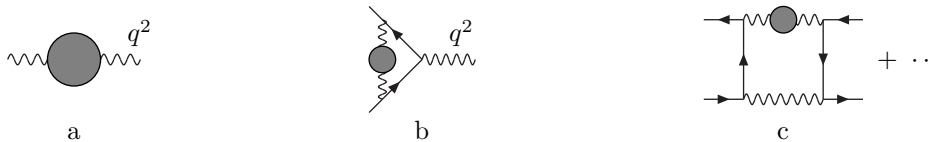

It is well known that contribution from the hadronic vacuum polarization 
can be directly evaluated by convoluting an appropriately chosen kernel 
with the familiar $R$-ratio ($R\equiv \sigma_{\rm had}/\sigma_{\rm pt}$) 
measured in electron-positron annihilation \cite{Cabibbo:1961sz}.
An arbitrary amplitude involving the hadronic vacuum polarization is 
obtained, by definition, from the original one by replacing the photon 
propagator as follows:
\bq
\frac{-\,i g_{\alpha\beta}}{q^2+i\ep} 
\quad\to\quad
\frac{-\,i g_{\alpha\delta}}{q^2+i\ep}\;
i\,(q^2 g^{\delta\ep} - q^\delta q^\ep )\,\Pi(q^2)\;
\frac{-\,i g_{\ep\beta}}{q^2+i\ep}.
\label{eq:prop1}
\eq
The renormalized vacuum polarization function $\Pi(q^2)$ is obtained 
from its absorptive part (essentially the $R$-ratio) by the subtracted 
dispersion relation:
\bq
\Pi(q^2)= 
-\,\frac{q^2}{\pi}\,\int_{4m^2}^{\infty}\!\!\frac{dz}{z}
\frac{{\rm Im}\Pi(z)}{q^2-z+i\ep},
\qquad
{\rm Im}\Pi(z)= -\,\frac{\alpha}{3}\,R(z),
\label{eq:Pi}
\eq
and has a cut for $q^2>4m^2$, with the threshold for hadron production 
at $4m^2$.
The $q^\delta q^\ep$ term in \eqn{eq:prop1} does not contribute and the photon 
propagator is effectively replaced as follows:
\bq
\frac{-\,i g_{\alpha\beta}}{q^2+i\ep} 
\quad\to\quad
\frac{-\,i g_{\alpha\beta}}{q^2+i\ep}\Pi(q^2)
=
-\,i g_{\alpha\beta}\,\frac{\alpha}{3\pi}
\int_{4m^2}^{\infty}\!\!\frac{dz}{z}
\frac{R(z)}{q^2-z+i\ep}.
\label{eq:prop2}
\eq
If $q^2$ is fixed by the external kinematics, applying the correction is 
equivalent to multiplication of the previous amplitude by $\Pi(q^2)$.
In higher orders, summing the one-particle reducible terms only, this 
corresponds to the replacement of the photon propagator by the dressed one:
\bq
\frac{-\,i g_{\alpha\beta}}{q^2} 
\quad\to\quad
\frac{-\,i g_{\alpha\beta}}{q^2}\frac{1}{1-\Pi(q^2)}.
\eq
However, if $q$ stands for a loop momentum, it is convenient to exchange 
the order of integration and evaluate in a first step the loop integral 
with a ficticious massive vector boson of mass $\sqrt{z}$, and to 
convolute subsequently this amplitude with the $R$-ratio, i.e. with 
$\frac{\alpha}{3\pi}\!\!\int_{{4m^2}}^{\infty}\!\!\frac{dz}{z}R(z)$.
In~\cite{Kniehl:1988id} this has been done for the Dirac form factor, 
assuming massless external fermions, and special emphasis has been put 
on the investigation of the limit, where the momentum transfer is far 
larger than $4m^2$.
A similar approach will be useful for the present case.
We will, in a first step, investigate the generic case valid for arbitrary 
$s,|t|,|u| \gg m_e^2$.
The weight functions needed to obtain the three building blocks, namely the 
vacuum polarization function $\Pi(q^2)$ (\fig{fig:buildingblock}a), the 
correction to the vertex function $V(q^2)$ (\fig{fig:buildingblock}b) 
and the amplitude arising from the box diagrams (\fig{fig:buildingblock}c), 
can be taken from the literature.

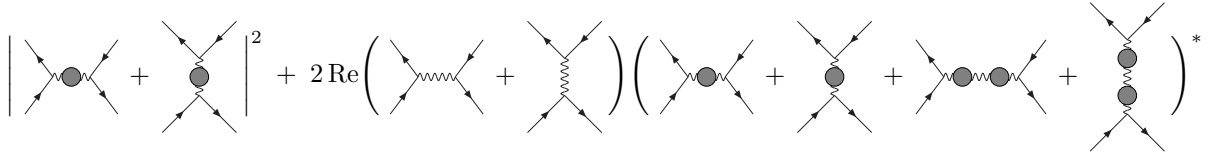
\begin{figure}[h]
$$
\Bigg|\,
\scalebox{0.7}{
\begin{picture}(50,20)(0,-3)
 \ArrowLine(15,0)(0,20)
 \ArrowLine(0,-20)(15,0)
 \Photon(15,0)(35,0){2}{5}
 \ArrowLine(50,20)(35,0)
 \ArrowLine(35,0)(50,-20)
 \GCirc(25,0){5}{0.5}
\end{picture}
}
+
\scalebox{0.7}{
\begin{picture}(40,40)(0,-3)
 \ArrowLine(20,10)(0,30)
 \ArrowLine(40,30)(20,10)
 \Photon(20,10)(20,-10){2}{5}
 \ArrowLine(0,-30)(20,-10)
 \ArrowLine(20,-10)(40,-30)
 \GCirc(20,0){5}{0.5}
\end{picture}
}
\Bigg|^2
\;
+
\;
2\,{\rm Re}
\Bigg(\,
\scalebox{0.7}{
\begin{picture}(50,20)(0,-3)
 \ArrowLine(15,0)(0,20)
 \ArrowLine(0,-20)(15,0)
 \Photon(15,0)(35,0){2}{5}
 \ArrowLine(50,20)(35,0)
 \ArrowLine(35,0)(50,-20)
\end{picture}
}
+
\scalebox{0.7}{
\begin{picture}(40,30)(0,-3)
 \ArrowLine(20,10)(0,30)
 \ArrowLine(40,30)(20,10)
 \Photon(20,10)(20,-10){2}{5}
 \ArrowLine(0,-30)(20,-10)
 \ArrowLine(20,-10)(40,-30)
\end{picture}
}
\Bigg)
\Bigg(\,
\scalebox{0.7}{
\begin{picture}(50,20)(0,-3)
 \ArrowLine(15,0)(0,20)
 \ArrowLine(0,-20)(15,0)
 \Photon(15,0)(35,0){2}{5}
 \ArrowLine(50,20)(35,0)
 \ArrowLine(35,0)(50,-20)
 \GCirc(25,0){5}{0.5}
\end{picture}
}
+
\scalebox{0.7}{
\begin{picture}(40,40)(0,-3)
 \ArrowLine(20,10)(0,30)
 \ArrowLine(40,30)(20,10)
 \Photon(20,10)(20,-10){2}{5}
 \ArrowLine(0,-30)(20,-10)
 \ArrowLine(20,-10)(40,-30)
 \GCirc(20,0){5}{0.5}
\end{picture}
}
+
\scalebox{0.7}{
\begin{picture}(70,20)(0,-3)
 \ArrowLine(15,0)(0,20)
 \ArrowLine(0,-20)(15,0)
 \Photon(15,0)(55,0){2}{10}
 \ArrowLine(70,20)(55,0)
 \ArrowLine(55,0)(70,-20)
 \GCirc(25,0){5}{0.5}
 \GCirc(45,0){5}{0.5}
\end{picture}
}
+
\scalebox{0.7}{
\begin{picture}(40,40)(0,-3)
 \ArrowLine(20,20)(0,40)
 \ArrowLine(40,40)(20,20)
 \Photon(20,20)(20,-20){2}{10}
 \ArrowLine(0,-40)(20,-20)
 \ArrowLine(20,-20)(40,-40)
 \GCirc(20,10){5}{0.5}
 \GCirc(20,-10){5}{0.5}
\end{picture}
}
\Bigg)^*
$$
\caption{Tree level diagrams with vacuum polarization insertion}
\label{fig:class1}
\end{figure}

\noindent
The complete hadron induced corrections are conveniently split into thre 
classes:
\bei
\item[1)]
Tree level diagrams, with two vacuum polarization insertions proportional 
to $\Pi(s)$ or $\Pi(t)$ where $\Pi$ originates from virtual hadrons, muons 
or electrons (\fig{fig:class1}).

\begin{figure}[h]
\bqaa
&&
2\,{\rm Re}
\Bigg(\,
\scalebox{0.7}{
\begin{picture}(50,20)(0,-3)
 \ArrowLine(15,0)(0,20)
 \ArrowLine(0,-20)(15,0)
 \Photon(15,0)(35,0){2}{5}
 \ArrowLine(50,20)(35,0)
 \ArrowLine(35,0)(50,-20)
\end{picture}
}
+
\scalebox{0.7}{
\begin{picture}(40,30)(0,-3)
 \ArrowLine(20,10)(0,30)
 \ArrowLine(40,30)(20,10)
 \Photon(20,10)(20,-10){2}{5}
 \ArrowLine(0,-30)(20,-10)
 \ArrowLine(20,-10)(40,-30)
\end{picture}
}
\Bigg)
\Bigg(\,
\scalebox{0.7}{
\begin{picture}(55,20)(0,-3)
 \ArrowLine(20,0)(0,20)
 \ArrowLine(0,-20)(20,0)
 \Photon(20,0)(40,0){2}{5}
 \ArrowLine(55,20)(40,0)
 \ArrowLine(40,0)(55,-20)
 \GCirc(30,0){5}{0.5}
 \Photon(5,15)(5,-15){1.5}{8}
\end{picture}
}
+
\scalebox{0.7}{
\begin{picture}(55,20)(0,-3)
 \ArrowLine(15,0)(0,20)
 \ArrowLine(0,-20)(15,0)
 \Photon(15,0)(35,0){2}{5}
 \ArrowLine(55,20)(35,0)
 \ArrowLine(35,0)(55,-20)
 \GCirc(25,0){5}{0.5}
 \Photon(50,15)(50,-15){1.5}{8}
\end{picture}
}
+
\scalebox{0.7}{
\begin{picture}(40,40)(0,-3)
 \ArrowLine(20,10)(0,30)
 \ArrowLine(40,30)(20,10)
 \Photon(20,10)(20,-10){2}{5}
 \ArrowLine(0,-30)(20,-10)
 \ArrowLine(20,-10)(40,-30)
 \GCirc(20,0){5}{0.5}
 \Photon(5,25)(35,25){1.5}{8}
\end{picture}
}
+
\scalebox{0.7}{
\begin{picture}(40,40)(0,-3)
 \ArrowLine(20,10)(0,30)
 \ArrowLine(40,30)(20,10)
 \Photon(20,10)(20,-10){2}{5}
 \ArrowLine(0,-30)(20,-10)
 \ArrowLine(20,-10)(40,-30)
 \GCirc(20,0){5}{0.5}
 \Photon(5,-25)(35,-25){1.5}{8}
\end{picture}
}
\Bigg)^*
\nl
&+&
2\,{\rm Re}
\Bigg(\,
\scalebox{0.7}{
\begin{picture}(50,20)(0,-3)
 \ArrowLine(15,0)(0,20)
 \ArrowLine(0,-20)(15,0)
 \Photon(15,0)(35,0){2}{5}
 \ArrowLine(50,20)(35,0)
 \ArrowLine(35,0)(50,-20)
 \GCirc(25,0){5}{0.5}
\end{picture}
}
+
\scalebox{0.7}{
\begin{picture}(40,40)(0,-3)
 \ArrowLine(20,10)(0,30)
 \ArrowLine(40,30)(20,10)
 \Photon(20,10)(20,-10){2}{5}
 \ArrowLine(0,-30)(20,-10)
 \ArrowLine(20,-10)(40,-30)
 \GCirc(20,0){5}{0.5}
\end{picture}
}
\Bigg)
\Bigg(\,
\scalebox{0.7}{
\begin{picture}(55,20)(0,-3)
 \ArrowLine(20,0)(0,20)
 \ArrowLine(0,-20)(20,0)
 \Photon(20,0)(40,0){2}{5}
 \ArrowLine(55,20)(40,0)
 \ArrowLine(40,0)(55,-20)
 \Photon(5,15)(5,-15){1.5}{8}
\end{picture}
}
+
\scalebox{0.7}{
\begin{picture}(55,20)(0,-3)
 \ArrowLine(15,0)(0,20)
 \ArrowLine(0,-20)(15,0)
 \Photon(15,0)(35,0){2}{5}
 \ArrowLine(55,20)(35,0)
 \ArrowLine(35,0)(55,-20)
 \Photon(50,15)(50,-15){1.5}{8}
\end{picture}
}
+
\scalebox{0.7}{
\begin{picture}(40,40)(0,-3)
 \ArrowLine(20,10)(0,30)
 \ArrowLine(40,30)(20,10)
 \Photon(20,10)(20,-10){2}{5}
 \ArrowLine(0,-30)(20,-10)
 \ArrowLine(20,-10)(40,-30)
 \Photon(5,25)(35,25){1.5}{8}
\end{picture}
}
+
\scalebox{0.7}{
\begin{picture}(40,40)(0,-3)
 \ArrowLine(20,10)(0,30)
 \ArrowLine(40,30)(20,10)
 \Photon(20,10)(20,-10){2}{5}
 \ArrowLine(0,-30)(20,-10)
 \ArrowLine(20,-10)(40,-30)
 \Photon(5,-25)(35,-25){1.5}{8}
\end{picture}
}
\nl
\nl
&&
\qquad\qquad\qquad\qquad\qquad\qquad\quad
+
\scalebox{0.7}{
\begin{picture}(60,20)(0,-3)
 \ArrowLine(15,15)(0,15)
 \ArrowLine(15,-15)(15,15)
 \ArrowLine(0,-15)(15,-15)
 \Photon(15,15)(45,15){2}{7}
 \Photon(15,-15)(45,-15){2}{7}
 \ArrowLine(60,15)(45,15)
 \ArrowLine(45,15)(45,-15)
 \ArrowLine(45,-15)(60,-15)
\end{picture}
}
+
\scalebox{0.7}{
\begin{picture}(60,20)(0,-3)
 \ArrowLine(15,15)(0,15)
 \ArrowLine(15,-15)(15,15)
 \ArrowLine(0,-15)(15,-15)
 \Photon(15,15)(45,-15){2}{7}
 \Photon(15,-15)(45,15){2}{7}
 \ArrowLine(60,15)(45,15)
 \ArrowLine(45,15)(45,-15)
 \ArrowLine(45,-15)(60,-15)
\end{picture}
}
+
\scalebox{0.7}{
\begin{picture}(60,20)(0,-3)
 \ArrowLine(15,15)(0,15)
 \ArrowLine(45,15)(15,15)
 \ArrowLine(60,15)(45,15)
 \Photon(15,15)(15,-15){2}{7}
 \Photon(45,15)(45,-15){2}{7}
 \ArrowLine(0,-15)(15,-15)
 \ArrowLine(15,-15)(45,-15)
 \ArrowLine(45,-15)(60,-15)
\end{picture}
}
+
\scalebox{0.7}{
\begin{picture}(60,20)(0,-3)
 \ArrowLine(15,15)(0,15)
 \ArrowLine(45,15)(15,15)
 \ArrowLine(60,15)(45,15)
 \Photon(15,-15)(45,15){2}{7}
 \Photon(15,15)(45,-15){2}{7}
 \ArrowLine(0,-15)(15,-15)
 \ArrowLine(15,-15)(45,-15)
 \ArrowLine(45,-15)(60,-15)
\end{picture}
}
\Bigg)^*
\eqaa
\caption{One-loop photonic diagrams in combination with a dressed 
photon propagator}
\label{fig:class2}
\end{figure}
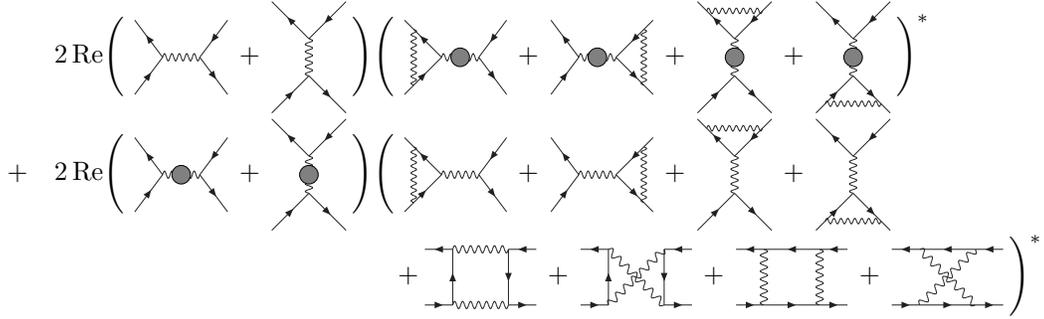

\noindent
These corrections to the amplitudes are proportional to $\Pi(s)^2$, 
$\Pi(t)^2$ or $\Pi(s)\Pi(t)$ (reducible quadratic $\Pi$-terms) and 
are directly obtained from the Born amplitude.
\item[2)]
Corrections which involve one-loop purely photonic corrections in combination 
with a dressed photon propagator in the $s$ or $t$ channel (\fig{fig:class2}).

\noindent
These are proportional to $\Pi(s)$ or $\Pi(t)$ (reducible linear $\Pi$-terms) 
and are directly obtained from the one-loop corrections to the Bhabha 
scattering.
They can again be separated into amplitudes resulting from one-loop vertex 
and box corrections, respectively.
Both are rendered infrared finite by adding soft real photon emission 
with $E_\gamma<\omega<<\sqrt{s}$.
In both cases this leads to a logarithmic $\omega$-dependence.

The photon vertex correction involves a collinear electron-mass 
singularity, which leads to the only $m_e$ dependence relevant for our 
investigation.

\begin{figure}[h]
\bqaa
&&
2\,{\rm Re}
\Bigg(\,
\scalebox{0.7}{
\begin{picture}(50,20)(0,-3)
 \ArrowLine(15,0)(0,20)
 \ArrowLine(0,-20)(15,0)
 \Photon(15,0)(35,0){2}{5}
 \ArrowLine(50,20)(35,0)
 \ArrowLine(35,0)(50,-20)
\end{picture}
}
+
\scalebox{0.7}{
\begin{picture}(40,40)(0,-3)
 \ArrowLine(20,10)(0,30)
 \ArrowLine(40,30)(20,10)
 \Photon(20,10)(20,-10){2}{5}
 \ArrowLine(0,-30)(20,-10)
 \ArrowLine(20,-10)(40,-30)
\end{picture}
}
\Bigg)
\Bigg(\,
\scalebox{0.7}{
\begin{picture}(55,20)(0,-3)
 \ArrowLine(20,0)(0,20)
 \ArrowLine(0,-20)(20,0)
 \Photon(20,0)(40,0){2}{5}
 \ArrowLine(55,20)(40,0)
 \ArrowLine(40,0)(55,-20)
 \Photon(5,15)(5,-15){1.5}{8}
 \GCirc(5,0){5}{0.5}
\end{picture}
}
+
\scalebox{0.7}{
\begin{picture}(55,20)(0,-3)
 \ArrowLine(15,0)(0,20)
 \ArrowLine(0,-20)(15,0)
 \Photon(15,0)(35,0){2}{5}
 \ArrowLine(55,20)(35,0)
 \ArrowLine(35,0)(55,-20)
 \Photon(50,15)(50,-15){1.5}{8}
 \GCirc(50,0){5}{0.5}
\end{picture}
}
+
\scalebox{0.7}{
\begin{picture}(40,40)(0,-3)
 \ArrowLine(20,10)(0,30)
 \ArrowLine(40,30)(20,10)
 \Photon(20,10)(20,-10){2}{5}
 \ArrowLine(0,-30)(20,-10)
 \ArrowLine(20,-10)(40,-30)
 \Photon(5,25)(35,25){1.5}{8}
 \GCirc(20,25){5}{0.5}
\end{picture}
}
+
\scalebox{0.7}{
\begin{picture}(40,40)(0,-3)
 \ArrowLine(20,10)(0,30)
 \ArrowLine(40,30)(20,10)
 \Photon(20,10)(20,-10){2}{5}
 \ArrowLine(0,-30)(20,-10)
 \ArrowLine(20,-10)(40,-30)
 \Photon(5,-25)(35,-25){1.5}{8}
 \GCirc(20,-25){5}{0.5}
\end{picture}
}
\nl
\nl
\nl
&&
\qquad\qquad\qquad\qquad\qquad\qquad\quad
+
\scalebox{0.7}{
\begin{picture}(60,20)(0,-3)
 \ArrowLine(15,15)(0,15)
 \ArrowLine(15,-15)(15,15)
 \ArrowLine(0,-15)(15,-15)
 \Photon(15,15)(45,15){2}{7}
 \Photon(15,-15)(45,-15){2}{7}
 \ArrowLine(60,15)(45,15)
 \ArrowLine(45,15)(45,-15)
 \ArrowLine(45,-15)(60,-15)
 \GCirc(30,15){5}{0.5}
\end{picture}
}
+
\scalebox{0.7}{
\begin{picture}(60,20)(0,-3)
 \ArrowLine(15,15)(0,15)
 \ArrowLine(15,-15)(15,15)
 \ArrowLine(0,-15)(15,-15)
 \Photon(15,15)(45,15){2}{7}
 \Photon(15,-15)(45,-15){2}{7}
 \ArrowLine(60,15)(45,15)
 \ArrowLine(45,15)(45,-15)
 \ArrowLine(45,-15)(60,-15)
 \GCirc(30,-15){5}{0.5}
\end{picture}
}
+
\scalebox{0.7}{
\begin{picture}(60,20)(0,-3)
 \ArrowLine(15,15)(0,15)
 \ArrowLine(15,-15)(15,15)
 \ArrowLine(0,-15)(15,-15)
 \Photon(15,15)(45,-15){2}{7}
 \Photon(15,-15)(45,15){2}{7}
 \ArrowLine(60,15)(45,15)
 \ArrowLine(45,15)(45,-15)
 \ArrowLine(45,-15)(60,-15)
 \GCirc(37.5,7.5){5}{0.5}
\end{picture}
}
+
\scalebox{0.7}{
\begin{picture}(60,20)(0,-3)
 \ArrowLine(15,15)(0,15)
 \ArrowLine(15,-15)(15,15)
 \ArrowLine(0,-15)(15,-15)
 \Photon(15,15)(45,-15){2}{7}
 \Photon(15,-15)(45,15){2}{7}
 \ArrowLine(60,15)(45,15)
 \ArrowLine(45,15)(45,-15)
 \ArrowLine(45,-15)(60,-15)
 \GCirc(37.5,-7.5){5}{0.5}
\end{picture}
}
\nl
\nl
&&
\qquad\qquad\qquad\qquad\qquad\qquad\quad
+
\scalebox{0.7}{
\begin{picture}(60,20)(0,-3)
 \ArrowLine(15,15)(0,15)
 \ArrowLine(45,15)(15,15)
 \ArrowLine(60,15)(45,15)
 \Photon(15,15)(15,-15){2}{7}
 \Photon(45,15)(45,-15){2}{7}
 \ArrowLine(0,-15)(15,-15)
 \ArrowLine(15,-15)(45,-15)
 \ArrowLine(45,-15)(60,-15)
 \GCirc(15,0){5}{0.5}
\end{picture}
}
+
\scalebox{0.7}{
\begin{picture}(60,20)(0,-3)
 \ArrowLine(15,15)(0,15)
 \ArrowLine(45,15)(15,15)
 \ArrowLine(60,15)(45,15)
 \Photon(15,15)(15,-15){2}{7}
 \Photon(45,15)(45,-15){2}{7}
 \ArrowLine(0,-15)(15,-15)
 \ArrowLine(15,-15)(45,-15)
 \ArrowLine(45,-15)(60,-15)
 \GCirc(45,0){5}{0.5}
\end{picture}
}
+
\scalebox{0.7}{
\begin{picture}(60,20)(0,-3)
 \ArrowLine(15,15)(0,15)
 \ArrowLine(45,15)(15,15)
 \ArrowLine(60,15)(45,15)
 \Photon(15,-15)(45,15){2}{7}
 \Photon(15,15)(45,-15){2}{7}
 \ArrowLine(0,-15)(15,-15)
 \ArrowLine(15,-15)(45,-15)
 \ArrowLine(45,-15)(60,-15)
 \GCirc(37.5,7.5){5}{0.5}
\end{picture}
}
+
\scalebox{0.7}{
\begin{picture}(60,20)(0,-3)
 \ArrowLine(15,15)(0,15)
 \ArrowLine(45,15)(15,15)
 \ArrowLine(60,15)(45,15)
 \Photon(15,-15)(45,15){2}{7}
 \Photon(15,15)(45,-15){2}{7}
 \ArrowLine(0,-15)(15,-15)
 \ArrowLine(15,-15)(45,-15)
 \ArrowLine(45,-15)(60,-15)
 \GCirc(37.5,-7.5){5}{0.5}
\end{picture}
}
\Bigg)^*
\eqaa
\caption{Irreducible two-loop diagrams}
\label{fig:class3}
\end{figure}
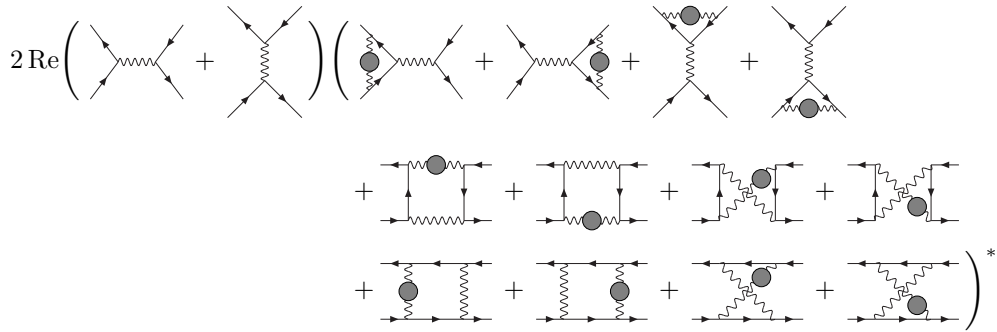

Also the $\gamma\gamma$-box amplitude, after interference with $s$- or 
$t$-channel dressed photon exchange, leads to a logarithmic dependence on 
$\omega$.
The electron mass, however, may safely be set to zero.
\item[3)]
As a third class we have to consider the irreducible two-loop contributions, 
i.e. amplitudes with dressed photon propagator in a loop (\fig{fig:class3}).

\noindent
The dressed propagator may be located either in a vertex or in a 
box, which will interfere with the two Born amplitudes from $s$- and 
$t$-channel exchange.
Irreducible vertex corrections are infrared finite and $m_e$ may be 
safely set to zero.
They are easily obtained from the Born amplitude by replacing the appropriate 
vertex by $V(s)$ or $V(t)$ as defined below.
The box amplitudes are again infrared divergent and must be made finite 
by combining with soft real radiation.
\eei
Let us emphasize that contributions from lepton loops follow the same 
classification\footnote{With the exception of the two-loop vacuum 
polarization (\fig{fig:se2}) whose absorptive part is in the hadronic 
case, by definition, part of the $R$-ratio.}.
The simple form of the result in the high energy limit, 
$s,|t|,|u| \gg 4m_\pi^2$, will allow for a convenient cross check of our 
calculation.

In addition to these corrections with virtual photons, one has to compute the 
corresponding emission of real photons (\fig{fig:real}) to compensate the 
infrared divergencies.
\begin{figure}[h]
\bqaa
&&
\!\!\!\!\!\!\!\!\!\!\!\!\!
\frac{1}{(2\pi)^3}\!\!\int_{\omega}\!\frac{d^3k}{2\,k_0}\;
2\,{\rm Re}
\Bigg(\,
\scalebox{0.7}{
\begin{picture}(50,20)(0,-3)
 \ArrowLine(15,0)(7.5,10)
 \ArrowLine(7.5,10)(0,20)
 \Photon(7.5,10)(20,20){2}{4}        \Text(25,16)[cb]{$k$}
 \ArrowLine(0,-20)(15,0)
 \Photon(15,0)(35,0){2}{5}
 \ArrowLine(50,20)(35,0)
 \ArrowLine(35,0)(50,-20)
\end{picture}
}
\!+\!
\scalebox{0.7}{
\begin{picture}(50,20)(0,-3)
 \ArrowLine(15,0)(0,20)
 \ArrowLine(0,-20)(7.5,-10)
 \ArrowLine(7.5,-10)(15,0)
 \Photon(7.5,-10)(20,-20){2}{4}        \Text(24,-22)[cb]{$k$}
 \Photon(15,0)(35,0){2}{5}
 \ArrowLine(50,20)(35,0)
 \ArrowLine(35,0)(50,-20)
\end{picture}
}
\!+\!
\scalebox{0.7}{
\begin{picture}(50,20)(0,-3)
 \ArrowLine(15,0)(0,20)
 \ArrowLine(0,-20)(15,0)
 \Photon(15,0)(35,0){2}{5}
 \ArrowLine(50,20)(35,0)
 \ArrowLine(35,0)(42.5,-10)
 \ArrowLine(42.5,-10)(50,-20)
 \Photon(42.5,-10)(35,-20){2}{4}        \Text(29.5,-22)[cb]{$k$}
\end{picture}
}
\!+\!
\scalebox{0.7}{
\begin{picture}(50,20)(0,-3)
 \ArrowLine(15,0)(0,20)
 \ArrowLine(0,-20)(15,0)
 \Photon(15,0)(35,0){2}{5}
 \ArrowLine(50,20)(42.5,10)
 \ArrowLine(42.5,10)(35,0)
 \Photon(42.5,10)(35,20){2}{4}        \Text(29.5,16)[cb]{$k$}
 \ArrowLine(35,0)(50,-20)
\end{picture}
}
\!+\!
\scalebox{0.7}{
\begin{picture}(40,40)(0,-3)
 \ArrowLine(20,10)(10,20)
 \ArrowLine(10,20)(0,30)
 \Photon(10,20)(20,30){2}{4}        \Text(25,26)[cb]{$k$}
 \ArrowLine(40,30)(20,10)
 \Photon(20,10)(20,-10){2}{5}
 \ArrowLine(0,-30)(20,-10)
 \ArrowLine(20,-10)(40,-30)
\end{picture}
}
\!+\!
\scalebox{0.7}{
\begin{picture}(40,40)(0,-3)
 \ArrowLine(20,10)(0,30)
 \ArrowLine(40,30)(20,10)
 \Photon(20,10)(20,-10){2}{5}
 \ArrowLine(0,-30)(10,-20)
 \ArrowLine(10,-20)(20,-10)
 \Photon(10,-20)(20,-30){2}{4}        \Text(24,-32)[cb]{$k$}
 \ArrowLine(20,-10)(40,-30)
\end{picture}
}
\!+\!
\scalebox{0.7}{
\begin{picture}(40,40)(0,-3)
 \ArrowLine(20,10)(0,30)
 \ArrowLine(40,30)(20,10)
 \Photon(20,10)(20,-10){2}{5}
 \ArrowLine(0,-30)(20,-10)
 \ArrowLine(20,-10)(30,-20)
 \ArrowLine(30,-20)(40,-30)
 \Photon(30,-20)(20,-30){2}{4}        \Text(15.5,-32)[cb]{$k$}
\end{picture}
}
\!+\!
\scalebox{0.7}{
\begin{picture}(40,40)(0,-3)
 \ArrowLine(20,10)(0,30)
 \ArrowLine(40,30)(30,20)
 \ArrowLine(30,20)(20,10)
 \Photon(30,20)(20,30){2}{4}        \Text(16,26)[cb]{$k$}
 \Photon(20,10)(20,-10){2}{5}
 \ArrowLine(0,-30)(20,-10)
 \ArrowLine(20,-10)(40,-30)
\end{picture}
}
\Bigg)
\nl
\nl
&&
\qquad\qquad\quad\;\,
\Bigg(\,
\scalebox{0.7}{
\begin{picture}(50,20)(0,-3)
 \ArrowLine(15,0)(7.5,10)
 \ArrowLine(7.5,10)(0,20)
 \Photon(7.5,10)(20,20){2}{4}        \Text(25,16)[cb]{$k$}
 \ArrowLine(0,-20)(15,0)
 \Photon(15,0)(35,0){2}{5}
 \ArrowLine(50,20)(35,0)
 \ArrowLine(35,0)(50,-20)
 \GCirc(25,0){5}{0.5}
\end{picture}
}
\!+\!
\scalebox{0.7}{
\begin{picture}(50,20)(0,-3)
 \ArrowLine(15,0)(0,20)
 \ArrowLine(0,-20)(7.5,-10)
 \ArrowLine(7.5,-10)(15,0)
 \Photon(7.5,-10)(20,-20){2}{4}        \Text(24,-22)[cb]{$k$}
 \Photon(15,0)(35,0){2}{5}
 \ArrowLine(50,20)(35,0)
 \ArrowLine(35,0)(50,-20)
 \GCirc(25,0){5}{0.5}
\end{picture}
}
\!+\!
\scalebox{0.7}{
\begin{picture}(50,20)(0,-3)
 \ArrowLine(15,0)(0,20)
 \ArrowLine(0,-20)(15,0)
 \Photon(15,0)(35,0){2}{5}
 \ArrowLine(50,20)(35,0)
 \ArrowLine(35,0)(42.5,-10)
 \ArrowLine(42.5,-10)(50,-20)
 \Photon(42.5,-10)(35,-20){2}{4}        \Text(29.5,-22)[cb]{$k$}
 \GCirc(25,0){5}{0.5}
\end{picture}
}
\!+\!
\scalebox{0.7}{
\begin{picture}(50,20)(0,-3)
 \ArrowLine(15,0)(0,20)
 \ArrowLine(0,-20)(15,0)
 \Photon(15,0)(35,0){2}{5}
 \ArrowLine(50,20)(42.5,10)
 \ArrowLine(42.5,10)(35,0)
 \Photon(42.5,10)(35,20){2}{4}        \Text(29.5,16)[cb]{$k$}
 \ArrowLine(35,0)(50,-20)
 \GCirc(25,0){5}{0.5}
\end{picture}
}
\!+\!
\scalebox{0.7}{
\begin{picture}(40,40)(0,-3)
 \ArrowLine(20,10)(10,20)
 \ArrowLine(10,20)(0,30)
 \Photon(10,20)(20,30){2}{4}        \Text(25,26)[cb]{$k$}
 \ArrowLine(40,30)(20,10)
 \Photon(20,10)(20,-10){2}{5}
 \ArrowLine(0,-30)(20,-10)
 \ArrowLine(20,-10)(40,-30)
 \GCirc(20,0){5}{0.5}
\end{picture}
}
\!+\!
\scalebox{0.7}{
\begin{picture}(40,40)(0,-3)
 \ArrowLine(20,10)(0,30)
 \ArrowLine(40,30)(20,10)
 \Photon(20,10)(20,-10){2}{5}
 \ArrowLine(0,-30)(10,-20)
 \ArrowLine(10,-20)(20,-10)
 \Photon(10,-20)(20,-30){2}{4}        \Text(24,-32)[cb]{$k$}
 \ArrowLine(20,-10)(40,-30)
 \GCirc(20,0){5}{0.5}
\end{picture}
}
\!+\!
\scalebox{0.7}{
\begin{picture}(40,40)(0,-3)
 \ArrowLine(20,10)(0,30)
 \ArrowLine(40,30)(20,10)
 \Photon(20,10)(20,-10){2}{5}
 \ArrowLine(0,-30)(20,-10)
 \ArrowLine(20,-10)(30,-20)
 \ArrowLine(30,-20)(40,-30)
 \Photon(30,-20)(20,-30){2}{4}        \Text(15.5,-32)[cb]{$k$}
 \GCirc(20,0){5}{0.5}
\end{picture}
}
\!+\!
\scalebox{0.7}{
\begin{picture}(40,40)(0,-3)
 \ArrowLine(20,10)(0,30)
 \ArrowLine(40,30)(30,20)
 \ArrowLine(30,20)(20,10)
 \Photon(30,20)(20,30){2}{4}        \Text(16,26)[cb]{$k$}
 \Photon(20,10)(20,-10){2}{5}
 \ArrowLine(0,-30)(20,-10)
 \ArrowLine(20,-10)(40,-30)
 \GCirc(20,0){5}{0.5}
\end{picture}
}
\Bigg)^*
\eqaa
\caption{
Diagrams with real photon emission.
The four-momentum of the real soft photon is $k=(k_0,\vec{k})$
and the integration is performed for $|\vec{k}|<\omega$. 
Here and below infrared regularization through a small photon mass $\lambda$ 
($\lambda^2=k_0^2-\vec{k}^2$) is implicitly understood.
}
\label{fig:real}
\end{figure}
\section{Details of the computation}
In this section we discuss these contributions in details.
In all equations containing products of Feynman diagrams,the  sum over the 
spins of the outgoing particles and the average over the spins of the incoming 
particles is implicit, as well as the conservation of the external momenta. 
In these formulas, the coefficient $c_{_{P\!S}}= (64s\pi^2)^{-1}$ comes from the 
integration of the phase-space of the outgoing electron and positron.
\subsection{Vacuum polarization insertion}
The Born cross section is obtained from the combination of $s$- and 
$t$-channel exchange:
\bq
\frac{d\sigma^0}{d\Omega} = 
\frac{\alpha^2}{s}\,\bigg[ 
  \bigg( \frac{1}{2} - x + x^2 \bigg) 
+ \frac{1}{x^2}\,\bigg( 1 - x + \frac{x^2}{2} \bigg) 
- \frac{1}{x}\,\big( 1 - 2\,x + x^2 \big) 
\bigg] =
\frac{\alpha^2}{s}\,\bigg( \frac{1-x+x^2}{x} \bigg)^2.
\eq
Replacing the photon propagator in the $s$- and $t$-channel by the dressed 
one, one obtains:
\bq
\frac{d\sigma^\Pi}{d\Omega} = 
\frac{\alpha^2}{s}\,\Bigg\{
  \frac{1 \!-\! 2x \!+\! 2x^2}{2}\,
  \bigg| \frac{1}{1\!-\!\Pi\,(s)} \bigg|^2
+ \,\frac{2 \!-\! 2x \!+\! x^2}{2\,x^2}\,
  \bigg| \frac{1}{1\!-\!\Pi\,(t)} \bigg|^2
- \,\frac{1 \!-\! 2x \!+\! x^2}{x}
  {\rm Re}\,\frac{1}{[1-\Pi\,(s)]\,[1\!-\!\Pi\,(t)]}
\Bigg\},
\label{eq:Pi1}
\eq
with $x=-t/s=(1-\cos\theta)/2$ and
\bq
\Pi(q^2)= \Pi_e(q^2) + \Pi_\mu(q^2) + \Pi_\tau(q^2) + \Pi_{\rm had}(q^2).
\eq
Expanding up to order $\alpha^2$, one easily recovers the Born contribution and 
the reducible one- and two-loop corrections:
\bqa
\frac{d\sigma^\Pi}{d\Omega} = 
\frac{\alpha^2}{s}\,\Bigg\{\!\!\!\! &&
  \frac{1 \!-\! 2x \!+\! 2x^2}{2}\Big[
    1 + 2{\rm Re}\Pi(s) + 3({\rm Re}\Pi(s))^2 - ({\rm Im}\Pi(s))^2 
  \Big]
+ \frac{2 \!-\! 2x \!+\! x^2}{2\,x^2}\,\Big[ 1 + 2\Pi(t) + 3\Pi(t)^2 \Big]
\nl
&&
- \,\frac{1 \!-\! 2x \!+\! x^2}{x}\Big[
    1 + {\rm Re}\Pi(s) + \Pi(t) 
  + ({\rm Re}\Pi(s))^2 - ({\rm Im}\Pi(s))^2 
  + {\rm Re}\Pi(s)\,\Pi(t) + \Pi(t)^2
  \Big]
\!\Bigg\}.
\label{eq:Pi2}
\eqa
Using for light leptons ($e$ and $\mu$)
\bq
\Pi_l(s)= 
\frac{\alpha}{3\pi}\bigg[ 
  \bigg( \ln\frac{s}{m_l^2} - \frac{5}{3} \bigg) - i\pi 
\bigg],
\qquad
\Pi_l(t)= \frac{\alpha}{3\pi}\bigg( \ln\frac{-t}{m_l^2} - \frac{5}{3} \bigg),
\qquad\quad
m_l= m_e,m_\mu,
\eq
the well known electron/muon induced one- and two-loop reducible contributions 
are easily recovered.
In the present context the two-loop terms involving hadrons arise from terms 
proportional to $\Pi_{\rm had}^2$ and $\Pi_{\rm had}\Pi_{\rm lept}$, with 
different combinations of real and imaginary parts.
\subsection{Reducible diagrams}
Contributions from one-loop photonic amplitudes, interfering with amplitudes 
with the dressed photon propagator in the s- or t-channel are infrared 
divergent and must be combined with real radiation.
For the amplitudes involving vertex corrections with have:
\bqa
\!\frac{d\sigma_{\rm red}^{\rm V\!,\,s}}{d\Omega}\!\!
&=&
c_{_{P\!S}}
2\,{\rm Re}
\Bigg[
\scalebox{0.55}{
\begin{picture}(50,20)(0,-3)
 \ArrowLine(15,0)(0,20)
 \ArrowLine(0,-20)(15,0)
 \Photon(15,0)(35,0){2}{5}
 \ArrowLine(50,20)(35,0)
 \ArrowLine(35,0)(50,-20)
\end{picture}
}
\Bigg(\!
\scalebox{0.55}{
\begin{picture}(55,20)(0,-3)
 \ArrowLine(20,0)(0,20)
 \ArrowLine(0,-20)(20,0)
 \Photon(20,0)(40,0){2}{5}
 \ArrowLine(55,20)(40,0)
 \ArrowLine(40,0)(55,-20)
 \GCirc(30,0){5}{0.5}
 \Photon(5,15)(5,-15){1.5}{8}
\end{picture}
}
+
\scalebox{0.55}{
\begin{picture}(55,20)(0,-3)
 \ArrowLine(15,0)(0,20)
 \ArrowLine(0,-20)(15,0)
 \Photon(15,0)(35,0){2}{5}
 \ArrowLine(55,20)(35,0)
 \ArrowLine(35,0)(55,-20)
 \GCirc(25,0){5}{0.5}
 \Photon(50,15)(50,-15){1.5}{8}
\end{picture}
}
\!\Bigg)^{\!\!*}
\!
+
\scalebox{0.55}{
\begin{picture}(50,20)(0,-3)
 \ArrowLine(15,0)(0,20)
 \ArrowLine(0,-20)(15,0)
 \Photon(15,0)(35,0){2}{5}
 \ArrowLine(50,20)(35,0)
 \ArrowLine(35,0)(50,-20)
 \GCirc(25,0){5}{0.5}
\end{picture}
}
\Bigg(\!
\scalebox{0.55}{
\begin{picture}(55,20)(0,-3)
 \ArrowLine(20,0)(0,20)
 \ArrowLine(0,-20)(20,0)
 \Photon(20,0)(40,0){2}{5}
 \ArrowLine(55,20)(40,0)
 \ArrowLine(40,0)(55,-20)
 \Photon(5,15)(5,-15){1.5}{8}
\end{picture}
}
+
\scalebox{0.55}{
\begin{picture}(55,20)(0,-3)
 \ArrowLine(15,0)(0,20)
 \ArrowLine(0,-20)(15,0)
 \Photon(15,0)(35,0){2}{5}
 \ArrowLine(55,20)(35,0)
 \ArrowLine(35,0)(55,-20)
 \Photon(50,15)(50,-15){1.5}{8}
\end{picture}
}
\!\Bigg)^{\!\!*}
\Bigg]
\nl
&&
\!+\,
\frac{c_{_{P\!S}}}{(2\pi)^3}\!\!\int_{\omega}\!\frac{d^3k}{2\,k_0}\;
2\,{\rm Re}
\Bigg[
\Bigg(\!\!
\scalebox{0.55}{
\begin{picture}(50,20)(0,-3)
 \ArrowLine(15,0)(7.5,10)
 \ArrowLine(7.5,10)(0,20)
 \Photon(7.5,10)(20,20){2}{4}        \Text(25,16)[cb]{$k$}
 \ArrowLine(0,-20)(15,0)
 \Photon(15,0)(35,0){2}{5}
 \ArrowLine(50,20)(35,0)
 \ArrowLine(35,0)(50,-20)
\end{picture}
}
\!\!+\!\!
\scalebox{0.55}{
\begin{picture}(50,20)(0,-3)
 \ArrowLine(15,0)(0,20)
 \ArrowLine(0,-20)(7.5,-10)
 \ArrowLine(7.5,-10)(15,0)
 \Photon(7.5,-10)(20,-20){2}{4}        \Text(24,-22)[cb]{$k$}
 \Photon(15,0)(35,0){2}{5}
 \ArrowLine(50,20)(35,0)
 \ArrowLine(35,0)(50,-20)
\end{picture}
}
\!\Bigg)\!\!
\Bigg(\!\!
\scalebox{0.55}{
\begin{picture}(50,20)(0,-3)
 \ArrowLine(15,0)(7.5,10)
 \ArrowLine(7.5,10)(0,20)
 \Photon(7.5,10)(20,20){2}{4}        \Text(25,16)[cb]{$k$}
 \ArrowLine(0,-20)(15,0)
 \Photon(15,0)(35,0){2}{5}
 \ArrowLine(50,20)(35,0)
 \ArrowLine(35,0)(50,-20)
 \GCirc(25,0){5}{0.5}
\end{picture}
}
\!\!+\!\!
\scalebox{0.55}{
\begin{picture}(50,20)(0,-3)
 \ArrowLine(15,0)(0,20)
 \ArrowLine(0,-20)(7.5,-10)
 \ArrowLine(7.5,-10)(15,0)
 \Photon(7.5,-10)(20,-20){2}{4}        \Text(24,-22)[cb]{$k$}
 \Photon(15,0)(35,0){2}{5}
 \ArrowLine(50,20)(35,0)
 \ArrowLine(35,0)(50,-20)
 \GCirc(25,0){5}{0.5}
\end{picture}
}
\!\Bigg)^{\!\!\!*}
\!\!+\!
\Bigg(\!\!
\scalebox{0.55}{
\begin{picture}(50,20)(0,-3)
 \ArrowLine(15,0)(0,20)
 \ArrowLine(0,-20)(15,0)
 \Photon(15,0)(35,0){2}{5}
 \ArrowLine(50,20)(35,0)
 \ArrowLine(35,0)(42.5,-10)
 \ArrowLine(42.5,-10)(50,-20)
 \Photon(42.5,-10)(35,-20){2}{4}        \Text(29.5,-22)[cb]{$k$}
\end{picture}
}
\!\!+\!\!
\scalebox{0.55}{
\begin{picture}(50,20)(0,-3)
 \ArrowLine(15,0)(0,20)
 \ArrowLine(0,-20)(15,0)
 \Photon(15,0)(35,0){2}{5}
 \ArrowLine(50,20)(42.5,10)
 \ArrowLine(42.5,10)(35,0)
 \Photon(42.5,10)(35,20){2}{4}        \Text(29.5,16)[cb]{$k$}
 \ArrowLine(35,0)(50,-20)
\end{picture}
}
\!\Bigg)\!\!
\Bigg(\!\!
\scalebox{0.55}{
\begin{picture}(50,20)(0,-3)
 \ArrowLine(15,0)(0,20)
 \ArrowLine(0,-20)(15,0)
 \Photon(15,0)(35,0){2}{5}
 \ArrowLine(50,20)(35,0)
 \ArrowLine(35,0)(42.5,-10)
 \ArrowLine(42.5,-10)(50,-20)
 \Photon(42.5,-10)(35,-20){2}{4}        \Text(29.5,-22)[cb]{$k$}
 \GCirc(25,0){5}{0.5}
\end{picture}
}
\!\!+\!\!
\scalebox{0.55}{
\begin{picture}(50,20)(0,-3)
 \ArrowLine(15,0)(0,20)
 \ArrowLine(0,-20)(15,0)
 \Photon(15,0)(35,0){2}{5}
 \ArrowLine(50,20)(42.5,10)
 \ArrowLine(42.5,10)(35,0)
 \Photon(42.5,10)(35,20){2}{4}        \Text(29.5,16)[cb]{$k$}
 \ArrowLine(35,0)(50,-20)
 \GCirc(25,0){5}{0.5}
\end{picture}
}
\!\Bigg)^{\!\!\!*}
\Bigg],
\qquad
\eqa
\bqa
\!\frac{d\sigma_{\rm red}^{\rm V\!,\,t}}{d\Omega}\!\!
&=&
c_{_{P\!S}}
2\,{\rm Re}
\Bigg[
\scalebox{0.55}{
\begin{picture}(40,30)(0,-3)
 \ArrowLine(20,10)(0,30)
 \ArrowLine(40,30)(20,10)
 \Photon(20,10)(20,-10){2}{5}
 \ArrowLine(0,-30)(20,-10)
 \ArrowLine(20,-10)(40,-30)
\end{picture}
}
\Bigg(\!
\scalebox{0.55}{
\begin{picture}(40,40)(0,-3)
 \ArrowLine(20,10)(0,30)
 \ArrowLine(40,30)(20,10)
 \Photon(20,10)(20,-10){2}{5}
 \ArrowLine(0,-30)(20,-10)
 \ArrowLine(20,-10)(40,-30)
 \GCirc(20,0){5}{0.5}
 \Photon(5,25)(35,25){1.5}{8}
\end{picture}
}
+
\scalebox{0.55}{
\begin{picture}(40,40)(0,-3)
 \ArrowLine(20,10)(0,30)
 \ArrowLine(40,30)(20,10)
 \Photon(20,10)(20,-10){2}{5}
 \ArrowLine(0,-30)(20,-10)
 \ArrowLine(20,-10)(40,-30)
 \GCirc(20,0){5}{0.5}
 \Photon(5,-25)(35,-25){1.5}{8}
\end{picture}
}
\!\Bigg)^{\!\!*}
+
\scalebox{0.55}{
\begin{picture}(40,40)(0,-3)
 \ArrowLine(20,10)(0,30)
 \ArrowLine(40,30)(20,10)
 \Photon(20,10)(20,-10){2}{5}
 \ArrowLine(0,-30)(20,-10)
 \ArrowLine(20,-10)(40,-30)
 \GCirc(20,0){5}{0.5}
\end{picture}
}
\Bigg(\!
\scalebox{0.55}{
\begin{picture}(40,40)(0,-3)
 \ArrowLine(20,10)(0,30)
 \ArrowLine(40,30)(20,10)
 \Photon(20,10)(20,-10){2}{5}
 \ArrowLine(0,-30)(20,-10)
 \ArrowLine(20,-10)(40,-30)
 \Photon(5,25)(35,25){1.5}{8}
\end{picture}
}
+
\scalebox{0.55}{
\begin{picture}(40,40)(0,-3)
 \ArrowLine(20,10)(0,30)
 \ArrowLine(40,30)(20,10)
 \Photon(20,10)(20,-10){2}{5}
 \ArrowLine(0,-30)(20,-10)
 \ArrowLine(20,-10)(40,-30)
 \Photon(5,-25)(35,-25){1.5}{8}
\end{picture}
}
\!\Bigg)^{\!\!*}
\Bigg]
\nl
&&
+\,
\frac{c_{_{P\!S}}}{(2\pi)^3}\!\!\int_{\omega}\!\frac{d^3k}{2\,k_0}\;
2\,{\rm Re}
\Bigg[
\Bigg(\!
\scalebox{0.55}{
\begin{picture}(40,40)(0,-3)
 \ArrowLine(20,10)(10,20)
 \ArrowLine(10,20)(0,30)
 \Photon(10,20)(20,30){2}{4}        \Text(25,26)[cb]{$k$}
 \ArrowLine(40,30)(20,10)
 \Photon(20,10)(20,-10){2}{5}
 \ArrowLine(0,-30)(20,-10)
 \ArrowLine(20,-10)(40,-30)
\end{picture}
}
\!\!+\!\!
\scalebox{0.55}{
\begin{picture}(40,40)(0,-3)
 \ArrowLine(20,10)(0,30)
 \ArrowLine(40,30)(30,20)
 \ArrowLine(30,20)(20,10)
 \Photon(30,20)(20,30){2}{4}        \Text(16,26)[cb]{$k$}
 \Photon(20,10)(20,-10){2}{5}
 \ArrowLine(0,-30)(20,-10)
 \ArrowLine(20,-10)(40,-30)
\end{picture}
}
\!\Bigg)\!
\Bigg(\!
\scalebox{0.55}{
\begin{picture}(40,40)(0,-3)
 \ArrowLine(20,10)(10,20)
 \ArrowLine(10,20)(0,30)
 \Photon(10,20)(20,30){2}{4}        \Text(25,26)[cb]{$k$}
 \ArrowLine(40,30)(20,10)
 \Photon(20,10)(20,-10){2}{5}
 \ArrowLine(0,-30)(20,-10)
 \ArrowLine(20,-10)(40,-30)
 \GCirc(20,0){5}{0.5}
\end{picture}
}
\!\!+\!\!
\scalebox{0.55}{
\begin{picture}(40,40)(0,-3)
 \ArrowLine(20,10)(0,30)
 \ArrowLine(40,30)(30,20)
 \ArrowLine(30,20)(20,10)
 \Photon(30,20)(20,30){2}{4}        \Text(16,26)[cb]{$k$}
 \Photon(20,10)(20,-10){2}{5}
 \ArrowLine(0,-30)(20,-10)
 \ArrowLine(20,-10)(40,-30)
 \GCirc(20,0){5}{0.5}
\end{picture}
}
\!\Bigg)^{\!\!*}
\!\!
+
\Bigg(\!
\scalebox{0.55}{
\begin{picture}(40,40)(0,-3)
 \ArrowLine(20,10)(0,30)
 \ArrowLine(40,30)(20,10)
 \Photon(20,10)(20,-10){2}{5}
 \ArrowLine(0,-30)(10,-20)
 \ArrowLine(10,-20)(20,-10)
 \Photon(10,-20)(20,-30){2}{4}        \Text(24,-32)[cb]{$k$}
 \ArrowLine(20,-10)(40,-30)
\end{picture}
}
\!\!+\!\!
\scalebox{0.55}{
\begin{picture}(40,40)(0,-3)
 \ArrowLine(20,10)(0,30)
 \ArrowLine(40,30)(20,10)
 \Photon(20,10)(20,-10){2}{5}
 \ArrowLine(0,-30)(20,-10)
 \ArrowLine(20,-10)(30,-20)
 \ArrowLine(30,-20)(40,-30)
 \Photon(30,-20)(20,-30){2}{4}        \Text(15.5,-32)[cb]{$k$}
\end{picture}
}
\!\Bigg)\!
\Bigg(\!
\scalebox{0.55}{
\begin{picture}(40,40)(0,-3)
 \ArrowLine(20,10)(0,30)
 \ArrowLine(40,30)(20,10)
 \Photon(20,10)(20,-10){2}{5}
 \ArrowLine(0,-30)(10,-20)
 \ArrowLine(10,-20)(20,-10)
 \Photon(10,-20)(20,-30){2}{4}        \Text(24,-32)[cb]{$k$}
 \ArrowLine(20,-10)(40,-30)
 \GCirc(20,0){5}{0.5}
\end{picture}
}
\!\!+\!\!
\scalebox{0.55}{
\begin{picture}(40,40)(0,-3)
 \ArrowLine(20,10)(0,30)
 \ArrowLine(40,30)(20,10)
 \Photon(20,10)(20,-10){2}{5}
 \ArrowLine(0,-30)(20,-10)
 \ArrowLine(20,-10)(30,-20)
 \ArrowLine(30,-20)(40,-30)
 \Photon(30,-20)(20,-30){2}{4}        \Text(15.5,-32)[cb]{$k$}
 \GCirc(20,0){5}{0.5}
\end{picture}
}
\!\Bigg)^{\!\!*}
\Bigg],
\eqa
\bqa
\!\frac{d\sigma_{\rm red}^{\rm V\!,\,st\!\!}}{d\Omega}\!\!
&=&
c_{_{\!P\!S}}
2 {\rm Re}
\Bigg[
\scalebox{0.55}{
\begin{picture}(45,20)(0,-3)
 \ArrowLine(15,0)(0,20)
 \ArrowLine(0,-20)(15,0)
 \Photon(15,0)(30,0){2}{4}
 \ArrowLine(45,20)(30,0)
 \ArrowLine(30,0)(45,-20)
\end{picture}
}\!\!
\Bigg(\!
\scalebox{0.55}{
\begin{picture}(40,40)(0,-3)
 \ArrowLine(20,10)(0,30)
 \ArrowLine(40,30)(20,10)
 \Photon(20,10)(20,-10){2}{5}
 \ArrowLine(0,-30)(20,-10)
 \ArrowLine(20,-10)(40,-30)
 \GCirc(20,0){5}{0.5}
 \Photon(5,25)(35,25){1.5}{8}
\end{picture}
}
\!\!\!\!+\!\!\!\!
\scalebox{0.55}{
\begin{picture}(40,40)(0,-3)
 \ArrowLine(20,10)(0,30)
 \ArrowLine(40,30)(20,10)
 \Photon(20,10)(20,-10){2}{5}
 \ArrowLine(0,-30)(20,-10)
 \ArrowLine(20,-10)(40,-30)
 \GCirc(20,0){5}{0.5}
 \Photon(5,-25)(35,-25){1.5}{8}
\end{picture}
}
\!\Bigg)^{\!\!\!*}
\!+\!\!\!\!
\scalebox{0.55}{
\begin{picture}(40,30)(0,-3)
 \ArrowLine(20,10)(0,30)
 \ArrowLine(40,30)(20,10)
 \Photon(20,10)(20,-10){2}{5}
 \ArrowLine(0,-30)(20,-10)
 \ArrowLine(20,-10)(40,-30)
\end{picture}
}\!\!\!
\Bigg(\!\!
\scalebox{0.55}{
\begin{picture}(55,20)(0,-3)
 \ArrowLine(20,0)(0,20)
 \ArrowLine(0,-20)(20,0)
 \Photon(20,0)(40,0){2}{5}
 \ArrowLine(55,20)(40,0)
 \ArrowLine(40,0)(55,-20)
 \GCirc(25,0){5}{0.5}
 \Photon(5,15)(5,-15){1.5}{8}
\end{picture}
}
\!\!\!+\!\!\!
\scalebox{0.55}{
\begin{picture}(55,20)(0,-3)
 \ArrowLine(15,0)(0,20)
 \ArrowLine(0,-20)(15,0)
 \Photon(15,0)(35,0){2}{5}
 \ArrowLine(55,20)(35,0)
 \ArrowLine(35,0)(55,-20)
 \GCirc(25,0){5}{0.5}
 \Photon(50,15)(50,-15){1.5}{8}
\end{picture}
}
\!\Bigg)^{\!\!\!*}
\!+\!\!
\scalebox{0.55}{
\begin{picture}(50,20)(0,-3)
 \ArrowLine(15,0)(0,20)
 \ArrowLine(0,-20)(15,0)
 \Photon(15,0)(35,0){2}{5}
 \ArrowLine(50,20)(35,0)
 \ArrowLine(35,0)(50,-20)
 \GCirc(25,0){5}{0.5}
\end{picture}
}\!\!
\Bigg(\!
\scalebox{0.55}{
\begin{picture}(40,40)(0,-3)
 \ArrowLine(20,10)(0,30)
 \ArrowLine(40,30)(20,10)
 \Photon(20,10)(20,-10){2}{5}
 \ArrowLine(0,-30)(20,-10)
 \ArrowLine(20,-10)(40,-30)
 \Photon(5,25)(35,25){1.5}{8}
\end{picture}
}
\!\!\!\!+\!\!\!\!
\scalebox{0.55}{
\begin{picture}(40,40)(0,-3)
 \ArrowLine(20,10)(0,30)
 \ArrowLine(40,30)(20,10)
 \Photon(20,10)(20,-10){2}{5}
 \ArrowLine(0,-30)(20,-10)
 \ArrowLine(20,-10)(40,-30)
 \Photon(5,-25)(35,-25){1.5}{8}
\end{picture}
}
\!\Bigg)^{\!\!\!*}
\!+\!\!\!\!
\scalebox{0.55}{
\begin{picture}(40,40)(0,-3)
 \ArrowLine(20,10)(0,30)
 \ArrowLine(40,30)(20,10)
 \Photon(20,10)(20,-10){2}{5}
 \ArrowLine(0,-30)(20,-10)
 \ArrowLine(20,-10)(40,-30)
 \GCirc(20,0){5}{0.5}
\end{picture}
}\!\!\!
\Bigg(\!\!
\scalebox{0.55}{
\begin{picture}(50,20)(0,-3)
 \ArrowLine(20,0)(0,20)
 \ArrowLine(0,-20)(20,0)
 \Photon(20,0)(35,0){2}{4}
 \ArrowLine(50,20)(35,0)
 \ArrowLine(35,0)(50,-20)
 \Photon(5,15)(5,-15){1.5}{8}
\end{picture}
}
\!\!\!+\!\!\!
\scalebox{0.55}{
\begin{picture}(50,20)(0,-3)
 \ArrowLine(15,0)(0,20)
 \ArrowLine(0,-20)(15,0)
 \Photon(15,0)(30,0){2}{4}
 \ArrowLine(50,20)(30,0)
 \ArrowLine(30,0)(50,-20)
 \Photon(45,15)(45,-15){1.5}{8}
\end{picture}
}
\!\Bigg)^{\!\!\!*}
\Bigg]
\nl
&&
+\,
\frac{c_{_{P\!S}}}{(2\pi)^3}\!\!\int_{\omega}\!\frac{d^3k}{2\,k_0}\;
{\rm Re}
\Bigg[
\Bigg(\!
\scalebox{0.55}{
\begin{picture}(50,20)(0,-3)
 \ArrowLine(15,0)(7.5,10)
 \ArrowLine(7.5,10)(0,20)
 \Photon(7.5,10)(20,20){2}{4}        \Text(25,16)[cb]{$k$}
 \ArrowLine(0,-20)(15,0)
 \Photon(15,0)(35,0){2}{5}
 \ArrowLine(50,20)(35,0)
 \ArrowLine(35,0)(50,-20)
\end{picture}
}
\!\!+\!\!
\scalebox{0.55}{
\begin{picture}(50,20)(0,-3)
 \ArrowLine(15,0)(0,20)
 \ArrowLine(0,-20)(15,0)
 \Photon(15,0)(35,0){2}{5}
 \ArrowLine(50,20)(42.5,10)
 \ArrowLine(42.5,10)(35,0)
 \Photon(42.5,10)(35,20){2}{4}        \Text(29.5,16)[cb]{$k$}
 \ArrowLine(35,0)(50,-20)
\end{picture}
}
\!\Bigg)
\Bigg(\!
\scalebox{0.55}{
\begin{picture}(40,40)(0,-3)
 \ArrowLine(20,10)(10,20)
 \ArrowLine(10,20)(0,30)
 \Photon(10,20)(20,30){2}{4}        \Text(25,26)[cb]{$k$}
 \ArrowLine(40,30)(20,10)
 \Photon(20,10)(20,-10){2}{5}
 \ArrowLine(0,-30)(20,-10)
 \ArrowLine(20,-10)(40,-30)
 \GCirc(20,0){5}{0.5}
\end{picture}
}
\!\!+\!\!
\scalebox{0.55}{
\begin{picture}(40,40)(0,-3)
 \ArrowLine(20,10)(0,30)
 \ArrowLine(40,30)(30,20)
 \ArrowLine(30,20)(20,10)
 \Photon(30,20)(20,30){2}{4}        \Text(16,26)[cb]{$k$}
 \Photon(20,10)(20,-10){2}{5}
 \ArrowLine(0,-30)(20,-10)
 \ArrowLine(20,-10)(40,-30)
 \GCirc(20,0){5}{0.5}
\end{picture}
}
\!\Bigg)^{\!\!*}
\!\!
+
\Bigg(\!
\scalebox{0.55}{
\begin{picture}(50,20)(0,-3)
 \ArrowLine(15,0)(0,20)
 \ArrowLine(0,-20)(7.5,-10)
 \ArrowLine(7.5,-10)(15,0)
 \Photon(7.5,-10)(20,-20){2}{4}        \Text(24,-22)[cb]{$k$}
 \Photon(15,0)(35,0){2}{5}
 \ArrowLine(50,20)(35,0)
 \ArrowLine(35,0)(50,-20)
\end{picture}
}
\!\!+\!\!
\scalebox{0.55}{
\begin{picture}(50,20)(0,-3)
 \ArrowLine(15,0)(0,20)
 \ArrowLine(0,-20)(15,0)
 \Photon(15,0)(35,0){2}{5}
 \ArrowLine(50,20)(35,0)
 \ArrowLine(35,0)(42.5,-10)
 \ArrowLine(42.5,-10)(50,-20)
 \Photon(42.5,-10)(35,-20){2}{4}        \Text(29.5,-22)[cb]{$k$}
\end{picture}
}
\!\Bigg)
\Bigg(\!
\scalebox{0.55}{
\begin{picture}(40,40)(0,-3)
 \ArrowLine(20,10)(0,30)
 \ArrowLine(40,30)(20,10)
 \Photon(20,10)(20,-10){2}{5}
 \ArrowLine(0,-30)(10,-20)
 \ArrowLine(10,-20)(20,-10)
 \Photon(10,-20)(20,-30){2}{4}        \Text(24,-32)[cb]{$k$}
 \ArrowLine(20,-10)(40,-30)
 \GCirc(20,0){5}{0.5}
\end{picture}
}
\!\!+\!\!
\scalebox{0.55}{
\begin{picture}(40,40)(0,-3)
 \ArrowLine(20,10)(0,30)
 \ArrowLine(40,30)(20,10)
 \Photon(20,10)(20,-10){2}{5}
 \ArrowLine(0,-30)(20,-10)
 \ArrowLine(20,-10)(30,-20)
 \ArrowLine(30,-20)(40,-30)
 \Photon(30,-20)(20,-30){2}{4}        \Text(15.5,-32)[cb]{$k$}
 \GCirc(20,0){5}{0.5}
\end{picture}
}
\!\Bigg)^{\!\!*}
\nl
&&
\qquad\qquad\qquad\qquad
+
\Bigg(\!
\scalebox{0.55}{
\begin{picture}(40,40)(0,-3)
 \ArrowLine(20,10)(10,20)
 \ArrowLine(10,20)(0,30)
 \Photon(10,20)(20,30){2}{4}        \Text(25,26)[cb]{$k$}
 \ArrowLine(40,30)(20,10)
 \Photon(20,10)(20,-10){2}{5}
 \ArrowLine(0,-30)(20,-10)
 \ArrowLine(20,-10)(40,-30)
\end{picture}
}
\!\!+\!\!
\scalebox{0.55}{
\begin{picture}(40,40)(0,-3)
 \ArrowLine(20,10)(0,30)
 \ArrowLine(40,30)(20,10)
 \Photon(20,10)(20,-10){2}{5}
 \ArrowLine(0,-30)(10,-20)
 \ArrowLine(10,-20)(20,-10)
 \Photon(10,-20)(20,-30){2}{4}        \Text(24,-32)[cb]{$k$}
 \ArrowLine(20,-10)(40,-30)
\end{picture}
}
\!\Bigg)
\Bigg(\!
\scalebox{0.55}{
\begin{picture}(50,20)(0,-3)
 \ArrowLine(15,0)(7.5,10)
 \ArrowLine(7.5,10)(0,20)
 \Photon(7.5,10)(20,20){2}{4}        \Text(25,16)[cb]{$k$}
 \ArrowLine(0,-20)(15,0)
 \Photon(15,0)(35,0){2}{5}
 \ArrowLine(50,20)(35,0)
 \ArrowLine(35,0)(50,-20)
 \GCirc(25,0){5}{0.5}
\end{picture}
}
\!\!+\!\!
\scalebox{0.55}{
\begin{picture}(50,20)(0,-3)
 \ArrowLine(15,0)(0,20)
 \ArrowLine(0,-20)(7.5,-10)
 \ArrowLine(7.5,-10)(15,0)
 \Photon(7.5,-10)(20,-20){2}{4}        \Text(24,-22)[cb]{$k$}
 \Photon(15,0)(35,0){2}{5}
 \ArrowLine(50,20)(35,0)
 \ArrowLine(35,0)(50,-20)
 \GCirc(25,0){5}{0.5}
\end{picture}
}
\!\Bigg)^{\!\!*}
\!\!
+
\Bigg(\!
\scalebox{0.55}{
\begin{picture}(40,40)(0,-3)
 \ArrowLine(20,10)(0,30)
 \ArrowLine(40,30)(20,10)
 \Photon(20,10)(20,-10){2}{5}
 \ArrowLine(0,-30)(20,-10)
 \ArrowLine(20,-10)(30,-20)
 \ArrowLine(30,-20)(40,-30)
 \Photon(30,-20)(20,-30){2}{4}        \Text(15.5,-32)[cb]{$k$}
\end{picture}
}
\!\!+\!\!
\scalebox{0.55}{
\begin{picture}(40,40)(0,-3)
 \ArrowLine(20,10)(0,30)
 \ArrowLine(40,30)(30,20)
 \ArrowLine(30,20)(20,10)
 \Photon(30,20)(20,30){2}{4}        \Text(16,26)[cb]{$k$}
 \Photon(20,10)(20,-10){2}{5}
 \ArrowLine(0,-30)(20,-10)
 \ArrowLine(20,-10)(40,-30)
\end{picture}
}
\!\Bigg)
\Bigg(\!
\scalebox{0.55}{
\begin{picture}(50,20)(0,-3)
 \ArrowLine(15,0)(0,20)
 \ArrowLine(0,-20)(15,0)
 \Photon(15,0)(35,0){2}{5}
 \ArrowLine(50,20)(35,0)
 \ArrowLine(35,0)(42.5,-10)
 \ArrowLine(42.5,-10)(50,-20)
 \Photon(42.5,-10)(35,-20){2}{4}        \Text(29.5,-22)[cb]{$k$}
 \GCirc(25,0){5}{0.5}
\end{picture}
}
\!\!+\!\!
\scalebox{0.55}{
\begin{picture}(50,20)(0,-3)
 \ArrowLine(15,0)(0,20)
 \ArrowLine(0,-20)(15,0)
 \Photon(15,0)(35,0){2}{5}
 \ArrowLine(50,20)(42.5,10)
 \ArrowLine(42.5,10)(35,0)
 \Photon(42.5,10)(35,20){2}{4}        \Text(29.5,16)[cb]{$k$}
 \ArrowLine(35,0)(50,-20)
 \GCirc(25,0){5}{0.5}
\end{picture}
}
\!\Bigg)^{\!\!*}
\nl
&&
\qquad\qquad\qquad\qquad
+
\Bigg(\!
\scalebox{0.55}{
\begin{picture}(40,40)(0,-3)
 \ArrowLine(20,10)(10,20)
 \ArrowLine(10,20)(0,30)
 \Photon(10,20)(20,30){2}{4}        \Text(25,26)[cb]{$k$}
 \ArrowLine(40,30)(20,10)
 \Photon(20,10)(20,-10){2}{5}
 \ArrowLine(0,-30)(20,-10)
 \ArrowLine(20,-10)(40,-30)
\end{picture}
}
\scalebox{0.55}{
\begin{picture}(40,40)(0,-3)
 \ArrowLine(20,10)(0,30)
 \ArrowLine(40,30)(30,20)
 \ArrowLine(30,20)(20,10)
 \Photon(30,20)(20,30){2}{4}        \Text(16,26)[cb]{$k$}
 \Photon(20,10)(20,-10){2}{5}
 \ArrowLine(0,-30)(20,-10)
 \ArrowLine(20,-10)(40,-30)
\end{picture}
}
\!\Bigg)
\Bigg(\!
\scalebox{0.55}{
\begin{picture}(50,20)(0,-3)
 \ArrowLine(15,0)(7.5,10)
 \ArrowLine(7.5,10)(0,20)
 \Photon(7.5,10)(20,20){2}{4}        \Text(25,16)[cb]{$k$}
 \ArrowLine(0,-20)(15,0)
 \Photon(15,0)(35,0){2}{5}
 \ArrowLine(50,20)(35,0)
 \ArrowLine(35,0)(50,-20)
 \GCirc(25,0){5}{0.5}
\end{picture}
}
\!\!+\!\!
\scalebox{0.55}{
\begin{picture}(50,20)(0,-3)
 \ArrowLine(15,0)(0,20)
 \ArrowLine(0,-20)(15,0)
 \Photon(15,0)(35,0){2}{5}
 \ArrowLine(50,20)(42.5,10)
 \ArrowLine(42.5,10)(35,0)
 \Photon(42.5,10)(35,20){2}{4}        \Text(29.5,16)[cb]{$k$}
 \ArrowLine(35,0)(50,-20)
 \GCirc(25,0){5}{0.5}
\end{picture}
}
\!\Bigg)^{\!\!*}
\!\!
+
\Bigg(\!
\scalebox{0.55}{
\begin{picture}(40,40)(0,-3)
 \ArrowLine(20,10)(0,30)
 \ArrowLine(40,30)(20,10)
 \Photon(20,10)(20,-10){2}{5}
 \ArrowLine(0,-30)(10,-20)
 \ArrowLine(10,-20)(20,-10)
 \Photon(10,-20)(20,-30){2}{4}        \Text(24,-32)[cb]{$k$}
 \ArrowLine(20,-10)(40,-30)
\end{picture}
}
\!\!+\!\!
\scalebox{0.55}{
\begin{picture}(40,40)(0,-3)
 \ArrowLine(20,10)(0,30)
 \ArrowLine(40,30)(20,10)
 \Photon(20,10)(20,-10){2}{5}
 \ArrowLine(0,-30)(20,-10)
 \ArrowLine(20,-10)(30,-20)
 \ArrowLine(30,-20)(40,-30)
 \Photon(30,-20)(20,-30){2}{4}        \Text(15.5,-32)[cb]{$k$}
\end{picture}
}
\!\Bigg)
\Bigg(\!
\scalebox{0.55}{
\begin{picture}(50,20)(0,-3)
 \ArrowLine(15,0)(0,20)
 \ArrowLine(0,-20)(7.5,-10)
 \ArrowLine(7.5,-10)(15,0)
 \Photon(7.5,-10)(20,-20){2}{4}        \Text(24,-22)[cb]{$k$}
 \Photon(15,0)(35,0){2}{5}
 \ArrowLine(50,20)(35,0)
 \ArrowLine(35,0)(50,-20)
 \GCirc(25,0){5}{0.5}
\end{picture}
}
\!\!+\!\!
\scalebox{0.55}{
\begin{picture}(50,20)(0,-3)
 \ArrowLine(15,0)(0,20)
 \ArrowLine(0,-20)(15,0)
 \Photon(15,0)(35,0){2}{5}
 \ArrowLine(50,20)(35,0)
 \ArrowLine(35,0)(42.5,-10)
 \ArrowLine(42.5,-10)(50,-20)
 \Photon(42.5,-10)(35,-20){2}{4}        \Text(29.5,-22)[cb]{$k$}
 \GCirc(25,0){5}{0.5}
\end{picture}
}
\!\Bigg)^{\!\!*}
\nl
&&
\qquad\qquad\qquad\qquad
+
\Bigg(\!
\scalebox{0.55}{
\begin{picture}(50,20)(0,-3)
 \ArrowLine(15,0)(7.5,10)
 \ArrowLine(7.5,10)(0,20)
 \Photon(7.5,10)(20,20){2}{4}        \Text(25,16)[cb]{$k$}
 \ArrowLine(0,-20)(15,0)
 \Photon(15,0)(35,0){2}{5}
 \ArrowLine(50,20)(35,0)
 \ArrowLine(35,0)(50,-20)
\end{picture}
}
\!\!+\!\!
\scalebox{0.55}{
\begin{picture}(50,20)(0,-3)
 \ArrowLine(15,0)(0,20)
 \ArrowLine(0,-20)(7.5,-10)
 \ArrowLine(7.5,-10)(15,0)
 \Photon(7.5,-10)(20,-20){2}{4}        \Text(24,-22)[cb]{$k$}
 \Photon(15,0)(35,0){2}{5}
 \ArrowLine(50,20)(35,0)
 \ArrowLine(35,0)(50,-20)
\end{picture}
}
\!\Bigg)
\Bigg(\!
\scalebox{0.55}{
\begin{picture}(40,40)(0,-3)
 \ArrowLine(20,10)(10,20)
 \ArrowLine(10,20)(0,30)
 \Photon(10,20)(20,30){2}{4}        \Text(25,26)[cb]{$k$}
 \ArrowLine(40,30)(20,10)
 \Photon(20,10)(20,-10){2}{5}
 \ArrowLine(0,-30)(20,-10)
 \ArrowLine(20,-10)(40,-30)
 \GCirc(20,0){5}{0.5}
\end{picture}
}
\!\!+\!\!
\scalebox{0.55}{
\begin{picture}(40,40)(0,-3)
 \ArrowLine(20,10)(0,30)
 \ArrowLine(40,30)(20,10)
 \Photon(20,10)(20,-10){2}{5}
 \ArrowLine(0,-30)(10,-20)
 \ArrowLine(10,-20)(20,-10)
 \Photon(10,-20)(20,-30){2}{4}        \Text(24,-32)[cb]{$k$}
 \ArrowLine(20,-10)(40,-30)
 \GCirc(20,0){5}{0.5}
\end{picture}
}
\!\Bigg)^{\!\!*}
\!\!
+
\Bigg(\!
\scalebox{0.55}{
\begin{picture}(50,20)(0,-3)
 \ArrowLine(15,0)(0,20)
 \ArrowLine(0,-20)(15,0)
 \Photon(15,0)(35,0){2}{5}
 \ArrowLine(50,20)(35,0)
 \ArrowLine(35,0)(42.5,-10)
 \ArrowLine(42.5,-10)(50,-20)
 \Photon(42.5,-10)(35,-20){2}{4}        \Text(29.5,-22)[cb]{$k$}
\end{picture}
}
\!\!+\!\!
\scalebox{0.55}{
\begin{picture}(50,20)(0,-3)
 \ArrowLine(15,0)(0,20)
 \ArrowLine(0,-20)(15,0)
 \Photon(15,0)(35,0){2}{5}
 \ArrowLine(50,20)(42.5,10)
 \ArrowLine(42.5,10)(35,0)
 \Photon(42.5,10)(35,20){2}{4}        \Text(29.5,16)[cb]{$k$}
 \ArrowLine(35,0)(50,-20)
\end{picture}
}
\!\Bigg)
\Bigg(\!
\scalebox{0.55}{
\begin{picture}(40,40)(0,-3)
 \ArrowLine(20,10)(0,30)
 \ArrowLine(40,30)(20,10)
 \Photon(20,10)(20,-10){2}{5}
 \ArrowLine(0,-30)(20,-10)
 \ArrowLine(20,-10)(30,-20)
 \ArrowLine(30,-20)(40,-30)
 \Photon(30,-20)(20,-30){2}{4}        \Text(15.5,-32)[cb]{$k$}
 \GCirc(20,0){5}{0.5}
\end{picture}
}
\!\!+\!\!
\scalebox{0.55}{
\begin{picture}(40,40)(0,-3)
 \ArrowLine(20,10)(0,30)
 \ArrowLine(40,30)(30,20)
 \ArrowLine(30,20)(20,10)
 \Photon(30,20)(20,30){2}{4}        \Text(16,26)[cb]{$k$}
 \Photon(20,10)(20,-10){2}{5}
 \ArrowLine(0,-30)(20,-10)
 \ArrowLine(20,-10)(40,-30)
 \GCirc(20,0){5}{0.5}
\end{picture}
}
\!\Bigg)^{\!\!*}
\Bigg].
\eqa

\noindent
The sum of these three contributions from s-channel, t-channel and their 
interference gives the differential cross section:
\bqa
\frac{d\sigma_{\rm red}^{\rm V}}{d\Omega} =
\frac{\alpha^3}{s\pi}\,\Bigg\{ &&
  \frac{1 \!-\! 2x \!+\! 2x^2}{2}\,\bigg[ 4\,V_s^\gamma\,{\rm Re}\,\Pi(s) \bigg]
+ \,\frac{2 \!-\! 2x \!+\! x^2}{2\,x^2}\, \bigg[ 4\,V_t^\gamma\,\Pi(t) \bigg]
\nl
&&\!\!\!
- \,\frac{1 \!-\! 2x \!+\! x^2}{x}
  \bigg[
    ( V_s^\gamma + V_t^\gamma )\Big( {\rm Re}\,\Pi(s) + \Pi(t) \Big)
  + \pi\Big( \ln\frac{\lambda^2}{s} + \frac{3}{2} \Big)\,{\rm Im}\,\Pi(s)
  \bigg]
\Bigg\},
\label{eq:redV}
\eqa
where $\lambda$ is the photon mass used as IR regulator.
We have also introduced
\bqa
V_s^\gamma &=&
  2\ln\!\frac{2\omega}{\sqrt{s}}\bigg(\! \ln\frac{s}{m_e^2} - 1 \!\bigg)
+ \frac{3}{2}\ln\frac{s}{m_e^2} + 2\,\zeta(2) - 2,
\nl
V_t^\gamma &=&
  2\ln\!\frac{2\omega}{\sqrt{s}}\bigg(\! \ln\frac{-t}{m_e^2} - 1 \!\bigg)
+ \frac{3}{2}\ln\frac{-t}{m_e^2} - \ln\frac{-t}{s}\ln\frac{-u}{s} 
- \li_2\left(\frac{-t}{s}\right) - 2.
\eqa
Just like for the one-loop corrections, a logarithmic dependence on $m_e$ 
from collinear singularities remains.
It is interesting to notice the presence of an infrared divergent term 
proportional to ${\rm Im}\,\Pi(s)$ surviving after the inclusion of the 
real soft photon emission. 
Remembering that these contributions can be easily obtained from the 
${\cal O}(\alpha)$ calculation it is clear that the ${\cal O}(\alpha)$ result 
is easily recovered by the substitution of $\Pi(s)$ and $\Pi(t)$ by 1.

A similar discussion applies to the photonic box diagrams, interfering 
with amplitudes with the dressed photon propagator in the $s$- or $t$-channel:
\bqa
\!\frac{d\sigma_{\rm red}^{\rm B\!,\,s}}{d\Omega}\!\!
&=&
c_{_{\!P\!S}}
2\,{\rm Re}
\scalebox{0.55}{
\begin{picture}(50,20)(0,-3)
 \ArrowLine(15,0)(0,20)
 \ArrowLine(0,-20)(15,0)
 \Photon(15,0)(35,0){2}{5}
 \ArrowLine(50,20)(35,0)
 \ArrowLine(35,0)(50,-20)
 \GCirc(25,0){5}{0.5}
\end{picture}
}
\Bigg(\!
\scalebox{0.55}{
\begin{picture}(60,20)(0,-3)
 \ArrowLine(15,15)(0,15)
 \ArrowLine(15,-15)(15,15)
 \ArrowLine(0,-15)(15,-15)
 \Photon(15,15)(45,15){2}{7}
 \Photon(15,-15)(45,-15){2}{7}
 \ArrowLine(60,15)(45,15)
 \ArrowLine(45,15)(45,-15)
 \ArrowLine(45,-15)(60,-15)
\end{picture}
}
+
\scalebox{0.55}{
\begin{picture}(60,20)(0,-3)
 \ArrowLine(15,15)(0,15)
 \ArrowLine(15,-15)(15,15)
 \ArrowLine(0,-15)(15,-15)
 \Photon(15,15)(45,-15){2}{7}
 \Photon(15,-15)(45,15){2}{7}
 \ArrowLine(60,15)(45,15)
 \ArrowLine(45,15)(45,-15)
 \ArrowLine(45,-15)(60,-15)
\end{picture}
}
\!\Bigg)^{\!\!*}
\nl
\nl
&&
\!+\,
\frac{c_{_{\!P\!S}}}{(2\pi)^3}\!\!\int_{\omega}\!\frac{d^3k}{2\,k_0}\;
{\rm Re}
\Bigg[
\Bigg(\!\!
\scalebox{0.55}{
\begin{picture}(50,20)(0,-3)
 \ArrowLine(15,0)(7.5,10)
 \ArrowLine(7.5,10)(0,20)
 \Photon(7.5,10)(20,20){2}{4}        \Text(25,16)[cb]{$k$}
 \ArrowLine(0,-20)(15,0)
 \Photon(15,0)(35,0){2}{5}
 \ArrowLine(50,20)(35,0)
 \ArrowLine(35,0)(50,-20)
\end{picture}
}
\!\!+\!\!
\scalebox{0.55}{
\begin{picture}(50,20)(0,-3)
 \ArrowLine(15,0)(0,20)
 \ArrowLine(0,-20)(7.5,-10)
 \ArrowLine(7.5,-10)(15,0)
 \Photon(7.5,-10)(20,-20){2}{4}        \Text(24,-22)[cb]{$k$}
 \Photon(15,0)(35,0){2}{5}
 \ArrowLine(50,20)(35,0)
 \ArrowLine(35,0)(50,-20)
\end{picture}
}
\!\Bigg)\!\!
\Bigg(\!\!
\scalebox{0.55}{
\begin{picture}(50,20)(0,-3)
 \ArrowLine(15,0)(0,20)
 \ArrowLine(0,-20)(15,0)
 \Photon(15,0)(35,0){2}{5}
 \ArrowLine(50,20)(35,0)
 \ArrowLine(35,0)(42.5,-10)
 \ArrowLine(42.5,-10)(50,-20)
 \Photon(42.5,-10)(35,-20){2}{4}        \Text(29.5,-22)[cb]{$k$}
 \GCirc(25,0){5}{0.5}
\end{picture}
}
\!\!+\!\!
\scalebox{0.55}{
\begin{picture}(50,20)(0,-3)
 \ArrowLine(15,0)(0,20)
 \ArrowLine(0,-20)(15,0)
 \Photon(15,0)(35,0){2}{5}
 \ArrowLine(50,20)(42.5,10)
 \ArrowLine(42.5,10)(35,0)
 \Photon(42.5,10)(35,20){2}{4}        \Text(29.5,16)[cb]{$k$}
 \ArrowLine(35,0)(50,-20)
 \GCirc(25,0){5}{0.5}
\end{picture}
}
\!\Bigg)^{\!\!\!*}
\!\!+\!
\Bigg(\!\!
\scalebox{0.55}{
\begin{picture}(50,20)(0,-3)
 \ArrowLine(15,0)(0,20)
 \ArrowLine(0,-20)(15,0)
 \Photon(15,0)(35,0){2}{5}
 \ArrowLine(50,20)(35,0)
 \ArrowLine(35,0)(42.5,-10)
 \ArrowLine(42.5,-10)(50,-20)
 \Photon(42.5,-10)(35,-20){2}{4}        \Text(29.5,-22)[cb]{$k$}
\end{picture}
}
\!\!+\!\!
\scalebox{0.55}{
\begin{picture}(50,20)(0,-3)
 \ArrowLine(15,0)(0,20)
 \ArrowLine(0,-20)(15,0)
 \Photon(15,0)(35,0){2}{5}
 \ArrowLine(50,20)(42.5,10)
 \ArrowLine(42.5,10)(35,0)
 \Photon(42.5,10)(35,20){2}{4}        \Text(29.5,16)[cb]{$k$}
 \ArrowLine(35,0)(50,-20)
\end{picture}
}
\!\Bigg)\!\!
\Bigg(\!\!
\scalebox{0.55}{
\begin{picture}(50,20)(0,-3)
 \ArrowLine(15,0)(7.5,10)
 \ArrowLine(7.5,10)(0,20)
 \Photon(7.5,10)(20,20){2}{4}        \Text(25,16)[cb]{$k$}
 \ArrowLine(0,-20)(15,0)
 \Photon(15,0)(35,0){2}{5}
 \ArrowLine(50,20)(35,0)
 \ArrowLine(35,0)(50,-20)
 \GCirc(25,0){5}{0.5}
\end{picture}
}
\!\!+\!\!
\scalebox{0.55}{
\begin{picture}(50,20)(0,-3)
 \ArrowLine(15,0)(0,20)
 \ArrowLine(0,-20)(7.5,-10)
 \ArrowLine(7.5,-10)(15,0)
 \Photon(7.5,-10)(20,-20){2}{4}        \Text(24,-22)[cb]{$k$}
 \Photon(15,0)(35,0){2}{5}
 \ArrowLine(50,20)(35,0)
 \ArrowLine(35,0)(50,-20)
 \GCirc(25,0){5}{0.5}
\end{picture}
}
\!\Bigg)^{\!\!\!*}
\Bigg],
\qquad
\eqa
\bqa
\!\frac{d\sigma_{\rm red}^{\rm B\!,\,t}}{d\Omega}\!\!
&=&
c_{_{\!P\!S}}2\,{\rm Re}
\scalebox{0.55}{
\begin{picture}(40,40)(0,-3)
 \ArrowLine(20,10)(0,30)
 \ArrowLine(40,30)(20,10)
 \Photon(20,10)(20,-10){2}{5}
 \ArrowLine(0,-30)(20,-10)
 \ArrowLine(20,-10)(40,-30)
 \GCirc(20,0){5}{0.5}
\end{picture}
}
\Bigg(\!
\scalebox{0.55}{
\begin{picture}(60,20)(0,-3)
 \ArrowLine(15,15)(0,15)
 \ArrowLine(45,15)(15,15)
 \ArrowLine(60,15)(45,15)
 \Photon(15,15)(15,-15){2}{7}
 \Photon(45,15)(45,-15){2}{7}
 \ArrowLine(0,-15)(15,-15)
 \ArrowLine(15,-15)(45,-15)
 \ArrowLine(45,-15)(60,-15)
\end{picture}
}
+
\scalebox{0.55}{
\begin{picture}(60,20)(0,-3)
 \ArrowLine(15,15)(0,15)
 \ArrowLine(45,15)(15,15)
 \ArrowLine(60,15)(45,15)
 \Photon(15,-15)(45,15){2}{7}
 \Photon(15,15)(45,-15){2}{7}
 \ArrowLine(0,-15)(15,-15)
 \ArrowLine(15,-15)(45,-15)
 \ArrowLine(45,-15)(60,-15)
\end{picture}
}
\!\Bigg)^{\!\!*}
\nl
&&
+\,
\frac{c_{_{\!P\!S}}}{(2\pi)^3}\!\!\int_{\omega}\!\frac{d^3k}{2\,k_0}\;
{\rm Re}
\Bigg[
\Bigg(\!
\scalebox{0.55}{
\begin{picture}(40,40)(0,-3)
 \ArrowLine(20,10)(10,20)
 \ArrowLine(10,20)(0,30)
 \Photon(10,20)(20,30){2}{4}        \Text(25,26)[cb]{$k$}
 \ArrowLine(40,30)(20,10)
 \Photon(20,10)(20,-10){2}{5}
 \ArrowLine(0,-30)(20,-10)
 \ArrowLine(20,-10)(40,-30)
\end{picture}
}
\!\!+\!\!
\scalebox{0.55}{
\begin{picture}(40,40)(0,-3)
 \ArrowLine(20,10)(0,30)
 \ArrowLine(40,30)(30,20)
 \ArrowLine(30,20)(20,10)
 \Photon(30,20)(20,30){2}{4}        \Text(16,26)[cb]{$k$}
 \Photon(20,10)(20,-10){2}{5}
 \ArrowLine(0,-30)(20,-10)
 \ArrowLine(20,-10)(40,-30)
\end{picture}
}
\!\Bigg)\!
\Bigg(\!
\scalebox{0.55}{
\begin{picture}(40,40)(0,-3)
 \ArrowLine(20,10)(0,30)
 \ArrowLine(40,30)(20,10)
 \Photon(20,10)(20,-10){2}{5}
 \ArrowLine(0,-30)(10,-20)
 \ArrowLine(10,-20)(20,-10)
 \Photon(10,-20)(20,-30){2}{4}        \Text(24,-32)[cb]{$k$}
 \ArrowLine(20,-10)(40,-30)
 \GCirc(20,0){5}{0.5}
\end{picture}
}
\!\!+\!\!
\scalebox{0.55}{
\begin{picture}(40,40)(0,-3)
 \ArrowLine(20,10)(0,30)
 \ArrowLine(40,30)(20,10)
 \Photon(20,10)(20,-10){2}{5}
 \ArrowLine(0,-30)(20,-10)
 \ArrowLine(20,-10)(30,-20)
 \ArrowLine(30,-20)(40,-30)
 \Photon(30,-20)(20,-30){2}{4}        \Text(15.5,-32)[cb]{$k$}
 \GCirc(20,0){5}{0.5}
\end{picture}
}
\!\Bigg)^{\!\!*}
\!+
\Bigg(\!
\scalebox{0.55}{
\begin{picture}(40,40)(0,-3)
 \ArrowLine(20,10)(0,30)
 \ArrowLine(40,30)(20,10)
 \Photon(20,10)(20,-10){2}{5}
 \ArrowLine(0,-30)(10,-20)
 \ArrowLine(10,-20)(20,-10)
 \Photon(10,-20)(20,-30){2}{4}        \Text(24,-32)[cb]{$k$}
 \ArrowLine(20,-10)(40,-30)
\end{picture}
}
\!\!+\!\!
\scalebox{0.55}{
\begin{picture}(40,40)(0,-3)
 \ArrowLine(20,10)(0,30)
 \ArrowLine(40,30)(20,10)
 \Photon(20,10)(20,-10){2}{5}
 \ArrowLine(0,-30)(20,-10)
 \ArrowLine(20,-10)(30,-20)
 \ArrowLine(30,-20)(40,-30)
 \Photon(30,-20)(20,-30){2}{4}        \Text(15.5,-32)[cb]{$k$}
\end{picture}
}
\!\Bigg)\!
\Bigg(\!
\scalebox{0.55}{
\begin{picture}(40,40)(0,-3)
 \ArrowLine(20,10)(10,20)
 \ArrowLine(10,20)(0,30)
 \Photon(10,20)(20,30){2}{4}        \Text(25,26)[cb]{$k$}
 \ArrowLine(40,30)(20,10)
 \Photon(20,10)(20,-10){2}{5}
 \ArrowLine(0,-30)(20,-10)
 \ArrowLine(20,-10)(40,-30)
 \GCirc(20,0){5}{0.5}
\end{picture}
}
\!\!+\!\!
\scalebox{0.55}{
\begin{picture}(40,40)(0,-3)
 \ArrowLine(20,10)(0,30)
 \ArrowLine(40,30)(30,20)
 \ArrowLine(30,20)(20,10)
 \Photon(30,20)(20,30){2}{4}        \Text(16,26)[cb]{$k$}
 \Photon(20,10)(20,-10){2}{5}
 \ArrowLine(0,-30)(20,-10)
 \ArrowLine(20,-10)(40,-30)
 \GCirc(20,0){5}{0.5}
\end{picture}
}
\!\Bigg)^{\!\!*}
\Bigg],
\eqa
\bqa
\!\frac{d\sigma_{\rm red}^{\rm B\!,\,st}}{d\Omega}\!\!
&=&
c_{_{\!P\!S}}2\,{\rm Re}
\Bigg[
\scalebox{0.55}{
\begin{picture}(50,20)(0,-3)
 \ArrowLine(15,0)(0,20)
 \ArrowLine(0,-20)(15,0)
 \Photon(15,0)(35,0){2}{5}
 \ArrowLine(50,20)(35,0)
 \ArrowLine(35,0)(50,-20)
 \GCirc(25,0){5}{0.5}
\end{picture}
}
\Bigg(\!
\scalebox{0.55}{
\begin{picture}(60,20)(0,-3)
 \ArrowLine(15,15)(0,15)
 \ArrowLine(45,15)(15,15)
 \ArrowLine(60,15)(45,15)
 \Photon(15,15)(15,-15){2}{7}
 \Photon(45,15)(45,-15){2}{7}
 \ArrowLine(0,-15)(15,-15)
 \ArrowLine(15,-15)(45,-15)
 \ArrowLine(45,-15)(60,-15)
\end{picture}
}
+
\scalebox{0.55}{
\begin{picture}(60,20)(0,-3)
 \ArrowLine(15,15)(0,15)
 \ArrowLine(45,15)(15,15)
 \ArrowLine(60,15)(45,15)
 \Photon(15,-15)(45,15){2}{7}
 \Photon(15,15)(45,-15){2}{7}
 \ArrowLine(0,-15)(15,-15)
 \ArrowLine(15,-15)(45,-15)
 \ArrowLine(45,-15)(60,-15)
\end{picture}
}
\!\Bigg)^{\!\!*}
+
\scalebox{0.55}{
\begin{picture}(40,40)(0,-3)
 \ArrowLine(20,10)(0,30)
 \ArrowLine(40,30)(20,10)
 \Photon(20,10)(20,-10){2}{5}
 \ArrowLine(0,-30)(20,-10)
 \ArrowLine(20,-10)(40,-30)
 \GCirc(20,0){5}{0.5}
\end{picture}
}
\Bigg(\!
\scalebox{0.55}{
\begin{picture}(60,20)(0,-3)
 \ArrowLine(15,15)(0,15)
 \ArrowLine(15,-15)(15,15)
 \ArrowLine(0,-15)(15,-15)
 \Photon(15,15)(45,15){2}{7}
 \Photon(15,-15)(45,-15){2}{7}
 \ArrowLine(60,15)(45,15)
 \ArrowLine(45,15)(45,-15)
 \ArrowLine(45,-15)(60,-15)
\end{picture}
}
+
\scalebox{0.55}{
\begin{picture}(60,20)(0,-3)
 \ArrowLine(15,15)(0,15)
 \ArrowLine(15,-15)(15,15)
 \ArrowLine(0,-15)(15,-15)
 \Photon(15,15)(45,-15){2}{7}
 \Photon(15,-15)(45,15){2}{7}
 \ArrowLine(60,15)(45,15)
 \ArrowLine(45,15)(45,-15)
 \ArrowLine(45,-15)(60,-15)
\end{picture}
}
\!\Bigg)^{\!\!*}
\Bigg]
\nl
&&
+\,
\frac{c_{_{\!P\!S}}}{(2\pi)^3}\!\!\int_{\omega}\!\frac{d^3k}{2\,k_0}\;
{\rm Re}
\Bigg[
\Bigg(\!
\scalebox{0.55}{
\begin{picture}(40,40)(0,-3)
 \ArrowLine(20,10)(10,20)
 \ArrowLine(10,20)(0,30)
 \Photon(10,20)(20,30){2}{4}        \Text(25,26)[cb]{$k$}
 \ArrowLine(40,30)(20,10)
 \Photon(20,10)(20,-10){2}{5}
 \ArrowLine(0,-30)(20,-10)
 \ArrowLine(20,-10)(40,-30)
\end{picture}
}
\!\!+\!\!
\scalebox{0.55}{
\begin{picture}(40,40)(0,-3)
 \ArrowLine(20,10)(0,30)
 \ArrowLine(40,30)(30,20)
 \ArrowLine(30,20)(20,10)
 \Photon(30,20)(20,30){2}{4}        \Text(16,26)[cb]{$k$}
 \Photon(20,10)(20,-10){2}{5}
 \ArrowLine(0,-30)(20,-10)
 \ArrowLine(20,-10)(40,-30)
\end{picture}
}
\!\Bigg)
\Bigg(\!
\scalebox{0.55}{
\begin{picture}(50,20)(0,-3)
 \ArrowLine(15,0)(0,20)
 \ArrowLine(0,-20)(7.5,-10)
 \ArrowLine(7.5,-10)(15,0)
 \Photon(7.5,-10)(20,-20){2}{4}        \Text(24,-22)[cb]{$k$}
 \Photon(15,0)(35,0){2}{5}
 \ArrowLine(50,20)(35,0)
 \ArrowLine(35,0)(50,-20)
 \GCirc(25,0){5}{0.5}
\end{picture}
}
\!\!+\!\!
\scalebox{0.55}{
\begin{picture}(50,20)(0,-3)
 \ArrowLine(15,0)(0,20)
 \ArrowLine(0,-20)(15,0)
 \Photon(15,0)(35,0){2}{5}
 \ArrowLine(50,20)(35,0)
 \ArrowLine(35,0)(42.5,-10)
 \ArrowLine(42.5,-10)(50,-20)
 \Photon(42.5,-10)(35,-20){2}{4}        \Text(29.5,-22)[cb]{$k$}
 \GCirc(25,0){5}{0.5}
\end{picture}
}
\!\Bigg)^{\!\!*}
\!\!
+
\Bigg(\!
\scalebox{0.55}{
\begin{picture}(40,40)(0,-3)
 \ArrowLine(20,10)(0,30)
 \ArrowLine(40,30)(20,10)
 \Photon(20,10)(20,-10){2}{5}
 \ArrowLine(0,-30)(10,-20)
 \ArrowLine(10,-20)(20,-10)
 \Photon(10,-20)(20,-30){2}{4}        \Text(24,-32)[cb]{$k$}
 \ArrowLine(20,-10)(40,-30)
\end{picture}
}
\!\!+\!\!
\scalebox{0.55}{
\begin{picture}(40,40)(0,-3)
 \ArrowLine(20,10)(0,30)
 \ArrowLine(40,30)(20,10)
 \Photon(20,10)(20,-10){2}{5}
 \ArrowLine(0,-30)(20,-10)
 \ArrowLine(20,-10)(30,-20)
 \ArrowLine(30,-20)(40,-30)
 \Photon(30,-20)(20,-30){2}{4}        \Text(15.5,-32)[cb]{$k$}
\end{picture}
}
\!\Bigg)
\Bigg(\!
\scalebox{0.55}{
\begin{picture}(50,20)(0,-3)
 \ArrowLine(15,0)(7.5,10)
 \ArrowLine(7.5,10)(0,20)
 \Photon(7.5,10)(20,20){2}{4}        \Text(25,16)[cb]{$k$}
 \ArrowLine(0,-20)(15,0)
 \Photon(15,0)(35,0){2}{5}
 \ArrowLine(50,20)(35,0)
 \ArrowLine(35,0)(50,-20)
 \GCirc(25,0){5}{0.5}
\end{picture}
}
\!\!+\!\!
\scalebox{0.55}{
\begin{picture}(50,20)(0,-3)
 \ArrowLine(15,0)(0,20)
 \ArrowLine(0,-20)(15,0)
 \Photon(15,0)(35,0){2}{5}
 \ArrowLine(50,20)(42.5,10)
 \ArrowLine(42.5,10)(35,0)
 \Photon(42.5,10)(35,20){2}{4}        \Text(29.5,16)[cb]{$k$}
 \ArrowLine(35,0)(50,-20)
 \GCirc(25,0){5}{0.5}
\end{picture}
}
\!\Bigg)^{\!\!*}
\nl
&&
\qquad\qquad\qquad\qquad
+
\Bigg(\!
\scalebox{0.55}{
\begin{picture}(50,20)(0,-3)
 \ArrowLine(15,0)(7.5,10)
 \ArrowLine(7.5,10)(0,20)
 \Photon(7.5,10)(20,20){2}{4}        \Text(25,16)[cb]{$k$}
 \ArrowLine(0,-20)(15,0)
 \Photon(15,0)(35,0){2}{5}
 \ArrowLine(50,20)(35,0)
 \ArrowLine(35,0)(50,-20)
\end{picture}
}
\!\!+\!\!
\scalebox{0.55}{
\begin{picture}(50,20)(0,-3)
 \ArrowLine(15,0)(0,20)
 \ArrowLine(0,-20)(7.5,-10)
 \ArrowLine(7.5,-10)(15,0)
 \Photon(7.5,-10)(20,-20){2}{4}        \Text(24,-22)[cb]{$k$}
 \Photon(15,0)(35,0){2}{5}
 \ArrowLine(50,20)(35,0)
 \ArrowLine(35,0)(50,-20)
\end{picture}
}
\!\Bigg)
\Bigg(\!
\scalebox{0.55}{
\begin{picture}(40,40)(0,-3)
 \ArrowLine(20,10)(0,30)
 \ArrowLine(40,30)(20,10)
 \Photon(20,10)(20,-10){2}{5}
 \ArrowLine(0,-30)(20,-10)
 \ArrowLine(20,-10)(30,-20)
 \ArrowLine(30,-20)(40,-30)
 \Photon(30,-20)(20,-30){2}{4}        \Text(15.5,-32)[cb]{$k$}
 \GCirc(20,0){5}{0.5}
\end{picture}
}
\!\!+\!\!
\scalebox{0.55}{
\begin{picture}(40,40)(0,-3)
 \ArrowLine(20,10)(0,30)
 \ArrowLine(40,30)(30,20)
 \ArrowLine(30,20)(20,10)
 \Photon(30,20)(20,30){2}{4}        \Text(16,26)[cb]{$k$}
 \Photon(20,10)(20,-10){2}{5}
 \ArrowLine(0,-30)(20,-10)
 \ArrowLine(20,-10)(40,-30)
 \GCirc(20,0){5}{0.5}
\end{picture}
}
\!\Bigg)^{\!\!*}
\!\!
+
\Bigg(\!
\scalebox{0.55}{
\begin{picture}(50,20)(0,-3)
 \ArrowLine(15,0)(0,20)
 \ArrowLine(0,-20)(15,0)
 \Photon(15,0)(35,0){2}{5}
 \ArrowLine(50,20)(35,0)
 \ArrowLine(35,0)(42.5,-10)
 \ArrowLine(42.5,-10)(50,-20)
 \Photon(42.5,-10)(35,-20){2}{4}        \Text(29.5,-22)[cb]{$k$}
\end{picture}
}
\!\!+\!\!
\scalebox{0.55}{
\begin{picture}(50,20)(0,-3)
 \ArrowLine(15,0)(0,20)
 \ArrowLine(0,-20)(15,0)
 \Photon(15,0)(35,0){2}{5}
 \ArrowLine(50,20)(42.5,10)
 \ArrowLine(42.5,10)(35,0)
 \Photon(42.5,10)(35,20){2}{4}        \Text(29.5,16)[cb]{$k$}
 \ArrowLine(35,0)(50,-20)
\end{picture}
}
\!\Bigg)
\Bigg(\!
\scalebox{0.55}{
\begin{picture}(40,40)(0,-3)
 \ArrowLine(20,10)(10,20)
 \ArrowLine(10,20)(0,30)
 \Photon(10,20)(20,30){2}{4}        \Text(25,26)[cb]{$k$}
 \ArrowLine(40,30)(20,10)
 \Photon(20,10)(20,-10){2}{5}
 \ArrowLine(0,-30)(20,-10)
 \ArrowLine(20,-10)(40,-30)
 \GCirc(20,0){5}{0.5}
\end{picture}
}
\!\!+\!\!
\scalebox{0.55}{
\begin{picture}(40,40)(0,-3)
 \ArrowLine(20,10)(0,30)
 \ArrowLine(40,30)(20,10)
 \Photon(20,10)(20,-10){2}{5}
 \ArrowLine(0,-30)(10,-20)
 \ArrowLine(10,-20)(20,-10)
 \Photon(10,-20)(20,-30){2}{4}        \Text(24,-32)[cb]{$k$}
 \ArrowLine(20,-10)(40,-30)
 \GCirc(20,0){5}{0.5}
\end{picture}
}
\!\Bigg)^{\!\!*}
\Bigg].
\qquad
\eqa
The differential cross section is then given by:
\bqa
\label{eq:redB}
\!\!\!
\frac{d\sigma_{\!\rm red}^{\rm B}}{d\Omega}\!\! = 
\frac{\alpha^3}{s\pi}\Bigg\{ &&
  \frac{1 \!-\! 2x \!+\! 2x^2}{2}\,\bigg[ 
  2\,B_s^\gamma\,{\rm Re}\,\Pi(s) + 2\pi\ln\!\frac{t}{u}\,{\rm Im}\,\Pi(s)
  \bigg]
- \,{\rm Re}\,\bigg[ \Big(\! B^\gamma(s,t) \!-\! B^\gamma(s,u) \Big)\,
                     \Pi^*\!(s) \bigg]
\\
\!\!\!
&&\!\!\!\!
+ \,\frac{2 \!-\! 2x \!+\! x^2}{2\,x^2}\,\bigg[ 2\,B_t^\gamma\,\Pi(t) \bigg]
- {\rm Re}\,\bigg[ \Big(\! B^\gamma(t,s) \!-\! B^\gamma(t,u) \Big)\,\Pi(t) \bigg]
\nl
\!\!\!
&&\!\!\!\!
- \,\frac{1 \!-\! 2x \!+\! x^2}{x}
  \bigg[
    B_t^\gamma{\rm Re}\Pi(s) + B_s^\gamma\Pi(t)
  - \pi\,\ln\frac{\,\lambda^2\!}{-t}\,{\rm Im}\Pi(s)
  \bigg]\!
+ {\rm Re}\bigg[ x B^\gamma(t,\!s)\Pi^*\!(s) 
                 + \frac{1}{x}B^\gamma(s,\!t)\Pi(t) \bigg]
\!\Bigg\}\!,
\nn
\eqa
where we have introduced
\bqa
B_s^\gamma &=&
  2\ln\!\frac{2\omega}{\sqrt{s}}\ln\frac{t}{u}
+ \frac{1}{2}\ln^2\frac{-t}{s} - \frac{1}{2}\ln^2\frac{-u}{s}
- \ln\frac{-t}{s}\ln\frac{-u}{s} - 2\,\li_2\left(\frac{-t}{s}\right) + \zeta(2),
\nl
B_t^\gamma &=&
- 2\ln\!\frac{2\omega}{\sqrt{s}}\ln\frac{-u}{s}
- \frac{1}{2}\ln^2\frac{-u}{s} + \ln\frac{-t}{s}\ln\frac{-u}{s} 
- \li_2\left(\frac{-t}{s}\right),
\nl
B^\gamma(a,b) &=&
- \frac{a+b}{2\,a}\ln\frac{b}{a+i\,\ep}
+ \frac{a+2b}{4\,a}\bigg( \ln^2\frac{b}{a+i\,\ep} + \pi^2 \bigg).
\eqa
In this case the electron mass can be safely set to zero.
Also in this case a logarithm of $\lambda^2$ survives and cancels exactly the one 
generated by the vertex reducible corrections, rendering the total cross section 
infrared finite.
\subsection{Irreducible diagrams}
Let us move to the third group consisting of the two irreducible contributions.
The vertex correction has been discussed in detail in~\cite{Kniehl:1988id} and 
can be cast into the following form:
\bq
\scalebox{0.7}{
\begin{picture}(40,20)(0,-3)
 \ArrowLine(20,0)(0,20)
 \ArrowLine(0,-20)(20,0)
 \Photon(20,0)(40,0){2}{5}          \Text(35,5)[cb]{$q^2$}
 \Photon(5,15)(5,-15){1.5}{8}
 \GCirc(5,0){5}{0.5}
\end{picture}
}
\qquad
\Longrightarrow
\qquad
V(q^2)= 
\frac{\alpha}{3\pi}\int_{4m^2}^{\infty}\!\!\frac{dz}{z}R(z)
\rho(q^2\!,z\!-\!i\ep),
\label{eq:V}
\eq
with
\bq
\rho(q^2,z)= 
- \frac{7}{8} - \frac{z}{2q^2} 
+ \frac{1}{2}\bigg( \frac{3}{2} + \frac{z}{q^2} \bigg)\ln\frac{-z}{q^2}
+ \frac{1}{2}\bigg( 1 + \frac{z}{q^2} \bigg)^{\!2}\,
  \bigg[ \zeta(2) - \li_2\!\left(\! 1 \!+\! \frac{z}{q^2} \right) \bigg].
\eq
The contribution to the cross section can be cast into a form closely related to 
the Born cross section.
\bqa
\!\frac{d\sigma^{\rm V\!,\,s}}{d\Omega}\!\!
&=&
c_{_{\!P\!S}}
2\,{\rm Re}
\scalebox{0.6}{
\begin{picture}(50,20)(0,-3)
 \ArrowLine(15,0)(0,20)
 \ArrowLine(0,-20)(15,0)
 \Photon(15,0)(35,0){2}{5}
 \ArrowLine(50,20)(35,0)
 \ArrowLine(35,0)(50,-20)
\end{picture}
}
\Bigg(\,
\scalebox{0.6}{
\begin{picture}(55,20)(0,-3)
 \ArrowLine(20,0)(0,20)
 \ArrowLine(0,-20)(20,0)
 \Photon(20,0)(40,0){2}{5}
 \ArrowLine(55,20)(40,0)
 \ArrowLine(40,0)(55,-20)
 \Photon(5,15)(5,-15){1.5}{8}
 \GCirc(5,0){5}{0.5}
\end{picture}
}
+
\scalebox{0.6}{
\begin{picture}(55,20)(0,-3)
 \ArrowLine(15,0)(0,20)
 \ArrowLine(0,-20)(15,0)
 \Photon(15,0)(35,0){2}{5}
 \ArrowLine(55,20)(35,0)
 \ArrowLine(35,0)(55,-20)
 \Photon(50,15)(50,-15){1.5}{8}
 \GCirc(50,0){5}{0.5}
\end{picture}
}
\Bigg)^*,
\qquad
\!\frac{d\sigma^{\rm V\!,\,t}}{d\Omega}\!\!
=
c_{_{\!P\!S}}
2\,{\rm Re}
\scalebox{0.6}{
\begin{picture}(40,40)(0,-3)
 \ArrowLine(20,10)(0,30)
 \ArrowLine(40,30)(20,10)
 \Photon(20,10)(20,-10){2}{5}
 \ArrowLine(0,-30)(20,-10)
 \ArrowLine(20,-10)(40,-30)
\end{picture}
}
\Bigg(\,
\scalebox{0.6}{
\begin{picture}(40,40)(0,-3)
 \ArrowLine(20,10)(0,30)
 \ArrowLine(40,30)(20,10)
 \Photon(20,10)(20,-10){2}{5}
 \ArrowLine(0,-30)(20,-10)
 \ArrowLine(20,-10)(40,-30)
 \Photon(5,25)(35,25){1.5}{8}
 \GCirc(20,25){5}{0.5}
\end{picture}
}
+
\scalebox{0.6}{
\begin{picture}(40,40)(0,-3)
 \ArrowLine(20,10)(0,30)
 \ArrowLine(40,30)(20,10)
 \Photon(20,10)(20,-10){2}{5}
 \ArrowLine(0,-30)(20,-10)
 \ArrowLine(20,-10)(40,-30)
 \Photon(5,-25)(35,-25){1.5}{8}
 \GCirc(20,-25){5}{0.5}
\end{picture}
}
\Bigg)^*,
\nl
\!\frac{d\sigma^{\rm V\!,\,st}}{d\Omega}\!\!
&=&
c_{_{\!P\!S}}
2\,{\rm Re}
\Bigg[
\scalebox{0.6}{
\begin{picture}(50,20)(0,-3)
 \ArrowLine(15,0)(0,20)
 \ArrowLine(0,-20)(15,0)
 \Photon(15,0)(35,0){2}{5}
 \ArrowLine(50,20)(35,0)
 \ArrowLine(35,0)(50,-20)
\end{picture}
}
\Bigg(\,
\scalebox{0.6}{
\begin{picture}(40,40)(0,-3)
 \ArrowLine(20,10)(0,30)
 \ArrowLine(40,30)(20,10)
 \Photon(20,10)(20,-10){2}{5}
 \ArrowLine(0,-30)(20,-10)
 \ArrowLine(20,-10)(40,-30)
 \Photon(5,25)(35,25){1.5}{8}
 \GCirc(20,25){5}{0.5}
\end{picture}
}
+
\scalebox{0.6}{
\begin{picture}(40,40)(0,-3)
 \ArrowLine(20,10)(0,30)
 \ArrowLine(40,30)(20,10)
 \Photon(20,10)(20,-10){2}{5}
 \ArrowLine(0,-30)(20,-10)
 \ArrowLine(20,-10)(40,-30)
 \Photon(5,-25)(35,-25){1.5}{8}
 \GCirc(20,-25){5}{0.5}
\end{picture}
}
\Bigg)^*
+
\scalebox{0.6}{
\begin{picture}(40,40)(0,-3)
 \ArrowLine(20,10)(0,30)
 \ArrowLine(40,30)(20,10)
 \Photon(20,10)(20,-10){2}{5}
 \ArrowLine(0,-30)(20,-10)
 \ArrowLine(20,-10)(40,-30)
\end{picture}
}
\Bigg(\,
\scalebox{0.6}{
\begin{picture}(55,20)(0,-3)
 \ArrowLine(20,0)(0,20)
 \ArrowLine(0,-20)(20,0)
 \Photon(20,0)(40,0){2}{5}
 \ArrowLine(55,20)(40,0)
 \ArrowLine(40,0)(55,-20)
 \Photon(5,15)(5,-15){1.5}{8}
 \GCirc(5,0){5}{0.5}
\end{picture}
}
+
\scalebox{0.6}{
\begin{picture}(55,20)(0,-3)
 \ArrowLine(15,0)(0,20)
 \ArrowLine(0,-20)(15,0)
 \Photon(15,0)(35,0){2}{5}
 \ArrowLine(55,20)(35,0)
 \ArrowLine(35,0)(55,-20)
 \Photon(50,15)(50,-15){1.5}{8}
 \GCirc(50,0){5}{0.5}
\end{picture}
}
\Bigg)^*
\Bigg],
\eqa
\bq
\frac{d\sigma^{\rm V}}{d\Omega} = 
\frac{\alpha^3}{s\,\pi}\,\Bigg\{ 
  \frac{1 \!-\! 2x \!+\! 2x^2}{2}\,\Big[4\,{\rm Re}\,V(s) \Big]
+ \,\frac{2 \!-\! 2x \!+\! x^2}{2\,x^2}\,\Big[4\,V(t) \Big]
- \,\frac{1 \!-\! 2x \!+\! x^2}{x}
  \Big[ 2\,V(t) + 2\,{\rm Re}\,V(s) \Big] 
\Bigg\}.
\eq
As stated above, $m_e$ has been set to zero and the result is obviously 
infrared finite.

Finally for the irreducible two-loop box contributions the kernels 
can again be directly taken from the literature \cite{Brown:1983jv}. 
The part of the kernel, which corresponds to the infrared divergent piece will 
be canceled by the proper combination of real soft radiation amplitudes which 
are also proportional to $\Pi(q^2)$, specifically:
\bqa
\!\frac{d\sigma^{\rm B\!,\,s}}{d\Omega}\!\!
&=&
c_{_{\!P\!S}}
2\,{\rm Re}
\scalebox{0.55}{
\begin{picture}(50,20)(0,-3)
 \ArrowLine(15,0)(0,20)
 \ArrowLine(0,-20)(15,0)
 \Photon(15,0)(35,0){2}{5}
 \ArrowLine(50,20)(35,0)
 \ArrowLine(35,0)(50,-20)
\end{picture}
}
\Bigg(\,
\scalebox{0.55}{
\begin{picture}(60,20)(0,-3)
 \ArrowLine(15,15)(0,15)
 \ArrowLine(15,-15)(15,15)
 \ArrowLine(0,-15)(15,-15)
 \Photon(15,15)(45,15){2}{7}
 \Photon(15,-15)(45,-15){2}{7}
 \ArrowLine(60,15)(45,15)
 \ArrowLine(45,15)(45,-15)
 \ArrowLine(45,-15)(60,-15)
 \GCirc(30,15){5}{0.5}
\end{picture}
}
+
\scalebox{0.55}{
\begin{picture}(60,20)(0,-3)
 \ArrowLine(15,15)(0,15)
 \ArrowLine(15,-15)(15,15)
 \ArrowLine(0,-15)(15,-15)
 \Photon(15,15)(45,15){2}{7}
 \Photon(15,-15)(45,-15){2}{7}
 \ArrowLine(60,15)(45,15)
 \ArrowLine(45,15)(45,-15)
 \ArrowLine(45,-15)(60,-15)
 \GCirc(30,-15){5}{0.5}
\end{picture}
}
+
\scalebox{0.55}{
\begin{picture}(60,20)(0,-3)
 \ArrowLine(15,15)(0,15)
 \ArrowLine(15,-15)(15,15)
 \ArrowLine(0,-15)(15,-15)
 \Photon(15,15)(45,-15){2}{7}
 \Photon(15,-15)(45,15){2}{7}
 \ArrowLine(60,15)(45,15)
 \ArrowLine(45,15)(45,-15)
 \ArrowLine(45,-15)(60,-15)
 \GCirc(37.5,7.5){5}{0.5}
\end{picture}
}
+
\scalebox{0.55}{
\begin{picture}(60,20)(0,-3)
 \ArrowLine(15,15)(0,15)
 \ArrowLine(15,-15)(15,15)
 \ArrowLine(0,-15)(15,-15)
 \Photon(15,15)(45,-15){2}{7}
 \Photon(15,-15)(45,15){2}{7}
 \ArrowLine(60,15)(45,15)
 \ArrowLine(45,15)(45,-15)
 \ArrowLine(45,-15)(60,-15)
 \GCirc(37.5,-7.5){5}{0.5}
\end{picture}
}
\Bigg)^*
\nl
\nl
&&
\!+\,
\frac{c_{_{\!P\!S}}}{(2\pi)^3}\!\!\int_{\omega}\!\frac{d^3k}{2\,k_0}\;
{\rm Re}
\Bigg[
\Bigg(\!\!
\scalebox{0.55}{
\begin{picture}(50,20)(0,-3)
 \ArrowLine(15,0)(7.5,10)
 \ArrowLine(7.5,10)(0,20)
 \Photon(7.5,10)(20,20){2}{4}        \Text(25,16)[cb]{$k$}
 \ArrowLine(0,-20)(15,0)
 \Photon(15,0)(35,0){2}{5}
 \ArrowLine(50,20)(35,0)
 \ArrowLine(35,0)(50,-20)
\end{picture}
}
\!\!+\!\!
\scalebox{0.55}{
\begin{picture}(50,20)(0,-3)
 \ArrowLine(15,0)(0,20)
 \ArrowLine(0,-20)(7.5,-10)
 \ArrowLine(7.5,-10)(15,0)
 \Photon(7.5,-10)(20,-20){2}{4}        \Text(24,-22)[cb]{$k$}
 \Photon(15,0)(35,0){2}{5}
 \ArrowLine(50,20)(35,0)
 \ArrowLine(35,0)(50,-20)
\end{picture}
}
\!\Bigg)\!\!
\Bigg(\!\!
\scalebox{0.55}{
\begin{picture}(50,20)(0,-3)
 \ArrowLine(15,0)(0,20)
 \ArrowLine(0,-20)(15,0)
 \Photon(15,0)(35,0){2}{5}
 \ArrowLine(50,20)(35,0)
 \ArrowLine(35,0)(42.5,-10)
 \ArrowLine(42.5,-10)(50,-20)
 \Photon(42.5,-10)(35,-20){2}{4}        \Text(29.5,-22)[cb]{$k$}
 \GCirc(25,0){5}{0.5}
\end{picture}
}
\!\!+\!\!
\scalebox{0.55}{
\begin{picture}(50,20)(0,-3)
 \ArrowLine(15,0)(0,20)
 \ArrowLine(0,-20)(15,0)
 \Photon(15,0)(35,0){2}{5}
 \ArrowLine(50,20)(42.5,10)
 \ArrowLine(42.5,10)(35,0)
 \Photon(42.5,10)(35,20){2}{4}        \Text(29.5,16)[cb]{$k$}
 \ArrowLine(35,0)(50,-20)
 \GCirc(25,0){5}{0.5}
\end{picture}
}
\!\Bigg)^{\!\!\!*}
\!\!+\!
\Bigg(\!\!
\scalebox{0.55}{
\begin{picture}(50,20)(0,-3)
 \ArrowLine(15,0)(0,20)
 \ArrowLine(0,-20)(15,0)
 \Photon(15,0)(35,0){2}{5}
 \ArrowLine(50,20)(35,0)
 \ArrowLine(35,0)(42.5,-10)
 \ArrowLine(42.5,-10)(50,-20)
 \Photon(42.5,-10)(35,-20){2}{4}        \Text(29.5,-22)[cb]{$k$}
\end{picture}
}
\!\!+\!\!
\scalebox{0.55}{
\begin{picture}(50,20)(0,-3)
 \ArrowLine(15,0)(0,20)
 \ArrowLine(0,-20)(15,0)
 \Photon(15,0)(35,0){2}{5}
 \ArrowLine(50,20)(42.5,10)
 \ArrowLine(42.5,10)(35,0)
 \Photon(42.5,10)(35,20){2}{4}        \Text(29.5,16)[cb]{$k$}
 \ArrowLine(35,0)(50,-20)
\end{picture}
}
\!\Bigg)\!\!
\Bigg(\!\!
\scalebox{0.55}{
\begin{picture}(50,20)(0,-3)
 \ArrowLine(15,0)(7.5,10)
 \ArrowLine(7.5,10)(0,20)
 \Photon(7.5,10)(20,20){2}{4}        \Text(25,16)[cb]{$k$}
 \ArrowLine(0,-20)(15,0)
 \Photon(15,0)(35,0){2}{5}
 \ArrowLine(50,20)(35,0)
 \ArrowLine(35,0)(50,-20)
 \GCirc(25,0){5}{0.5}
\end{picture}
}
\!\!+\!\!
\scalebox{0.55}{
\begin{picture}(50,20)(0,-3)
 \ArrowLine(15,0)(0,20)
 \ArrowLine(0,-20)(7.5,-10)
 \ArrowLine(7.5,-10)(15,0)
 \Photon(7.5,-10)(20,-20){2}{4}        \Text(24,-22)[cb]{$k$}
 \Photon(15,0)(35,0){2}{5}
 \ArrowLine(50,20)(35,0)
 \ArrowLine(35,0)(50,-20)
 \GCirc(25,0){5}{0.5}
\end{picture}
}
\!\Bigg)^{\!\!\!*}
\Bigg],
\qquad
\eqa
\bqa
\!\frac{d\sigma^{\rm B\!,\,t}}{d\Omega}\!\!
&=&
c_{_{\!P\!S}}
2\,{\rm Re}
\scalebox{0.55}{
\begin{picture}(40,40)(0,-3)
 \ArrowLine(20,10)(0,30)
 \ArrowLine(40,30)(20,10)
 \Photon(20,10)(20,-10){2}{5}
 \ArrowLine(0,-30)(20,-10)
 \ArrowLine(20,-10)(40,-30)
\end{picture}
}
\Bigg(\,
\scalebox{0.55}{
\begin{picture}(60,20)(0,-3)
 \ArrowLine(15,15)(0,15)
 \ArrowLine(45,15)(15,15)
 \ArrowLine(60,15)(45,15)
 \Photon(15,15)(15,-15){2}{7}
 \Photon(45,15)(45,-15){2}{7}
 \ArrowLine(0,-15)(15,-15)
 \ArrowLine(15,-15)(45,-15)
 \ArrowLine(45,-15)(60,-15)
 \GCirc(15,0){5}{0.5}
\end{picture}
}
+
\scalebox{0.55}{
\begin{picture}(60,20)(0,-3)
 \ArrowLine(15,15)(0,15)
 \ArrowLine(45,15)(15,15)
 \ArrowLine(60,15)(45,15)
 \Photon(15,15)(15,-15){2}{7}
 \Photon(45,15)(45,-15){2}{7}
 \ArrowLine(0,-15)(15,-15)
 \ArrowLine(15,-15)(45,-15)
 \ArrowLine(45,-15)(60,-15)
 \GCirc(45,0){5}{0.5}
\end{picture}
}
+
\scalebox{0.55}{
\begin{picture}(60,20)(0,-3)
 \ArrowLine(15,15)(0,15)
 \ArrowLine(45,15)(15,15)
 \ArrowLine(60,15)(45,15)
 \Photon(15,-15)(45,15){2}{7}
 \Photon(15,15)(45,-15){2}{7}
 \ArrowLine(0,-15)(15,-15)
 \ArrowLine(15,-15)(45,-15)
 \ArrowLine(45,-15)(60,-15)
 \GCirc(37.5,7.5){5}{0.5}
\end{picture}
}
+
\scalebox{0.55}{
\begin{picture}(60,20)(0,-3)
 \ArrowLine(15,15)(0,15)
 \ArrowLine(45,15)(15,15)
 \ArrowLine(60,15)(45,15)
 \Photon(15,-15)(45,15){2}{7}
 \Photon(15,15)(45,-15){2}{7}
 \ArrowLine(0,-15)(15,-15)
 \ArrowLine(15,-15)(45,-15)
 \ArrowLine(45,-15)(60,-15)
 \GCirc(37.5,-7.5){5}{0.5}
\end{picture}
}
\Bigg)^*
\nl
&&
+\,
\frac{c_{_{\!P\!S}}}{(2\pi)^3}\!\!\int_{\omega}\!\frac{d^3k}{2\,k_0}\;
{\rm Re}
\Bigg[
\Bigg(\!
\scalebox{0.55}{
\begin{picture}(40,40)(0,-3)
 \ArrowLine(20,10)(10,20)
 \ArrowLine(10,20)(0,30)
 \Photon(10,20)(20,30){2}{4}        \Text(25,26)[cb]{$k$}
 \ArrowLine(40,30)(20,10)
 \Photon(20,10)(20,-10){2}{5}
 \ArrowLine(0,-30)(20,-10)
 \ArrowLine(20,-10)(40,-30)
\end{picture}
}
\!\!+\!\!
\scalebox{0.55}{
\begin{picture}(40,40)(0,-3)
 \ArrowLine(20,10)(0,30)
 \ArrowLine(40,30)(30,20)
 \ArrowLine(30,20)(20,10)
 \Photon(30,20)(20,30){2}{4}        \Text(16,26)[cb]{$k$}
 \Photon(20,10)(20,-10){2}{5}
 \ArrowLine(0,-30)(20,-10)
 \ArrowLine(20,-10)(40,-30)
\end{picture}
}
\!\Bigg)\!
\Bigg(\!
\scalebox{0.55}{
\begin{picture}(40,40)(0,-3)
 \ArrowLine(20,10)(0,30)
 \ArrowLine(40,30)(20,10)
 \Photon(20,10)(20,-10){2}{5}
 \ArrowLine(0,-30)(10,-20)
 \ArrowLine(10,-20)(20,-10)
 \Photon(10,-20)(20,-30){2}{4}        \Text(24,-32)[cb]{$k$}
 \ArrowLine(20,-10)(40,-30)
 \GCirc(20,0){5}{0.5}
\end{picture}
}
\!\!+\!\!
\scalebox{0.55}{
\begin{picture}(40,40)(0,-3)
 \ArrowLine(20,10)(0,30)
 \ArrowLine(40,30)(20,10)
 \Photon(20,10)(20,-10){2}{5}
 \ArrowLine(0,-30)(20,-10)
 \ArrowLine(20,-10)(30,-20)
 \ArrowLine(30,-20)(40,-30)
 \Photon(30,-20)(20,-30){2}{4}        \Text(15.5,-32)[cb]{$k$}
 \GCirc(20,0){5}{0.5}
\end{picture}
}
\!\Bigg)^{\!\!*}
\!+
\Bigg(\!
\scalebox{0.55}{
\begin{picture}(40,40)(0,-3)
 \ArrowLine(20,10)(0,30)
 \ArrowLine(40,30)(20,10)
 \Photon(20,10)(20,-10){2}{5}
 \ArrowLine(0,-30)(10,-20)
 \ArrowLine(10,-20)(20,-10)
 \Photon(10,-20)(20,-30){2}{4}        \Text(24,-32)[cb]{$k$}
 \ArrowLine(20,-10)(40,-30)
\end{picture}
}
\!\!+\!\!
\scalebox{0.55}{
\begin{picture}(40,40)(0,-3)
 \ArrowLine(20,10)(0,30)
 \ArrowLine(40,30)(20,10)
 \Photon(20,10)(20,-10){2}{5}
 \ArrowLine(0,-30)(20,-10)
 \ArrowLine(20,-10)(30,-20)
 \ArrowLine(30,-20)(40,-30)
 \Photon(30,-20)(20,-30){2}{4}        \Text(15.5,-32)[cb]{$k$}
\end{picture}
}
\!\Bigg)\!
\Bigg(\!
\scalebox{0.55}{
\begin{picture}(40,40)(0,-3)
 \ArrowLine(20,10)(10,20)
 \ArrowLine(10,20)(0,30)
 \Photon(10,20)(20,30){2}{4}        \Text(25,26)[cb]{$k$}
 \ArrowLine(40,30)(20,10)
 \Photon(20,10)(20,-10){2}{5}
 \ArrowLine(0,-30)(20,-10)
 \ArrowLine(20,-10)(40,-30)
 \GCirc(20,0){5}{0.5}
\end{picture}
}
\!\!+\!\!
\scalebox{0.55}{
\begin{picture}(40,40)(0,-3)
 \ArrowLine(20,10)(0,30)
 \ArrowLine(40,30)(30,20)
 \ArrowLine(30,20)(20,10)
 \Photon(30,20)(20,30){2}{4}        \Text(16,26)[cb]{$k$}
 \Photon(20,10)(20,-10){2}{5}
 \ArrowLine(0,-30)(20,-10)
 \ArrowLine(20,-10)(40,-30)
 \GCirc(20,0){5}{0.5}
\end{picture}
}
\!\Bigg)^{\!\!*}
\Bigg],
\eqa
\bqa
\!\frac{d\sigma^{\rm B\!,\,st\!\!\!\!\!}}{d\Omega}\!
&=&
\!c_{_{\!P\!S}}
2{\rm Re}
\Bigg[
\scalebox{0.55}{
\begin{picture}(45,20)(0,-3)
 \ArrowLine(15,0)(0,20)
 \ArrowLine(0,-20)(15,0)
 \Photon(15,0)(30,0){2}{4}
 \ArrowLine(45,20)(30,0)
 \ArrowLine(30,0)(45,-20)
\end{picture}
}\!\!
\bigg(\!
\scalebox{0.55}{
\begin{picture}(60,20)(0,-3)
 \ArrowLine(15,15)(0,15)
 \ArrowLine(45,15)(15,15)
 \ArrowLine(60,15)(45,15)
 \Photon(15,15)(15,-15){2}{7}
 \Photon(45,15)(45,-15){2}{7}
 \ArrowLine(0,-15)(15,-15)
 \ArrowLine(15,-15)(45,-15)
 \ArrowLine(45,-15)(60,-15)
 \GCirc(15,0){5}{0.5}
\end{picture}
}
\!\!+\!\!
\scalebox{0.55}{
\begin{picture}(60,20)(0,-3)
 \ArrowLine(15,15)(0,15)
 \ArrowLine(45,15)(15,15)
 \ArrowLine(60,15)(45,15)
 \Photon(15,15)(15,-15){2}{7}
 \Photon(45,15)(45,-15){2}{7}
 \ArrowLine(0,-15)(15,-15)
 \ArrowLine(15,-15)(45,-15)
 \ArrowLine(45,-15)(60,-15)
 \GCirc(45,0){5}{0.5}
\end{picture}
}
\!\!+\!\!
\scalebox{0.55}{
\begin{picture}(60,20)(0,-3)
 \ArrowLine(15,15)(0,15)
 \ArrowLine(45,15)(15,15)
 \ArrowLine(60,15)(45,15)
 \Photon(15,-15)(45,15){2}{7}
 \Photon(15,15)(45,-15){2}{7}
 \ArrowLine(0,-15)(15,-15)
 \ArrowLine(15,-15)(45,-15)
 \ArrowLine(45,-15)(60,-15)
 \GCirc(37.5,7.5){5}{0.5}
\end{picture}
}
\!\!+\!\!
\scalebox{0.55}{
\begin{picture}(60,20)(0,-3)
 \ArrowLine(15,15)(0,15)
 \ArrowLine(45,15)(15,15)
 \ArrowLine(60,15)(45,15)
 \Photon(15,-15)(45,15){2}{7}
 \Photon(15,15)(45,-15){2}{7}
 \ArrowLine(0,-15)(15,-15)
 \ArrowLine(15,-15)(45,-15)
 \ArrowLine(45,-15)(60,-15)
 \GCirc(37.5,-7.5){5}{0.5}
\end{picture}
}
\!\bigg)^{\!\!\!*}
\!+\!\!\!\!
\scalebox{0.55}{
\begin{picture}(40,40)(0,-3)
 \ArrowLine(20,10)(0,30)
 \ArrowLine(40,30)(20,10)
 \Photon(20,10)(20,-10){2}{5}
 \ArrowLine(0,-30)(20,-10)
 \ArrowLine(20,-10)(40,-30)
\end{picture}
}\!\!\!\!
\bigg(\!
\scalebox{0.55}{
\begin{picture}(60,20)(0,-3)
 \ArrowLine(15,15)(0,15)
 \ArrowLine(15,-15)(15,15)
 \ArrowLine(0,-15)(15,-15)
 \Photon(15,15)(45,15){2}{7}
 \Photon(15,-15)(45,-15){2}{7}
 \ArrowLine(60,15)(45,15)
 \ArrowLine(45,15)(45,-15)
 \ArrowLine(45,-15)(60,-15)
 \GCirc(30,15){5}{0.5}
\end{picture}
}
\!\!+\!\!
\scalebox{0.55}{
\begin{picture}(60,20)(0,-3)
 \ArrowLine(15,15)(0,15)
 \ArrowLine(15,-15)(15,15)
 \ArrowLine(0,-15)(15,-15)
 \Photon(15,15)(45,15){2}{7}
 \Photon(15,-15)(45,-15){2}{7}
 \ArrowLine(60,15)(45,15)
 \ArrowLine(45,15)(45,-15)
 \ArrowLine(45,-15)(60,-15)
 \GCirc(30,-15){5}{0.5}
\end{picture}
}
\!\!+\!\!
\scalebox{0.55}{
\begin{picture}(60,20)(0,-3)
 \ArrowLine(15,15)(0,15)
 \ArrowLine(15,-15)(15,15)
 \ArrowLine(0,-15)(15,-15)
 \Photon(15,15)(45,-15){2}{7}
 \Photon(15,-15)(45,15){2}{7}
 \ArrowLine(60,15)(45,15)
 \ArrowLine(45,15)(45,-15)
 \ArrowLine(45,-15)(60,-15)
 \GCirc(37.5,7.5){5}{0.5}
\end{picture}
}
\!\!+\!\!
\scalebox{0.55}{
\begin{picture}(60,20)(0,-3)
 \ArrowLine(15,15)(0,15)
 \ArrowLine(15,-15)(15,15)
 \ArrowLine(0,-15)(15,-15)
 \Photon(15,15)(45,-15){2}{7}
 \Photon(15,-15)(45,15){2}{7}
 \ArrowLine(60,15)(45,15)
 \ArrowLine(45,15)(45,-15)
 \ArrowLine(45,-15)(60,-15)
 \GCirc(37.5,-7.5){5}{0.5}
\end{picture}
}\!
\bigg)^{\!\!\!*}
\Bigg]
\nl
&&
+\,
\frac{c_{_{\!P\!S}}}{(2\pi)^3}\!\!\int_{\omega}\!\frac{d^3k}{2\,k_0}\;
{\rm Re}
\Bigg[
\Bigg(\!
\scalebox{0.55}{
\begin{picture}(50,20)(0,-3)
 \ArrowLine(15,0)(7.5,10)
 \ArrowLine(7.5,10)(0,20)
 \Photon(7.5,10)(20,20){2}{4}        \Text(25,16)[cb]{$k$}
 \ArrowLine(0,-20)(15,0)
 \Photon(15,0)(35,0){2}{5}
 \ArrowLine(50,20)(35,0)
 \ArrowLine(35,0)(50,-20)
\end{picture}
}
\!\!+\!\!
\scalebox{0.55}{
\begin{picture}(50,20)(0,-3)
 \ArrowLine(15,0)(0,20)
 \ArrowLine(0,-20)(15,0)
 \Photon(15,0)(35,0){2}{5}
 \ArrowLine(50,20)(42.5,10)
 \ArrowLine(42.5,10)(35,0)
 \Photon(42.5,10)(35,20){2}{4}        \Text(29.5,16)[cb]{$k$}
 \ArrowLine(35,0)(50,-20)
\end{picture}
}
\!\Bigg)
\Bigg(\!
\scalebox{0.55}{
\begin{picture}(40,40)(0,-3)
 \ArrowLine(20,10)(0,30)
 \ArrowLine(40,30)(20,10)
 \Photon(20,10)(20,-10){2}{5}
 \ArrowLine(0,-30)(10,-20)
 \ArrowLine(10,-20)(20,-10)
 \Photon(10,-20)(20,-30){2}{4}        \Text(24,-32)[cb]{$k$}
 \ArrowLine(20,-10)(40,-30)
 \GCirc(20,0){5}{0.5}
\end{picture}
}
\!\!+\!\!
\scalebox{0.55}{
\begin{picture}(40,40)(0,-3)
 \ArrowLine(20,10)(0,30)
 \ArrowLine(40,30)(20,10)
 \Photon(20,10)(20,-10){2}{5}
 \ArrowLine(0,-30)(20,-10)
 \ArrowLine(20,-10)(30,-20)
 \ArrowLine(30,-20)(40,-30)
 \Photon(30,-20)(20,-30){2}{4}        \Text(15.5,-32)[cb]{$k$}
 \GCirc(20,0){5}{0.5}
\end{picture}
}
\!\Bigg)^{\!\!*}
\!\!
+
\Bigg(\!
\scalebox{0.55}{
\begin{picture}(50,20)(0,-3)
 \ArrowLine(15,0)(0,20)
 \ArrowLine(0,-20)(7.5,-10)
 \ArrowLine(7.5,-10)(15,0)
 \Photon(7.5,-10)(20,-20){2}{4}        \Text(24,-22)[cb]{$k$}
 \Photon(15,0)(35,0){2}{5}
 \ArrowLine(50,20)(35,0)
 \ArrowLine(35,0)(50,-20)
\end{picture}
}
\!\!+\!\!
\scalebox{0.55}{
\begin{picture}(50,20)(0,-3)
 \ArrowLine(15,0)(0,20)
 \ArrowLine(0,-20)(15,0)
 \Photon(15,0)(35,0){2}{5}
 \ArrowLine(50,20)(35,0)
 \ArrowLine(35,0)(42.5,-10)
 \ArrowLine(42.5,-10)(50,-20)
 \Photon(42.5,-10)(35,-20){2}{4}        \Text(29.5,-22)[cb]{$k$}
\end{picture}
}
\!\Bigg)
\Bigg(\!
\scalebox{0.55}{
\begin{picture}(40,40)(0,-3)
 \ArrowLine(20,10)(10,20)
 \ArrowLine(10,20)(0,30)
 \Photon(10,20)(20,30){2}{4}        \Text(25,26)[cb]{$k$}
 \ArrowLine(40,30)(20,10)
 \Photon(20,10)(20,-10){2}{5}
 \ArrowLine(0,-30)(20,-10)
 \ArrowLine(20,-10)(40,-30)
 \GCirc(20,0){5}{0.5}
\end{picture}
}
\!\!+\!\!
\scalebox{0.55}{
\begin{picture}(40,40)(0,-3)
 \ArrowLine(20,10)(0,30)
 \ArrowLine(40,30)(30,20)
 \ArrowLine(30,20)(20,10)
 \Photon(30,20)(20,30){2}{4}        \Text(16,26)[cb]{$k$}
 \Photon(20,10)(20,-10){2}{5}
 \ArrowLine(0,-30)(20,-10)
 \ArrowLine(20,-10)(40,-30)
 \GCirc(20,0){5}{0.5}
\end{picture}
}
\!\Bigg)^{\!\!*}
\nl
&&
\qquad\qquad\qquad\qquad
+
\Bigg(\!
\scalebox{0.55}{
\begin{picture}(40,40)(0,-3)
 \ArrowLine(20,10)(10,20)
 \ArrowLine(10,20)(0,30)
 \Photon(10,20)(20,30){2}{4}        \Text(25,26)[cb]{$k$}
 \ArrowLine(40,30)(20,10)
 \Photon(20,10)(20,-10){2}{5}
 \ArrowLine(0,-30)(20,-10)
 \ArrowLine(20,-10)(40,-30)
\end{picture}
}
\!\!+\!\!
\scalebox{0.55}{
\begin{picture}(40,40)(0,-3)
 \ArrowLine(20,10)(0,30)
 \ArrowLine(40,30)(20,10)
 \Photon(20,10)(20,-10){2}{5}
 \ArrowLine(0,-30)(10,-20)
 \ArrowLine(10,-20)(20,-10)
 \Photon(10,-20)(20,-30){2}{4}        \Text(24,-32)[cb]{$k$}
 \ArrowLine(20,-10)(40,-30)
\end{picture}
}
\!\Bigg)
\Bigg(\!
\scalebox{0.55}{
\begin{picture}(50,20)(0,-3)
 \ArrowLine(15,0)(0,20)
 \ArrowLine(0,-20)(15,0)
 \Photon(15,0)(35,0){2}{5}
 \ArrowLine(50,20)(35,0)
 \ArrowLine(35,0)(42.5,-10)
 \ArrowLine(42.5,-10)(50,-20)
 \Photon(42.5,-10)(35,-20){2}{4}        \Text(29.5,-22)[cb]{$k$}
 \GCirc(25,0){5}{0.5}
\end{picture}
}
\!\!+\!\!
\scalebox{0.55}{
\begin{picture}(50,20)(0,-3)
 \ArrowLine(15,0)(0,20)
 \ArrowLine(0,-20)(15,0)
 \Photon(15,0)(35,0){2}{5}
 \ArrowLine(50,20)(42.5,10)
 \ArrowLine(42.5,10)(35,0)
 \Photon(42.5,10)(35,20){2}{4}        \Text(29.5,16)[cb]{$k$}
 \ArrowLine(35,0)(50,-20)
 \GCirc(25,0){5}{0.5}
\end{picture}
}
\!\Bigg)^{\!\!*}
\!\!
+
\Bigg(\!
\scalebox{0.55}{
\begin{picture}(40,40)(0,-3)
 \ArrowLine(20,10)(0,30)
 \ArrowLine(40,30)(20,10)
 \Photon(20,10)(20,-10){2}{5}
 \ArrowLine(0,-30)(20,-10)
 \ArrowLine(20,-10)(30,-20)
 \ArrowLine(30,-20)(40,-30)
 \Photon(30,-20)(20,-30){2}{4}        \Text(15.5,-32)[cb]{$k$}
\end{picture}
}
\!\!+\!\!
\scalebox{0.55}{
\begin{picture}(40,40)(0,-3)
 \ArrowLine(20,10)(0,30)
 \ArrowLine(40,30)(30,20)
 \ArrowLine(30,20)(20,10)
 \Photon(30,20)(20,30){2}{4}        \Text(16,26)[cb]{$k$}
 \Photon(20,10)(20,-10){2}{5}
 \ArrowLine(0,-30)(20,-10)
 \ArrowLine(20,-10)(40,-30)
\end{picture}
}
\!\Bigg)
\Bigg(\!
\scalebox{0.55}{
\begin{picture}(50,20)(0,-3)
 \ArrowLine(15,0)(7.5,10)
 \ArrowLine(7.5,10)(0,20)
 \Photon(7.5,10)(20,20){2}{4}        \Text(25,16)[cb]{$k$}
 \ArrowLine(0,-20)(15,0)
 \Photon(15,0)(35,0){2}{5}
 \ArrowLine(50,20)(35,0)
 \ArrowLine(35,0)(50,-20)
 \GCirc(25,0){5}{0.5}
\end{picture}
}
\!\!+\!\!
\scalebox{0.55}{
\begin{picture}(50,20)(0,-3)
 \ArrowLine(15,0)(0,20)
 \ArrowLine(0,-20)(7.5,-10)
 \ArrowLine(7.5,-10)(15,0)
 \Photon(7.5,-10)(20,-20){2}{4}        \Text(24,-22)[cb]{$k$}
 \Photon(15,0)(35,0){2}{5}
 \ArrowLine(50,20)(35,0)
 \ArrowLine(35,0)(50,-20)
 \GCirc(25,0){5}{0.5}
\end{picture}
}
\!\Bigg)^{\!\!*}
\Bigg].
\qquad
\eqa
In total we find:
\bqa
\frac{d\sigma^{\rm B}}{d\Omega} = 
\frac{\alpha^3}{s\pi}\,\Bigg\{ &&
  \frac{1 \!-\! 2x \!+\! 2x^2}{2}\bigg[ 2\,B_s\,{\rm Re}\,\Pi(s) \bigg]
- {\rm Re}\,\bigg[ B(s,t,u) \!-\! B(s,u,t) \bigg]
\nl
&&
+ \,\frac{2 \!-\! 2x \!+\! x^2}{2\,x^2}\bigg[ 2\,B_t\,\Pi(t) \bigg]
- {\rm Re}\,\bigg[ B(t,s,u) \!-\! B(t,u,s) \bigg]
\nl
&&
- \,\frac{1 \!-\! 2x \!+\! x^2}{x}
  \bigg[ B_t\,{\rm Re}\,\Pi(s) + B_s\,\Pi(t) \bigg]
+ {\rm Re}\,\bigg[ x\,B(t,s,u) \!+\! \frac{1}{x}\,B(s,t,u) \bigg]
\Bigg\},
\eqa
where
\bq
B_s =
  2\ln\!\frac{2\omega}{\sqrt{s}}\ln\frac{t}{u}
- \ln\frac{-t}{s}\ln\frac{-u}{s} - 2\,\li_2\left(\frac{-t}{s}\right) + \zeta(2),
\qquad
B_t =
- 2\ln\!\frac{2\omega}{\sqrt{s}}\ln\frac{-u}{s}
- \li_2\left(\frac{-t}{s}\right) + 3\,\zeta(2),
\eq
and
\bq
B(a,b,c) = B_A(a,b,c) + B_B(a,b,c),
\qquad
B_j(a,b,c) = 
\frac{\alpha}{3\pi}
\int_{4m^2}^{\infty}\!\!\frac{dz}{z}R(z)\,\xi_j(a,b,c,z\!-\!i\ep),
\qquad\quad
j= A,B;
\label{eq:B}
\eq
\bqaa
\xi_A(a,b,c,z) &=&
\frac{c^2}{a\,(z\!-\!a)}\Bigg[ 
  2\ln\!\frac{c}{b\!+\!i\ep}\ln\!\bigg(\! 1 \!-\! \frac{a}{z} \!\bigg)
- \li_2\bigg(\! 1 \!+\! \frac{b}{z} \!\bigg)
+ \li_2\bigg(\! 1 \!+\! \frac{c}{z} \!\bigg)
\Bigg],
\nl
\xi_B(a,b,c,z) &=& 
  \frac{c}{a}\Bigg[ 
    \left( \frac{z}{a} \!-\! 1 \!\right)
    \ln\!\bigg(\! 1 \!-\! \frac{a}{z} \!\bigg)
  + \ln\!\frac{-b}{z}
  \Bigg]
+ \;\frac{c-b-z}{a}\Bigg[ 
  \ln\frac{b\!+\!i\ep}{-a}\ln\!\bigg(\! 1 \!-\! \frac{a}{z} \!\bigg)
- \li_2\bigg(\! 1 \!-\! \frac{a}{z} \!\bigg)
+ \li_2\bigg(\! 1 \!+\! \frac{b}{z} \!\bigg)
\Bigg].
\eqaa
The part proportional to $\ln(2\omega/\sqrt{s})$ has been displayed 
separately and, as stated before, is proportional to $\Pi(s)$ or $\Pi(t)$.
As discussed in the introduction, the functions $\xi_A$ and $\xi_B$, corresponding
to the box diagram with a massive and a massless vector boson, can be directly 
read off from the literature \cite{Brown:1983jv}.

\vspace{0.5cm}
For the two-loop hadronic contributions to cross section we find (without 
the trivial vacuum polarization)
\bqa
\frac{d\sigma_{\rm had}}{d\Omega} =
\;
\frac{d\sigma_{\rm red}^{\rm V}}{d\Omega} 
\!\!\!\!
&&
\;
\;+ \;\,\frac{d\sigma_{\rm red}^{\rm B}}{d\Omega} 
\;+ \;\frac{d\sigma^{\rm V}}{d\Omega} 
\;+ \;\frac{d\sigma^{\rm B}}{d\Omega} 
\nl
=
\frac{\alpha^3}{s\pi}\,\Bigg\{\!\!\!\! &&
  \frac{1 \!-\! 2x \!+\! 2x^2}{2}\bigg[
    2\,\Big( 2\,V_s^\gamma + B_s^\gamma + B_s \!\Big)\,{\rm Re}\,\Pi(s)
  + 2\,\pi\ln\!\frac{t}{u}\,{\rm Im}\,\Pi(s) + 4\,{\rm Re}\,V(s)
  \bigg]
\nl
&&
+ \,\frac{2 \!-\! 2x \!+\! x^2}{2\,x^2}\,\bigg[ 
      2\,\Big( 2\,V_t^\gamma + B_t^\gamma + B_t \!\Big)\,\Pi(t) + 4\,V(t)
    \bigg]
\nl
&&
- \,\frac{1 \!-\! 2x \!+\! x^2}{x}\bigg[
      \Big( V_s^\gamma + V_t^\gamma + B_t^\gamma + B_t \Big)\,{\rm Re}\,\Pi(s)
    + \Big( V_s^\gamma + V_t^\gamma + B_s^\gamma + B_s \Big)\,\Pi(t)
\nl
&&
\qquad\qquad\qquad
    + \,\pi\Big(\! \ln\!\frac{-t}{s} + \frac{3}{2} \,\Big)\,{\rm Im}\,\Pi(s)
    + 2\,V(t) + 2\,{\rm Re}\,V(s)
  \bigg]
\nl
&&
- \,{\rm Re}\,\bigg[ 
    \Big(\! 
    B^\gamma(s,t) \!-\! B^\gamma(s,u) \!+\! \frac{t}{s} B^\gamma(t,s) 
    \!\Big)\,\Pi^*\!(s)
  + \Big(\! 
    B^\gamma(t,s) \!-\! B^\gamma(t,u) \!+\! \frac{s}{t}B^\gamma(s,t) 
    \!\Big)\,\Pi(t) 
\nl
&&
\qquad\qquad
  - \frac{u}{t}\,B(s,t,u) - B(s,u,t) - \frac{u}{s}B(t,s,u) - B(t,u,s)
  \bigg]
\,\Bigg\}.
\label{eq:res}
\eqa
\section{Evaluation of the dispersion integrals}
In \eqn{eq:res}, the total cross section is written in terms of the 
building blocks $\Pi(q^2)$, $V(q^2)$ and $B(a,b,c)$  defined 
in~\eqn{eq:Pi}, \eqn{eq:V} and \eqn{eq:B} respectively.
In the hadronic case, given a suitable parametrization of $R(s)$, these 
dispersion integrals have to be integrated numerically.
Therefore, before attempting the evaluation of the cross section, all 
sources of numerical instability must be cured.

The expression for $\Pi(q^2)$ in~\eqn{eq:Pi} is very simple, but reveals the 
presence of a pole of the integrand at $z=q^2+i\ep$. 
The simplest way to get rid of it is to add and subtract $R(q^2)$ in the 
integrand for $q^2>0$ ($s$-channel). 
After the useful change of variable $z= 4m^2/y$, we get:
\bq
\Pi(t)\!= 
\frac{\alpha}{3\pi}\!\!\int_0^1\!\!\!\!dy\frac{t}{y t\!-\!4m^{\!2}}
R\bigg(\!\frac{4m^{\!2}}{y}\!\bigg),
\quad\,
\Pi(s)\!= 
\frac{\alpha}{3\pi}\Bigg\{
  \!\ln\!\!\left(\! 1 \!-\! \frac{s}{4m^{\!2}\!-\!i\ep} \right)\!\!R(s)
+ \!\int_0^1\!\!\!\!dy\frac{s}{y s\!-\!4m^{\!2}}
  \bigg[\! R\bigg(\!\frac{4m^{\!2}}{y}\!\bigg) \!-\! R(s) \bigg]
\Bigg\}.
\eq
The integral for $V(q^2)$ given in  \eqn{eq:V} does not show any pole in the 
integration domain and is directly accessible to a numerical evaluation.
However, its convergence in the high energy integration region can be improved 
introducing the asymptotic, approximately constant value $R(\infty)$ of the 
$R$-ratio.
To this purpose, let us recall the results from~\cite{Kniehl:1988id} for the 
vertex $V$, which can be rewritten in the form:
\bq
V(q^2)= 
\frac{\alpha}{3\pi}\Bigg\{
  R(\infty)\int_0^1\!\frac{dy}{y}\rho\,(q^2\!,\frac{4m^2}{y})
+ \int_0^1\!\frac{dy}{y}\rho\,(q^2\!,\frac{4m^2}{y})
  \bigg[ R\bigg(\frac{4m^2}{y}\bigg) - R(\infty) \bigg]
\Bigg\}.
\eq
The first one of these integrals can be solved exactly:
\bqa
I_\rho(r) = 
\int_0^1\!\frac{dy}{y}\rho(q^2\!,\frac{4m^2}{y})
&=&
- \,\frac{1}{12}\ln^3(-r) 
- \ln(-r)\bigg[ 
    \zeta(2) + \frac{7}{8} + \frac{1}{4r}
  + \frac{1}{2}\li_2\left(-\frac{1}{r}\right)
  \bigg]
\nl
&&
+ \bigg( \frac{3}{4} + \frac{1}{r} + \frac{1}{4r^2} \bigg)
  \big[ \zeta(2) - \li_2(1+r) \big]
+ \frac{15}{16} + \frac{1}{4r} - \li_3\left(-\frac{1}{r}\right),
\eqa
where $r=q^2/(4m^2-i\ep)$, and the second integral converges well in the 
large momentum region.

A similar approach can be adopted for integrating the kernel from the box 
diagram $B(a,b,c)$ defined in \eqn{eq:B}. 
The function $\xi_A$ has a good high energy behaviour, but has a pole 
(for $a=s>0$) in $z=a+i\ep$ and can be treated in the same way as $\Pi(q^2)$: 
\bqa
B_A(t,b,c) &=&
\frac{\alpha}{3\pi}\,\int_0^1\!\frac{dy}{y}\,
\xi_A(t,b,c,\frac{4m^2}{y})\,R\bigg(\frac{4m^2}{y}\bigg),
\nl
B_A(s,b,c) &=&
\frac{\alpha}{3\pi}\,\Bigg\{
  R(s)\!\int_0^1\!\frac{dy}{y}\,\xi_A(s,b,c,\frac{4m^2}{y})
+ \int_0^1\!\frac{dy}{y}\,\xi_A(s,b,c,\frac{4m^2}{y})
  \bigg[ R\bigg(\frac{4m^2}{y}\bigg) - R(s) \bigg]
\Bigg\}.
\eqa
The first integral in the expression for $B_A(s,b,c)$ is then given by:
\bqa
I_A(s,b,c) = 
\!\!
\int_0^1\!\frac{dy}{y}\xi_A(s,b,c,\frac{4m^{\!2}}{y})\! 
&=&
\frac{c^2}{s^2}\,\bigg[\,
  \ln\frac{c}{b\!+\!i\ep}\,\ln^2\frac{4m^{\!2}\!-\!s}{4m^2}
+  J_A\bigg( -\frac{b}{s} \bigg)
-  J_A\bigg( -\frac{c}{s} \bigg)
\,\bigg],
\eqa
where
\bqa
\!\!
J_A(x)\! &=&
  \frac{1}{6}\ln^3(-xr)
+ \frac{1}{6}\ln^3\!\frac{xr}{-\bar{r}}
- \frac{1}{2}\ln x\ln^2(-xr)
- \frac{1}{2}\ln(\bar{x}r)\ln^{\!2}(1\!-\!\bar{x}r)
+ \frac{1}{2}\ln\!\frac{1\!-\!\bar{x}r}{x}
  \ln^2\!\frac{xr}{-\bar{r}}
\nl
&-& \ln\bar{r}\,\li_2(1\!-\!\bar{x}r)
 +  \ln\!\frac{xr}{-\bar{r}}\,
    \li_2\bigg(\! \frac{\bar{r}\bar{x}}{-x} \!\bigg)
 +  \ln(\bar{x}r)\bigg[ 
          \li_2(r)
    \!-\! \li_2(\bar{x}r)
    \!-\! \li_2\bigg(\! \frac{xr}{1\!-\!\bar{x}r} \!\bigg)
    \!\bigg]
\nl
&+& \ln(-xr)\bigg[
          \li_2(r)
    \!-\! \li_2(\bar{x}r)
    \!-\! \li_2\bigg(\! \frac{-\bar{x}}{x} \!\bigg)
    \!\bigg]
 -  \li_3\bigg(\! \frac{-\bar{x}}{x} \!\bigg)
 +  \li_3(\bar{x}r)
 +  \li_3\bigg(\! \frac{\bar{r}\bar{x}}{-x} \!\bigg)
 +  S_{1\!2}\bigg(\! \frac{1\!-\!\bar{x}r}{\bar{r}} \!\bigg).
\eqa
In the last expression we have introduced $\bar{x}= 1-x$, $r=s/(4m^2-i\ep)$, 
$\bar{r}= 1-r$.
On the other hand, $B_B$ can be computed following the same procedure used 
for $V(q^2)$:
\bqa
B_B(a,b,c) &=&
\frac{\alpha}{3\pi}\,\Bigg\{
  R(\infty)\!\int_0^1\!\frac{dy}{y}\,\xi_B(a,b,c,\frac{4m^2}{y})
+ \int_0^1\!\frac{dy}{y}\,\xi_B(a,b,c,\frac{4m^2}{y})
  \bigg[ R\bigg(\frac{4m^2}{y}\bigg) - R(\infty) \bigg]
\Bigg\}.
\eqa
where for the first integral we have:
\bqa
I_B(a,b,c) &=& 
\int_0^1\!\!\frac{dy}{y}\xi_B(a,\!b,\!c,\frac{4m^{\!2}}{y})\! 
\nl
&=&
  \frac{c}{a}\,\bigg[ 
  \li_2\bigg(\! \frac{a}{4m^{\!2}} \!\bigg) - \ln\!\frac{-b}{4m^{\!2}} 
  \bigg]
+ \frac{c-\!b}{a}\Bigg\{\! 
    \ln\!\frac{-b}{4m^2} \bigg[
      \li_2\!\bigg(\! \frac{-b}{4m^{\!2}} \!\bigg)
    - \li_2\!\bigg(\! \frac{a}{4m^{\!2}} \!\bigg)
    \bigg]
  + 2\,\li_3\!\bigg(\! \frac{a}{4m^{\!2}} \!\bigg) 
  - 2\,\li_3\!\bigg(\! \frac{-b}{4m^{\!2}} \!\bigg)
  \!\Bigg\}
\nl
&&
 + \,\frac{4m^{\!2}\!-\!a}{a}\bigg[\!
      \bigg(\!\! \ln\!\frac{-b}{4m^{\!2}} \!-\! \frac{c}{a} \bigg)\!
      \ln\!\frac{4m^{\!2}\!-\!a}{4m^2}
    \!+\! \li_2\bigg(\! \frac{a}{4m^{\!2}} \!\bigg)
    \!\bigg]
 - \frac{4m^{\!2}\!+\!b}{a}\bigg[\!
      \ln\!\frac{-b}{4m^{\!2}}\ln\!\frac{4m^{\!2}\!+\!b}{4m^2}
    \!+\! \li_2\bigg(\! \frac{-b}{4m^{\!2}} \!\bigg)
    \!\bigg].
\eqa
\subsection{High energy limit}
In the high energy limit, i.e. for $\sqrt{s}$ and $\sqrt{-t}$ far larger 
than the energy above which $R(s)$ approaches (sufficiently rapidly) 
$R(\infty)$, the building blocks $\Pi(q^2)$, $V(q^2)$ and $B(a,b,c)$ can be 
expressed in terms of the moments $R_n$ defined through:
\bq
R_n= 
\int_0^1\!\frac{dx}{x}\frac{\ln^n x}{n!}
\bigg[ R\bigg(\frac{4m^2}{x}\bigg) - R(\infty) \bigg].
\eq
The large $q^2$ behaviour of the vacuum polarization is then given by:
\bq
\Pi(q^2)= 
\frac{\alpha}{3\pi}\,\bigg( R(\infty)\ln\frac{-q^2}{4m^2-i\ep} + R_0 \bigg),
\label{eq:Pi-he}
\eq
and $V$ takes the following form:
\bqa
V(q^2) 
&=&
\frac{\alpha}{3\pi}\Bigg\{
R(\infty)I_\rho(r)
+ R_0\bigg[ 
  - \frac{1}{4}\ln^2(-r) + \frac{3}{4}\ln(-r) - \zeta(2) - \frac{7}{8}
  \bigg]
+ R_1\bigg[ - \frac{1}{2}\ln(-r) + \frac{3}{4} \bigg]
- \frac{1}{2}\,R_2
\Bigg\},
\nl
I_\rho(r) &=& 
- \,\frac{1}{12}\ln^3(-r) 
+ \frac{3}{8}\ln^2(-r)
- \bigg[ \zeta(2) + \frac{7}{8} \bigg]\ln(-r)
+ \frac{3}{2}\,\zeta(2) + \frac{15}{16}
+ {\cal O}(|r|^{-1}).
\label{eq:V-he}
\eqa
Similarly, the building block $B(a,b,c)$, in the high energy limit 
$s,|t|,|u| \gg 4\,m^2$ is given by:
\bq
B(a,\!b,\!c)
=
\frac{\alpha}{3\pi}\bigg\{
  R(\infty)\,I_\xi(a,\!b,\!c)\,
+ \,R_0\bigg[
    \frac{c^{\!2}}{a^{\!2}}L_{b\,c}
    \bigg(\!
    \ln\!\frac{-a}{4m^{\!2}} - \frac{L_{b} \!+\! L_{c}}{2}
    \!\bigg)
  - \frac{c\!-\!b}{2a}\Big(\! L_{b}^{\!2} + 6\zeta(2) \!\Big)
  + \frac{c}{a}L_{b}
  \bigg]\,
+ \,R_1\frac{c^2}{a^2}L_{b\,c}
\bigg\},
\,
\eq
where
\bqa
\!\!\!\!
I_\xi(a,b,c)
&=&
  \frac{c^2}{a^2}\,\bigg\{
    L_{b\,c}\,
    \bigg[\,
      \frac{1}{2}\ln^{\!2}\!\frac{-a}{4m^{\!2}}
    - \frac{1}{2}\ln\!\frac{-a}{4m^{\!2}}\,\big( L_b + L_c \big)
    + \zeta(2)
    \bigg]
  - J(b,c)
  + J(c,b)
  \bigg\}
\nl
&&
- \,\frac{c-\!b}{2a}\,\Big[ L_b^2 + 6\zeta(2) \Big]\!
  \bigg(\! \ln\!\frac{-a}{4m^{\!2}} - 1 + \frac{L_b}{3} \bigg)
+ \frac{c}{a}\,\bigg[
    L_b\bigg(\!\! \ln\!\frac{-a}{4m^{\!2}} \!-\! 1 \!\bigg)
  - \frac{L_b^2}{2}
  - 5\,\zeta(2)
  \bigg]
+ {\cal O}(m^{\!2}).
\eqa
In the last equations we have introduced:
\bq
L_b= \ln\frac{b\!+\!i\ep}{a},
\qquad
L_c= \ln\frac{c\!+\!i\ep}{a},
\qquad
L_{b\,c}= \ln\frac{b\!+\!i\ep}{c},
\qquad
J(x,y)
=
  S_{1\!2}\bigg(\! \frac{-x}{y\!-\!i\ep} \bigg) 
+ i\,\pi\,\li_2\bigg(\! \frac{-x}{y\!-\!i\ep} \bigg).
\label{eq:J}
\eq
\subsection{The leptonic contribution}
With these ingredients the two-loop result induced by massive and light 
leptons is easily recovered. 
\bq
\frac{d\sigma_l}{d\Omega} =
  \frac{d\sigma_{{\rm red}\!,\,l}^{\rm V}}{d\Omega} 
+ \frac{d\sigma_{{\rm red}\!,\,l}^{\rm B}}{d\Omega} 
+ \frac{d\sigma_{l}^{\rm V}}{d\Omega} 
+ \frac{d\sigma_{l}^{\rm B}}{d\Omega}.
\label{eq:lept}
\eq
All ingredients are obtained from the corresponding expressions for 
the hadronic case using the $R$-ratio:
\bq
R_l(z)= \bigg( 1 + \frac{4m_l^2}{2\,z} \bigg)\sqrt{1 - \frac{4m_l^2}{z}}.
\label{eq:Rl}
\eq
Numerical evaluations for the muon and $\tau$-lepton will be presented below. 
In the high energy limit we will use the moments~\cite{Kniehl:1988id}:
\bqa
R_l(\infty)&=& 1,
\qquad
R_{l\!,0}= \ln\!4 - \frac{5}{3},
\qquad
R_{l,1}= \frac{1}{2}\ln^2\!4 - \frac{5}{3}\ln\!4 + \frac{28}{9} - \zeta(2),
\nl
R_{l,2} &=& 
  \frac{1}{6}\ln^3\!4 - \frac{5}{6}\ln^2\!4
+ 2 \Big[ \frac{28}{9} - \zeta(2) \Big]\ln\!4
+ 2\zeta(3) + \frac{5}{3}\zeta(2) - \frac{164}{27}.
\eqa
The integral over $z$ can then be analytically evaluated, giving for the 
building blocks:
\bqa
\Pi_l(q^2)
&=&
- \frac{\alpha}{3\pi}\bigg( \ln\!\frac{-q^2}{m_l^2} - \frac{5}{3} \bigg),
\nl
V_l(q^2)
&=&
- \frac{\alpha}{3\pi}\bigg\{\,
  \frac{1}{12}\ln^{3}\!\frac{-q^2}{m_l^2}
- \frac{19}{24}\!\ln^{2}\!\frac{-q^2}{m_l^2}
+ \frac{1}{2}\bigg[ \zeta(2) + \frac{265}{36} \bigg]\ln\!\frac{-q^2}{m_l^2}
+ \zeta(3) 
- \frac{19}{12}\,\zeta(2) 
- \frac{3355}{432}
\,\bigg\},
\nl
B_l(a,\!b,\!c)
&=&
- \frac{\alpha}{3\pi}\bigg\{
  - \,\frac{c^2}{a^2}\bigg[\,
         \frac{1}{2}L_{b\,c}\bigg(\!
            \ln^{\!2}\frac{-a}{m^{\!2}}
          - \ln\!\frac{-a}{m^{\!2}}
            \bigg(\! L_b \!+\! L_c \!+\! \frac{10}{3} \bigg)
          + \frac{5}{3}\big( L_b \!+\! L_c \big)
          + \frac{56}{9}
          \bigg)
    - J(b,\!c)
    + J(c,\!b)
    \bigg]
\nl
&&
\qquad\quad
  + \,\frac{c-\!b}{2a}\,\Big[ L_b^2 + 6\,\zeta(2) \Big]
    \bigg(
    \ln\!\frac{-a}{m_l^{\!2}} - \frac{8}{3} + \frac{1}{3}\,L_b
    \bigg)
  - \frac{c}{a}\,\bigg[\,
    L_b\bigg( \ln\!\frac{-a}{m_l^{\!2}} - \frac{8}{3} \bigg)
    - \frac{1}{2}\,L_b^2
    - 5\,\zeta(2)
    \bigg]
\bigg\},
\qquad
\label{eq:lept-hea}
\eqa
where 
$L_b$, $L_c$, $L_{b\,c}$ and $J(x,y)$ were defined in \eqn{eq:J}.
In the previous formula the $i\ep$ prescription is implicit in the squared 
lepton mass ($m_l^2 \to m_l^2-i\ep$) and gives the rule to extract the proper 
imaginary part of the logarithms.
For electron loops the vertex correction differs by a 
constant~\cite{Burgers:1985qg}:
\bq
V_e(q^2)=
- \frac{\alpha}{3\pi}\bigg[\,
  \frac{1}{12}\ln^{3}\!\frac{-q^2}{m_e^2}
- \frac{19}{24}\!\ln^{2}\!\frac{-q^2}{m_e^2}
+ \frac{1}{2}\bigg[ \zeta(2) + \frac{265}{36} \bigg]\ln\!\frac{-q^2}{m_e^2}
+ \frac{3}{4}\,\zeta(2) 
- \frac{383}{36}
\,\bigg],
\eq
the remaining corrections are identical.

In order to obtain the total leptonic corrections, the contributions 
of the one- and two-loop vacuum polarization function have to be added:
\bq
\frac{d\sigma_l^{\rm tot}}{d\Omega} =
  \frac{d\sigma_l}{d\Omega} 
+ \frac{d\sigma^\Pi_l}{d\Omega} 
+ \frac{d\sigma_{l}^{\rm S}}{d\Omega}.
\eq
The second term can be obtained from \eqn{eq:Pi} with the substitution 
$\Pi \to \Pi_l$, while last term can be computed taking from the 
literature the expression for the leptonic contribution to the 
two-loop vacuum polarization function (\fig{fig:se2}):
\bq
\nl
\frac{d\sigma_{l}^{\rm S}}{d\Omega} = 
\frac{\alpha^2}{s}\,\Bigg\{
  \frac{1 \!-\! 2x \!+\! 2x^2}{2}\,\Big[ 2\,{\rm Re}\Pi_l^{(2)}(s) \Big]
+ \frac{2 \!-\! 2x \!+\! x^2}{2\,x^2}\,\Big[ 2\,\Pi_l^{(2)}(t) \Big]
- \,\frac{1 \!-\! 2x \!+\! x^2}{x}\,
  \Big[ {\rm Re}\Pi_l^{(2)}(s) + \Pi_l^{(2)}(s) \Big]
\!\Bigg\},
\eq
where in the high-energy limit\footnote{For general $m_l^2/|q^2|$ the 
result can be found in~\cite{Kallen:1955fb}}
\bq
\Pi_l^{(2)}(q^2)= 
\frac{\alpha^2}{4\pi^2}\,\bigg[
\ln\frac{-\,q^2}{m_l^2-i\ep} - \frac{5}{6} + 4\,\zeta(3)
\bigg].
\eq
Comparing our analytical result with~\cite{Becher:2007cu}, we find 
perfect agreement.
\begin{figure}[!h]
$$
\frac{d\sigma_l^{\rm S}}{d\Omega} =
c_{_{P\!S}}
2\,{\rm Re}
\Bigg(\,
\scalebox{0.7}{
\begin{picture}(50,20)(0,-3)
 \ArrowLine(15,0)(0,20)
 \ArrowLine(0,-20)(15,0)
 \Photon(15,0)(35,0){2}{5}
 \ArrowLine(50,20)(35,0)
 \ArrowLine(35,0)(50,-20)
\end{picture}
}
+
\scalebox{0.7}{
\begin{picture}(40,40)(0,-3)
 \ArrowLine(20,10)(0,30)
 \ArrowLine(40,30)(20,10)
 \Photon(20,10)(20,-10){2}{5}
 \ArrowLine(0,-30)(20,-10)
 \ArrowLine(20,-10)(40,-30)
\end{picture}
}
\Bigg)
\Bigg(\,
\scalebox{0.7}{
\begin{picture}(70,20)(0,-3)
 \ArrowLine(15,0)(0,20)
 \ArrowLine(0,-20)(15,0)
 \Photon(15,0)(25,0){2}{3}
 \ArrowArc(35,0)(10,0,180)
 \PhotonArc(35,10)(10,210,330){1}{7}
 \ArrowArc(35,0)(10,180,360)
 \Photon(45,0)(55,0){2}{3}
 \ArrowLine(70,20)(55,0)
 \ArrowLine(55,0)(70,-20)
 \Text(35,15)[cb]{$l$}
\end{picture}
}
+
\scalebox{0.7}{
\begin{picture}(40,40)(0,-3)
 \ArrowLine(20,20)(0,40)
 \ArrowLine(40,40)(20,20)
 \Photon(20,20)(20,10){2}{3}
 \ArrowArc(20,0)(10,-90,90)
 \PhotonArc(30,0)(10,120,240){1}{7}
 \ArrowArc(20,0)(10,90,270)
 \Photon(20,-10)(20,-20){2}{3}
 \ArrowLine(0,-40)(20,-20)
 \ArrowLine(20,-20)(40,-40)
 \Text(36,-2)[cb]{$l$}
\end{picture}
}
+
\scalebox{0.7}{
\begin{picture}(70,20)(0,-3)
 \ArrowLine(15,0)(0,20)
 \ArrowLine(0,-20)(15,0)
 \Photon(15,0)(25,0){2}{3}
 \ArrowArc(35,0)(10,0,180)
 \Photon(28,7)(42,-7){2}{5}
 \ArrowArc(35,0)(10,180,360)
 \Photon(45,0)(55,0){2}{3}
 \ArrowLine(70,20)(55,0)
 \ArrowLine(55,0)(70,-20)
 \Text(35,15)[cb]{$l$}
\end{picture}
}
+
\scalebox{0.7}{
\begin{picture}(40,40)(0,-3)
 \ArrowLine(20,20)(0,40)
 \ArrowLine(40,40)(20,20)
 \Photon(20,20)(20,10){2}{3}
 \ArrowArc(20,0)(10,-90,90)
 \Photon(13,7)(27,-7){2}{5}
 \ArrowArc(20,0)(10,90,270)
 \Photon(20,-10)(20,-20){2}{3}
 \ArrowLine(0,-40)(20,-20)
 \ArrowLine(20,-20)(40,-40)
 \Text(36,-2)[cb]{$l$}
\end{picture}
}
\Bigg)^*
$$
\caption{Contributions involving the two-loop vacuum polarization from a 
lepton $l$.}
\label{fig:se2}
\end{figure}
\section{Numerical analysis}
\begin{figure}[h]
\vspace{-0.6cm}
\begin{center}
\epsfig{file=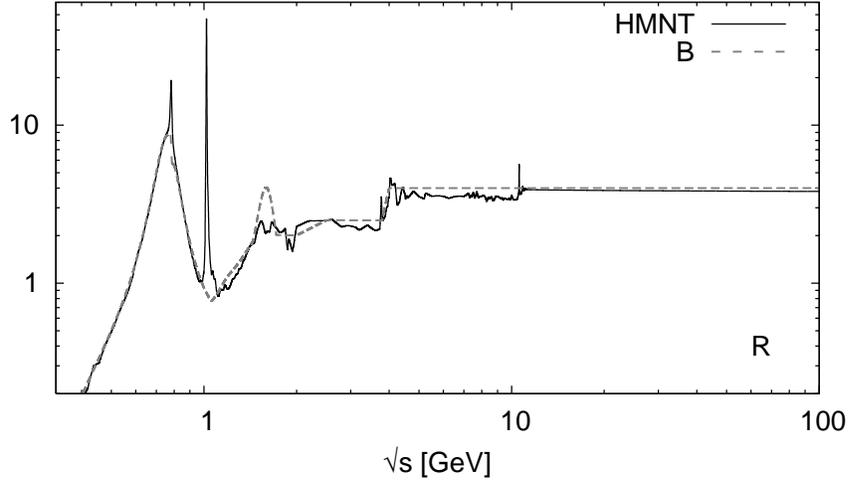, scale=0.95}
\end{center}
\vspace{-0.6cm}
\caption{The two parametrization B (dashed) and HMNT (solid) for $R(s)$ 
without narrow resonances.}
\vspace{-0.1cm}
\label{fig:R0}
\end{figure}

To arrive at a numerical result we adopt the following parametrizations for 
$R(s)$:
For the comparison with earlier work~\cite{Actis:2007fs}  
we take the function provided by H.Burkhardt~\cite{Burkhardt:1981jk}.
This parametrization (denoted by B) is simple and efficient for the 
integration, however, it includes only data more than 20 years old.
A newer parametrization (denoted by HMNT) is based on the most recent and 
accurate data and will be used for most of our detailed predictions.
The two parametrizations for $R(s)$ are shown in~\fig{fig:R0}.

The contributions from narrow resonances are incorporated using:
\bq
R_{\rm res}(s)= 
\frac{9\pi}{\alpha^2(M_{\rm res})}M_{\rm res}\Gamma_{e^+e^-}
\delta(s-M_{\rm res}^2).
\eq
For the parametrization HMNT we take $J/\Psi$, $\psi(2S)$, $\Upsilon(1S)$, 
$\Upsilon(2S)$ and $\Upsilon(3S)$ as narrow resonances with the parameters 
listed in Table~\ref{tab:res}, thus replacing their rapidly varying cross 
section governed by a narrow Breit-Wigner shape with an easy to be 
integrated delta function.
\begin{table}[!h]
\vspace{0.1cm}
\begin{center}
{
\begin{tabular}{c|c|c|c|c|c}
  \hline
                       & $J/\Psi$       & $\Psi(2S)$     & $\Upsilon(1S)$ &
                         $\Upsilon(2S)$ & $\Upsilon(3S)$ 
\\\hline                
$M$(GeV)               & 3.096916(11)   & 3.686093(34)   & 9.46030(26)    & 
                         10.02326(31)   & 10.3552(5)     
\\                      
$\Gamma_{ee}$(keV)     & 5.55(14)       & 2.48(6)        &  1.340(18)     & 
                         0.612(11)      & 0.443(8)       
\\
$(\alpha/\alpha(M))^2$ & 0.957785       & 0.95554        & 0.932069       & 
                         0.93099        & 0.930811       
\\\hline
\end{tabular}
}
\caption{\label{tab:res} Masses and electronic widths
of the narrow resonances and effective electromagnetic 
coupling at the appropriate scales.}
\end{center}
\vspace{-0.6cm}
\end{table}
Parametrization B uses slightly different values and includes in addition 
$\omega(782)$, $\Phi(1020)$, 
$\psi(3770)$, $\psi(4040)$, $\psi(4160)$, $\psi(4415)$, 
$\Upsilon(4S)$, $\Upsilon(10860)$ and $\Upsilon(11020)$ 
as narrow resonances and we adopt the parameter values listed in the 
code~\cite{Burkhardt:1981jk}.
For later use we also give the results for the moments $R(\infty)$, $R_0$, 
$R_1$ and $R_2$ based on parametrization B:
\bq
R(\infty)= 4.0,
\qquad
R_0= - 8.31,
\qquad
R_1= 13.1,
\qquad
R_2= - 15.6.
\eq

\begin{figure}[!h]
\vspace{-0.2cm}
\begin{tabular}{cc}
\!\!\!\!\!\!\!\!\!\!\!\!\!
\epsfig{file=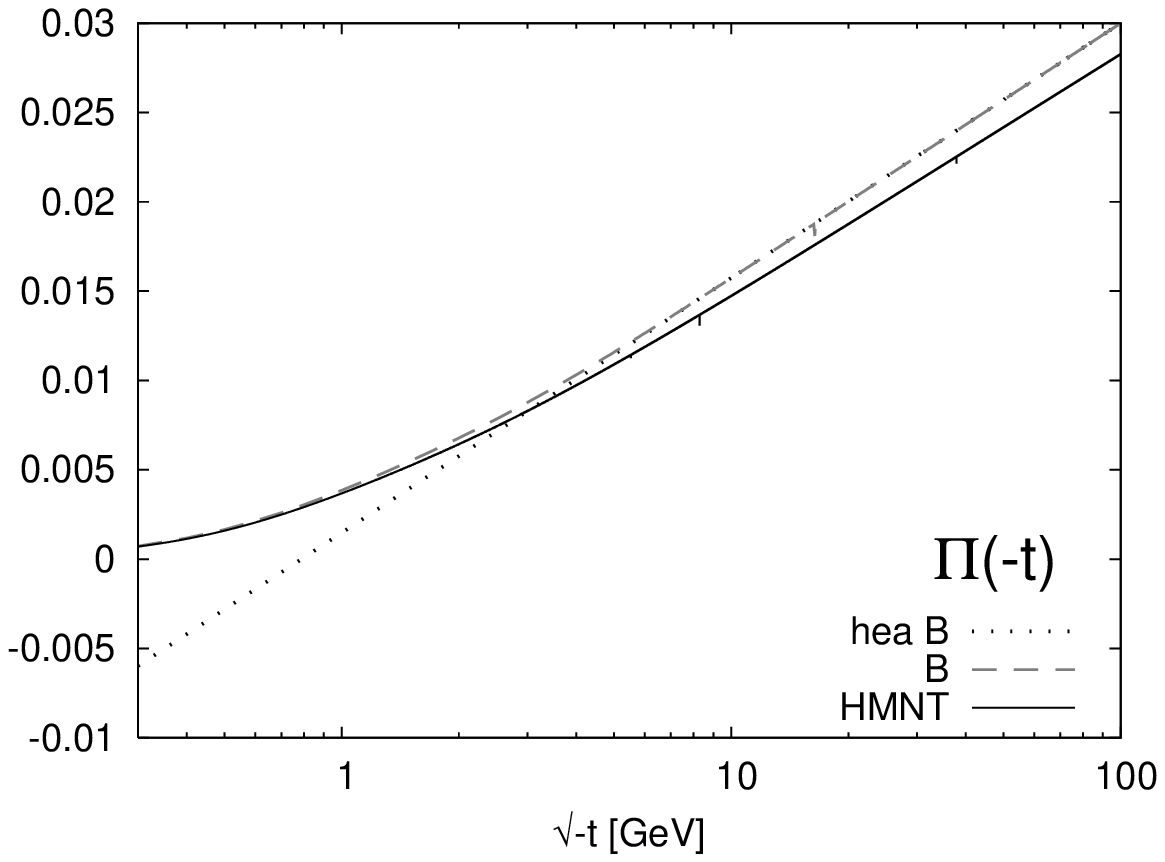, scale=0.65}
&
\!\!\!\!\!\!\!\!\!\!\!\!\!\!\!\!\!\!
\epsfig{file=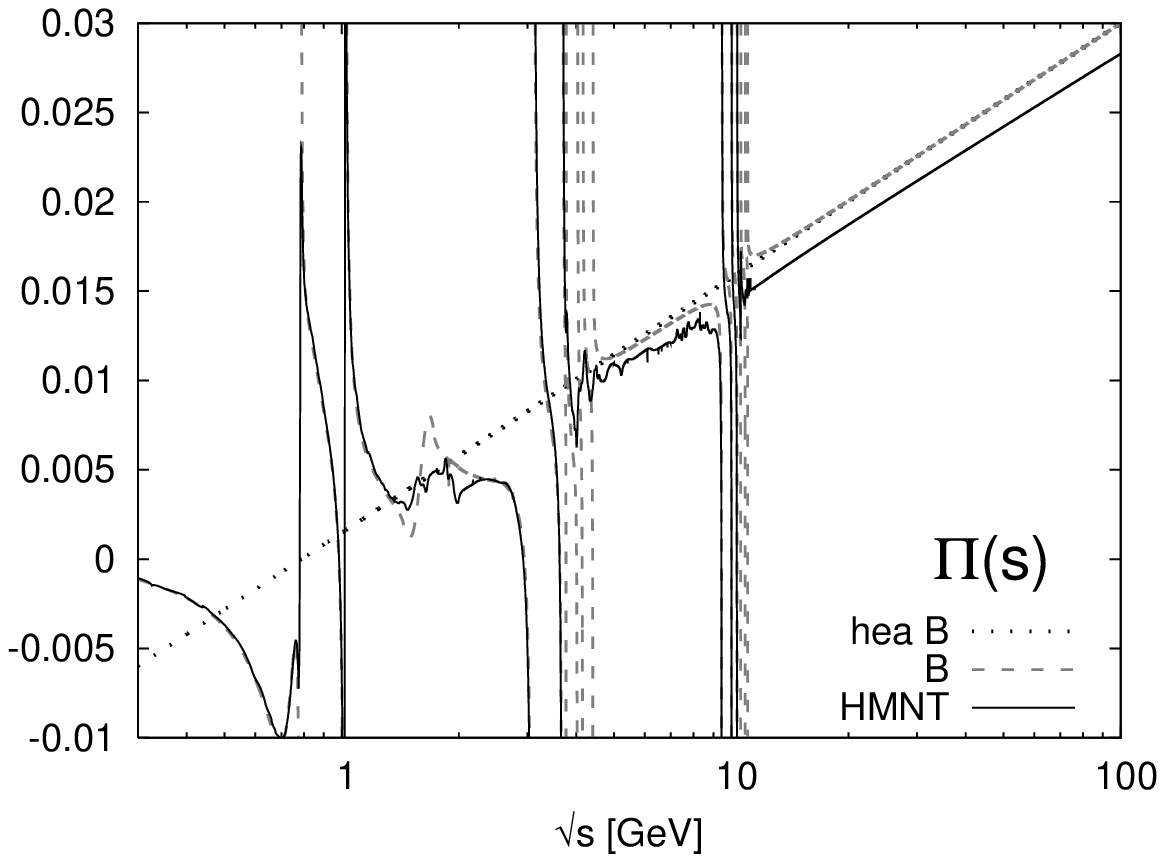, scale=0.65}
\\[-0.2cm]
\!\!\!\!\!\!\!\!\!\!\!\!\!
\epsfig{file=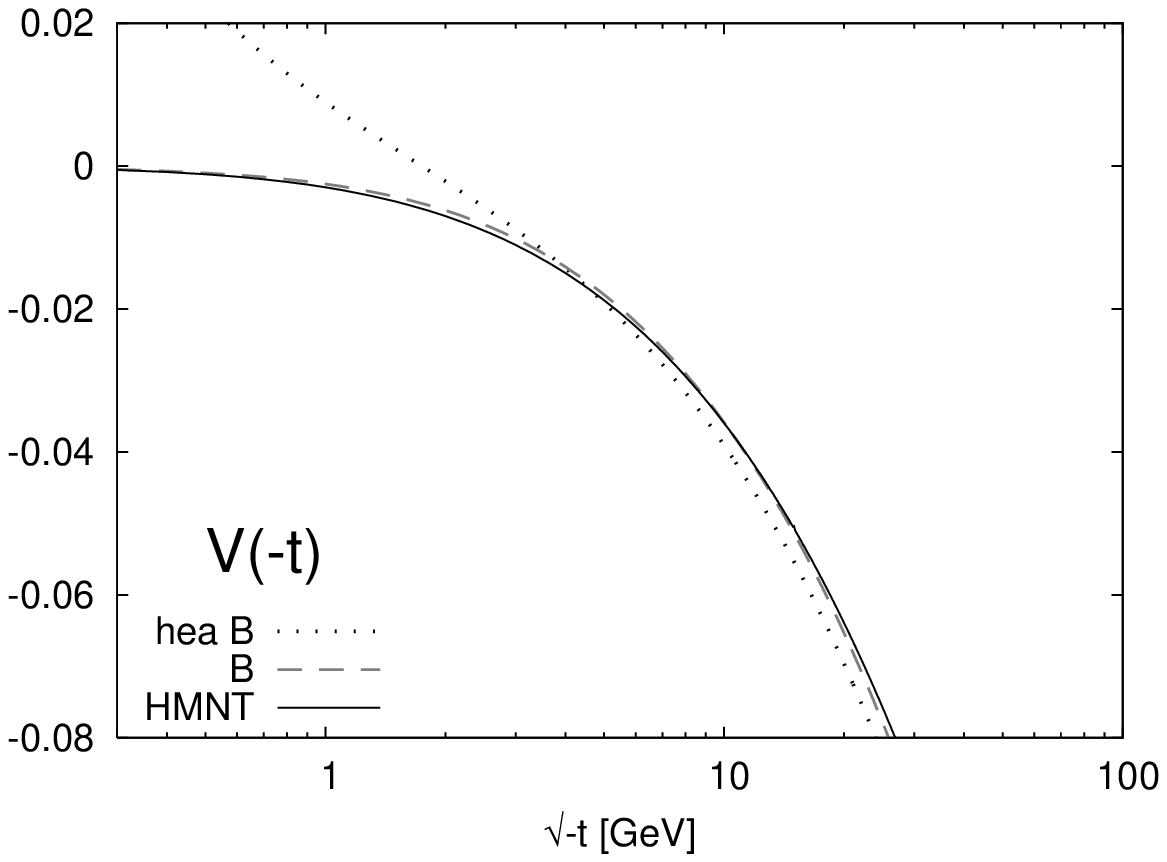, scale=0.65}
&
\!\!\!\!\!\!\!\!\!\!\!\!\!\!\!\!\!\!
\epsfig{file=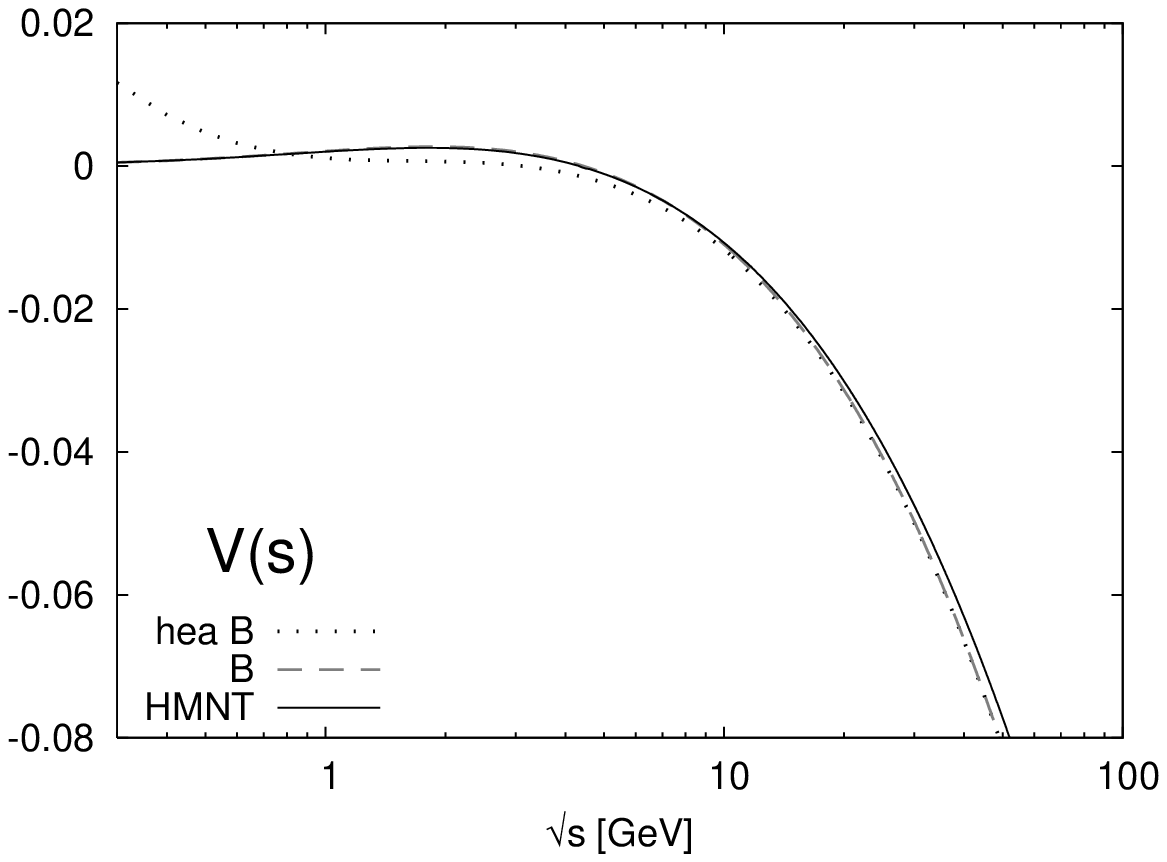, scale=0.65}
\end{tabular}
\vspace{-0.4cm}
\caption{Vacuum polarization $\Pi(q^2)$ and vertex correction $V(q^2)$ for 
spacelike (-t) and timelike (s) momenta for the parametrizations B (dashed) 
and HMNT (solid) and the high energy approximation (dotted)}
\label{fig:pi+vertex}
\vspace{-0.4cm}
\end{figure}
\begin{figure}[h]
\vspace{-0.6cm}
\epsfig{file=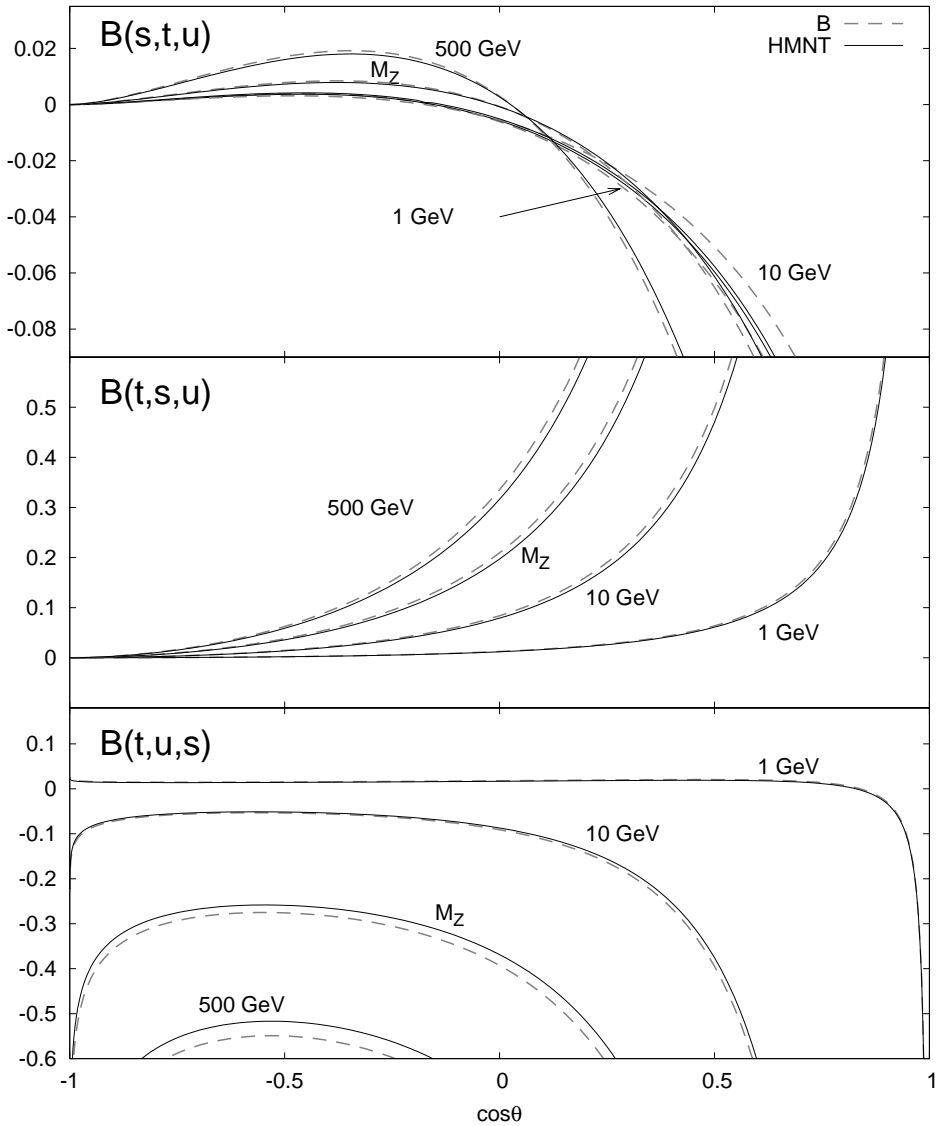, scale=0.75}
\vspace{-0.8cm}
\caption{The functions B(a,b,c) defined in \eqn{eq:B} for different 
kinematical regions versus $\cos\theta$ using the parametrizations 
B (dashed) and HMNT (solid).}
\label{fig:B}
\end{figure}

The results for the vacuum polarization $\Pi(q^2)$ and the vertex correction
$V(q^2)$ for space-like and time-like $q^2$ are shown in 
\fig{fig:pi+vertex} as functions of $q^2$.
We display the predictions based on both parametrization B (dashed) and 
HMNT (solid). 
For comparison we also show the behaviour in the high 
energy approximation (dotted) of eq.(\ref{eq:Pi-he}-\ref{eq:V-he}), 
for parametrization B only.
As expected from the comparison in \fig{fig:R0}, the difference between 
the two parametrization leads to differences in $\Pi(q^2)$ and $V(q^2)$ 
of less than 10\% which are unimportant for the two-loop analysis 
(the present uncertainty for HMNT amounts to typically one to two percent).
For $\Pi(t)$ and $V(q^2)$ the high energy approximation starts to 
deviate significantly from the full result for energies below 3 GeV, while 
for $\Pi(s)$ the resonant behaviour cannot be reproduced by this 
approximation.
The corresponding results for the functions $B(a,b,c)$, which govern the 
behaviour of the irreducible box contribution are shown in \fig{fig:B} 
for a set of representative energies as functions of the scattering 
angle $\theta$. 
The result for $B(s,u,t)$ is obtained from $B(s,t,u)$ through the 
substitution $\cos\theta \to -\cos\theta$.

As expected, the predictions for $\Pi$, $V$ and $B$ based on the two 
paratrizations B and HMNT are always quite close, hence the following 
discussion will be based on HMNT only.

\begin{figure}[h]
\begin{tabular}{cc}
\!\!\!\!\!\!\!\!\!\!\!\!\!
\epsfig{file=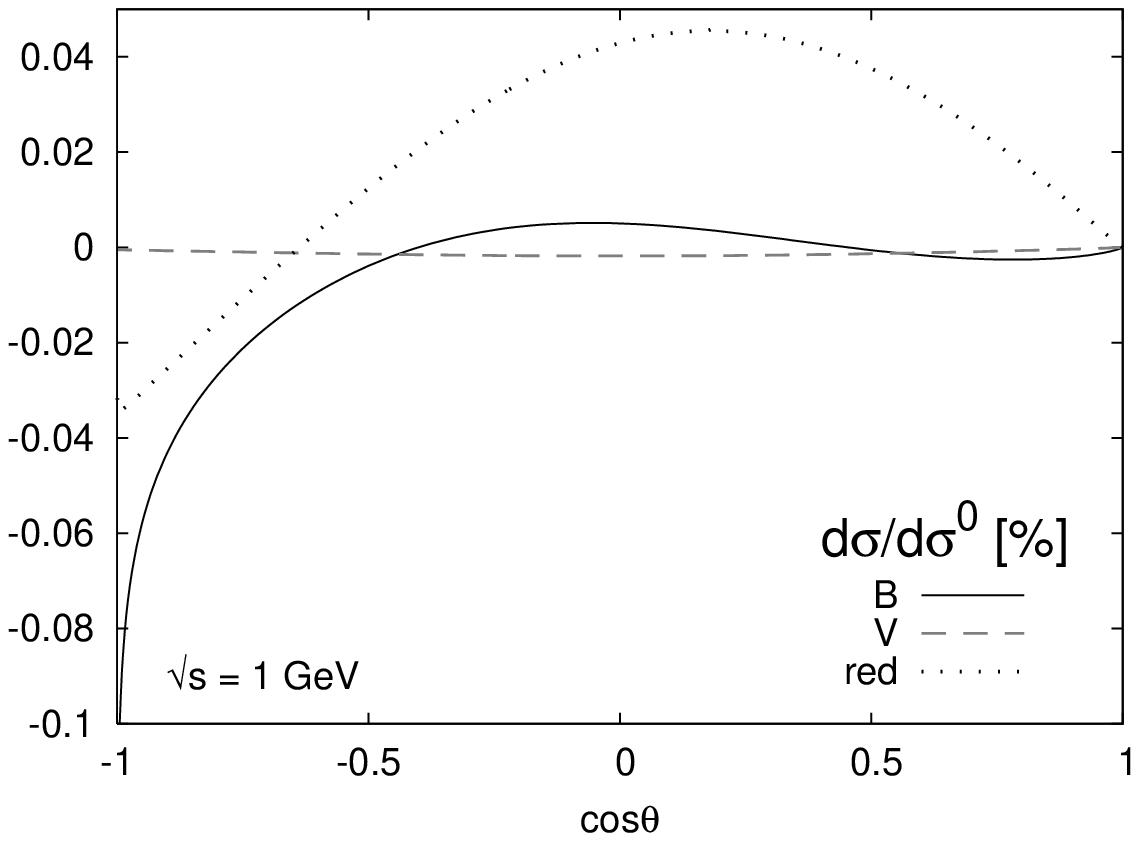, scale=0.7}
&
\!\!\!\!\!\!\!\!\!\!\!\!\!\!\!\!\!\!
\epsfig{file=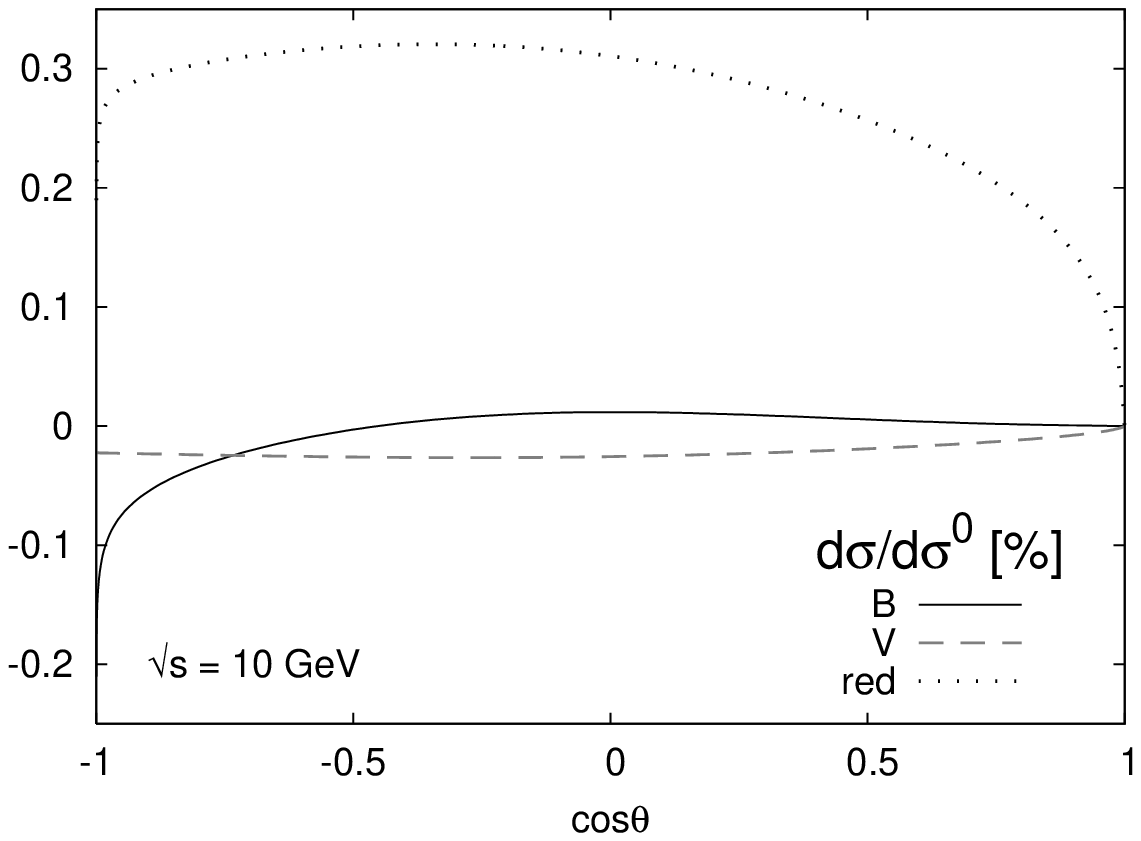, scale=0.7}
\\[-0.1cm]
\!\!\!\!\!\!\!\!\!\!\!\!\!
\epsfig{file=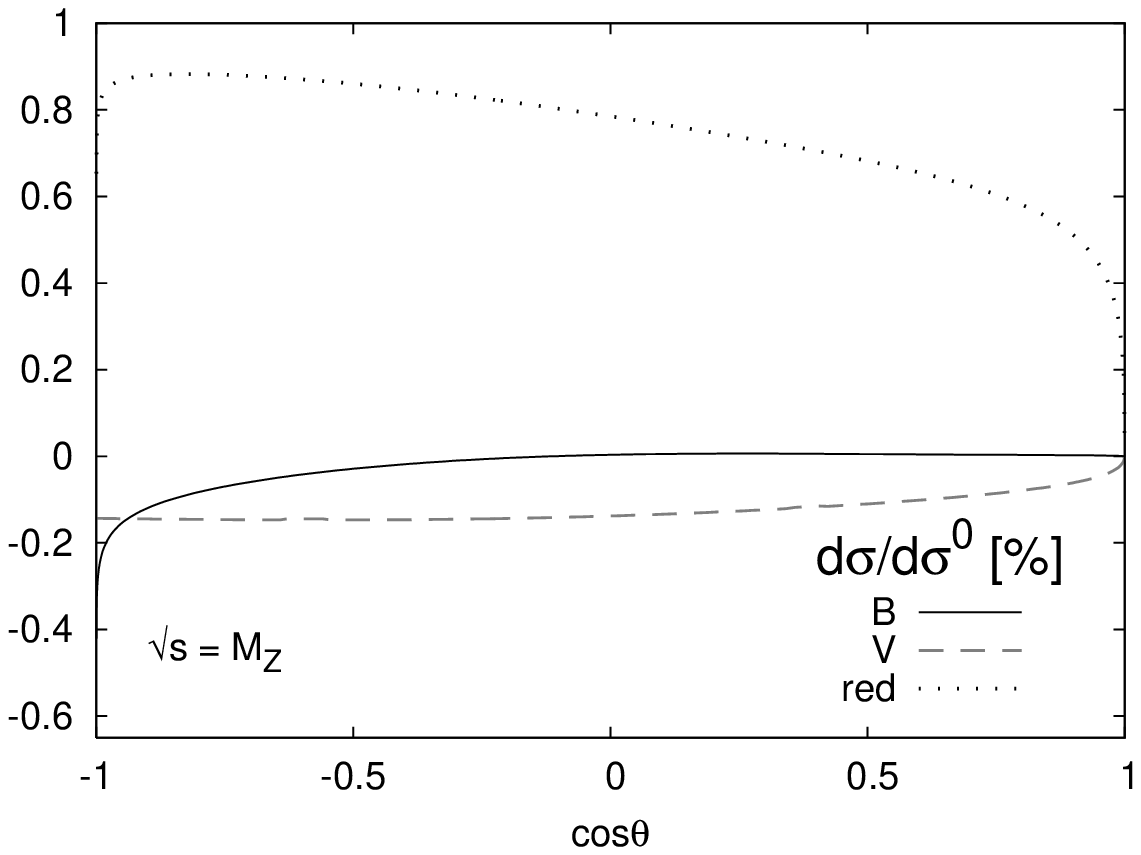, scale=0.7}
&
\!\!\!\!\!\!\!\!\!\!\!\!\!\!\!\!\!\!
\epsfig{file=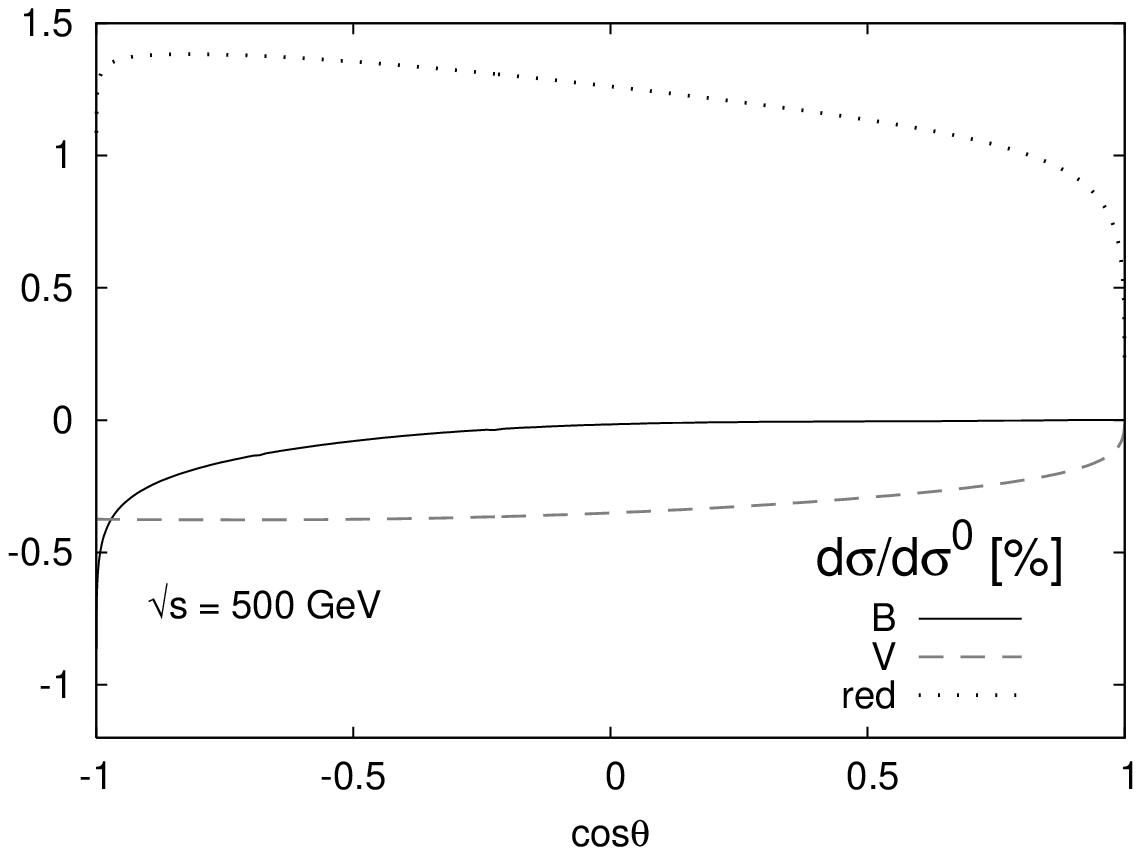, scale=0.7}
\end{tabular}
\vspace{-0.3cm}
\caption{
Relative corrections to the cross section from irreducible 
boxes $d\sigma^{\!B}$ (solid), vertices $d\sigma^{\!V}$ (dashed) and 
reducible contribution 
$d\sigma^{\rm red}= d\sigma^{\!B}_{\!\rm red} + d\sigma^{\!V}_{\!\rm red}$ 
(dotted) for four characteristic energies using parametrization HMNT.}
\label{fig:had}
\vspace{0.3cm}
\end{figure}

The corrections for the differential distributions are shown in 
\fig{fig:had} for four characteristic energies, normalized relative 
to the Born prediction\footnote{We do not present the two-loop vacuum 
polarization insertions of \eqn{eq:Pi2} which are best combined with 
the one-loop and Born contribution in the resummed form of \eqn{eq:Pi1}.
It is clear that $\Pi_{\rm had}(q^2)$ must be known with a relative 
precision of about one percent, if one aims at luminosity determination 
with an error significantly below one per mille.}.
They are separated into those from reducible diagrams 
($d\sigma^{\rm red}= d\sigma_{\!\rm red}^{\rm V} + d\sigma_{\!\rm red}^{\rm B}$), 
irreducible vertex ($d\sigma^{\rm V}$) and box ($d\sigma^{\rm B}$) 
diagrams\footnote{Here and below the infrared-sensitive contributions 
proportional to $\ln(2\omega/\sqrt{s})$ are set to zero.}.
In most of the cases the reducible ones are significantly larger than 
the irreducible ones, a consequence of their enhancement by the large 
logarithm $\ln(s/m_e^2)$.

\begin{figure}[!h]
\begin{tabular}{cc}
\!\!\!\!\!\!\!\!\!\!\!\!\!
\epsfig{file=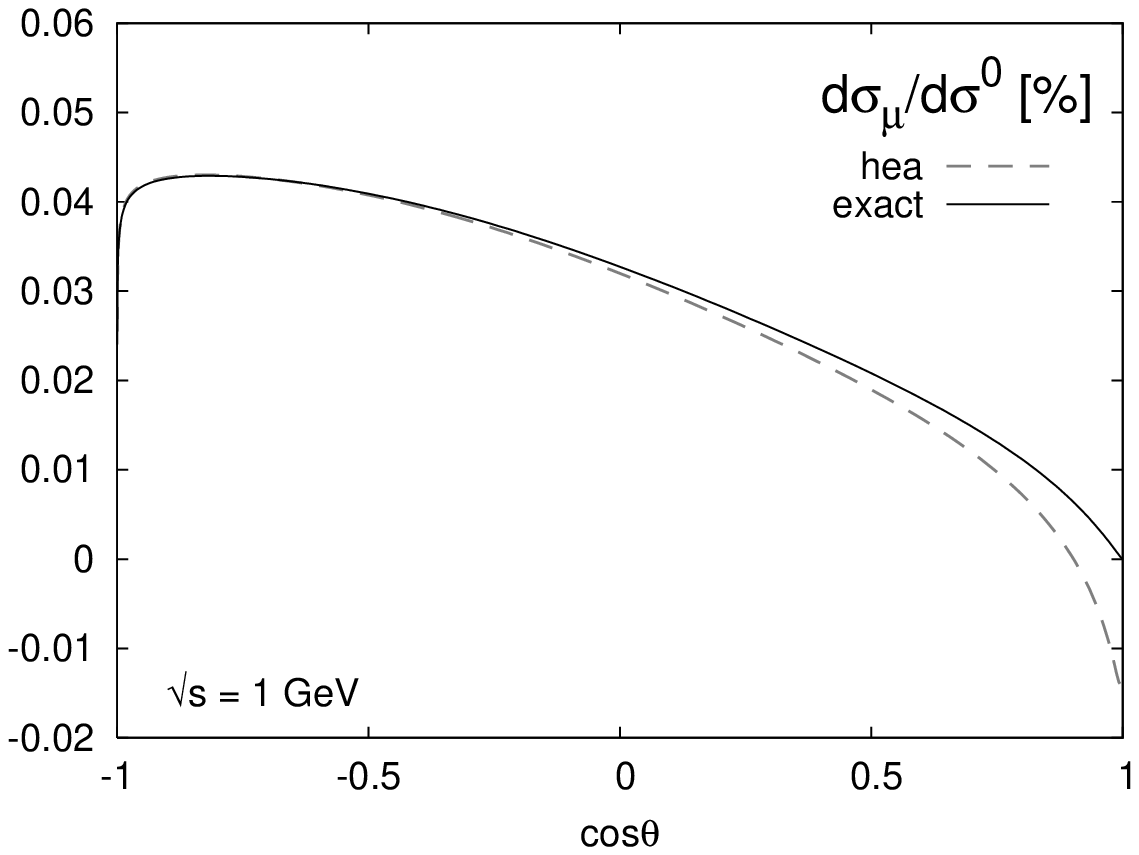, scale=0.7}
&
\!\!\!\!\!\!\!\!\!\!\!\!\!\!\!\!\!\!
\epsfig{file=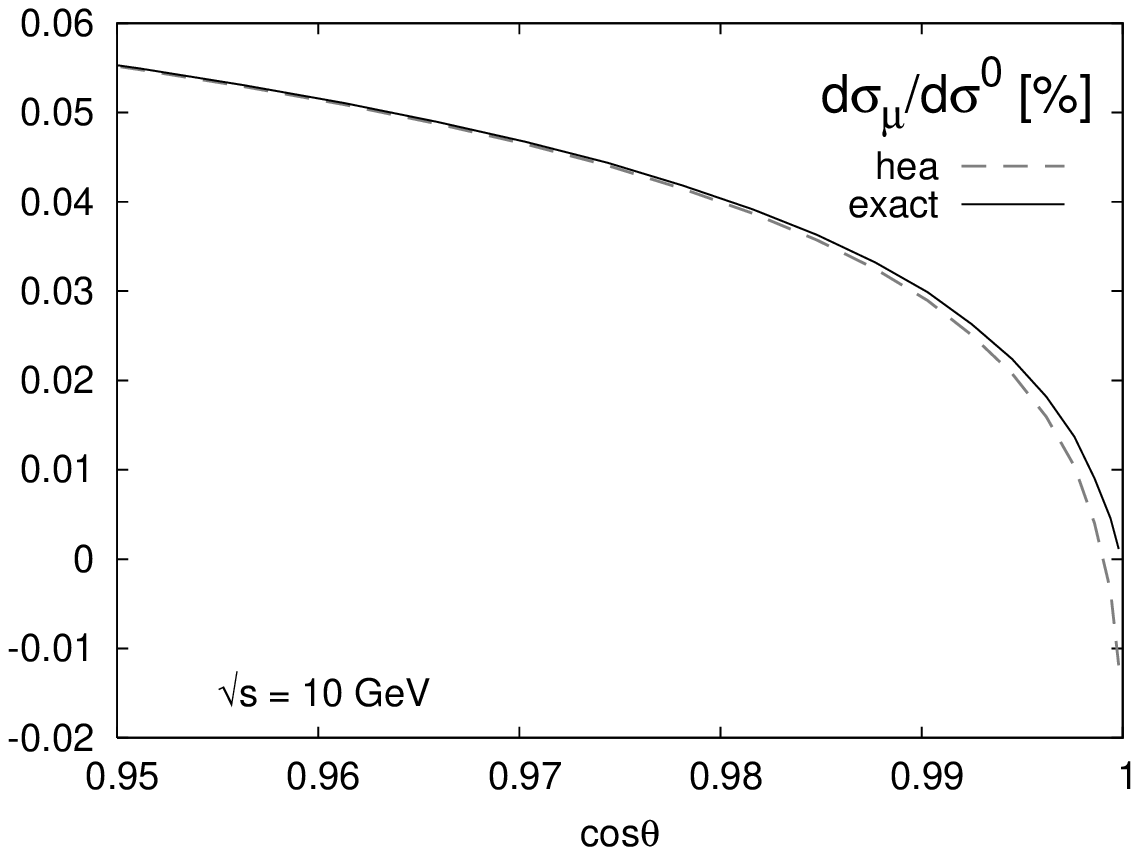, scale=0.7}
\\
\!\!\!\!\!\!\!\!\!\!\!\!\!
\epsfig{file=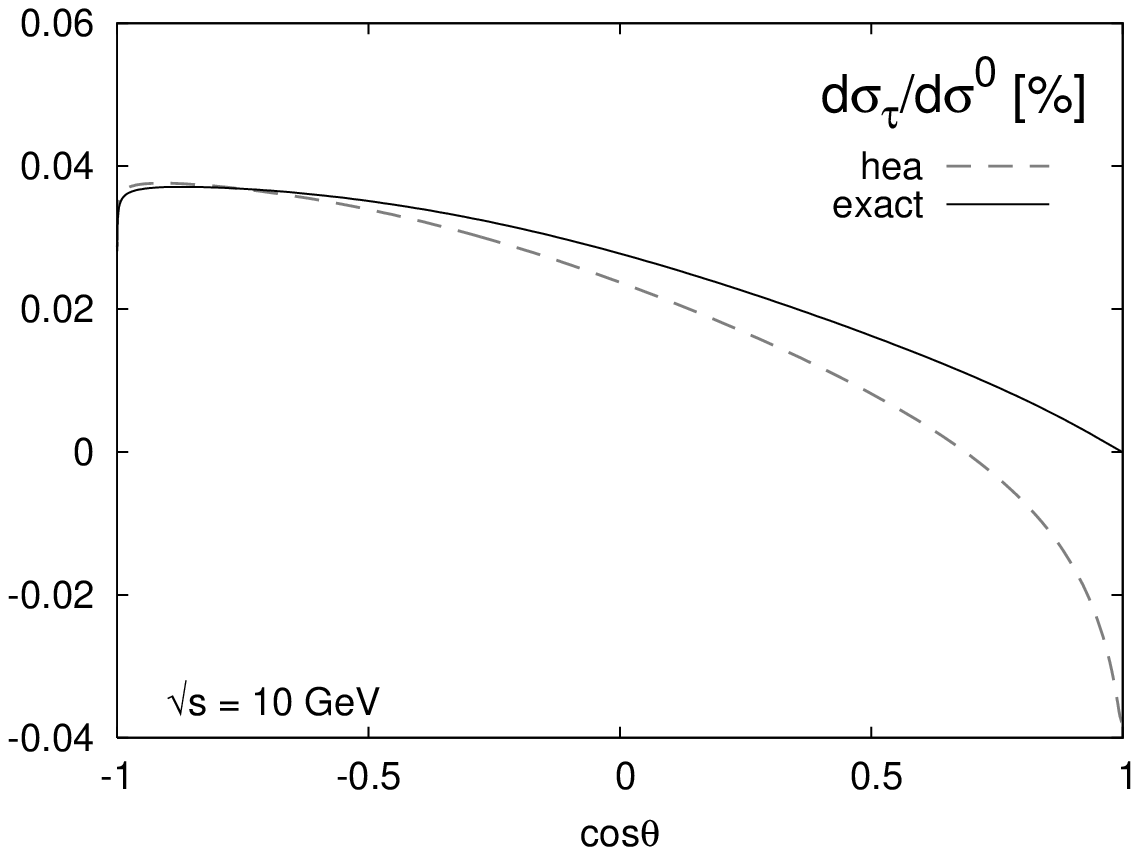, scale=0.7}
&
\!\!\!\!\!\!\!\!\!\!\!\!\!\!\!\!\!\!
\epsfig{file=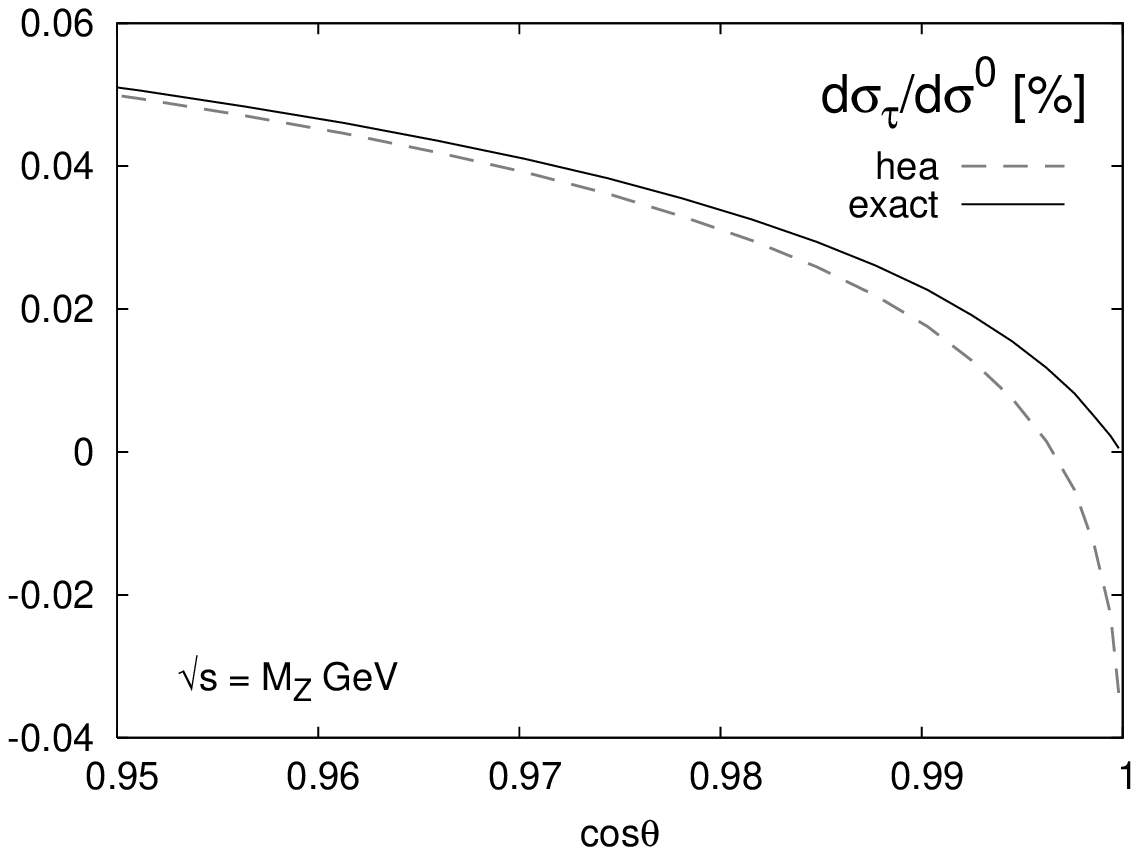, scale=0.7}
\end{tabular}
\caption{Leptonic cross section $d\sigma_l$ from muons (upper) 
and $\tau$-leptons (lower) (\eqn{eq:lept}) relative to the Born cross 
section. 
Comparison is shown between the high energy approximation (hea) of 
\eqn{eq:lept-hea} (dashed) and the exact result (solid) obtained from 
numerical integration using $R_l$ as defined in \eqn{eq:Rl}.}
\label{fig:lept}
\vspace{0.2cm}
\end{figure}

In \fig{fig:lept} we display the corresponding contributions from muons 
(solid line) and $\tau$ leptons (dashed line), which can be evaluated 
similar to the hadronic ones.
It is interesting to observe that the high energy approximation (hea) for the 
muon case, $ m_\mu^2 \ll s,|t|,|u|$, (dotted) fails quite badly for small 
angles at $\sqrt{s} = 1$ GeV, a fact that could be anticipated already from 
\fig{fig:pi+vertex}, which shows the pour quality of this approximation 
for $\sqrt{-t} < 3$ GeV in the hadronic case.
For high energies, the quality of the approximation should be sufficient 
for all practical purposes (\fig{fig:lept}).

\begin{figure}[!h]
\begin{tabular}{cc}
\!\!\!\!\!\!\!\!\!\!\!\!\!
\epsfig{file=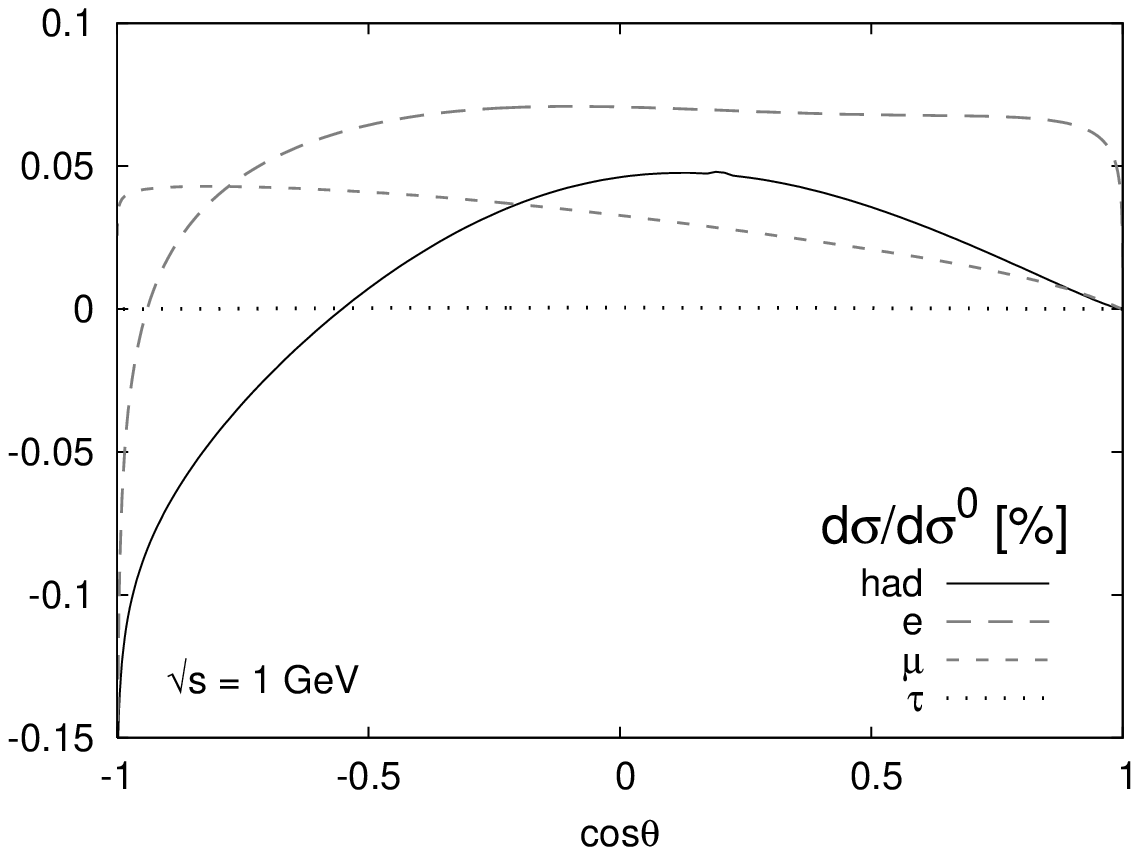, scale=0.7}
&
\!\!\!\!\!\!\!\!\!\!\!\!\!\!\!\!\!\!
\epsfig{file=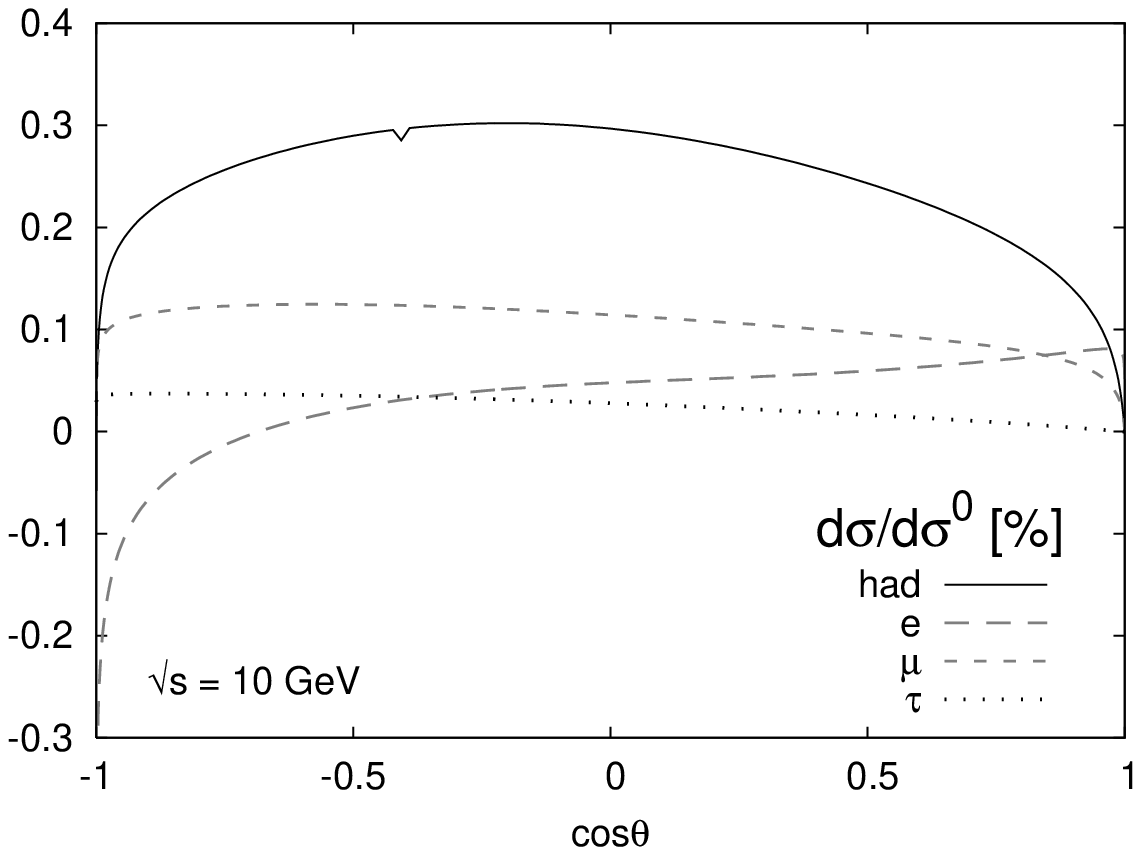, scale=0.7}
\\
\!\!\!\!\!\!\!\!\!\!\!\!\!
\epsfig{file=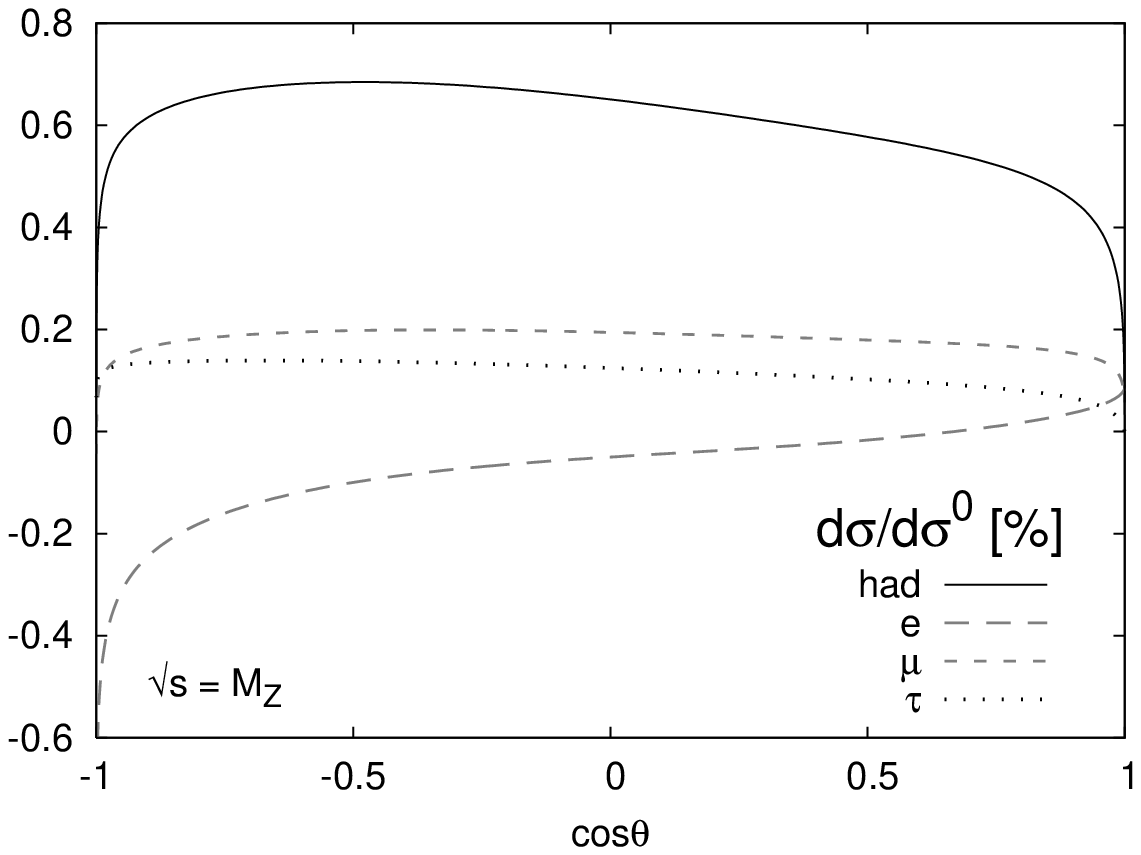, scale=0.7}
&
\!\!\!\!\!\!\!\!\!\!\!\!\!\!\!\!\!\!
\epsfig{file=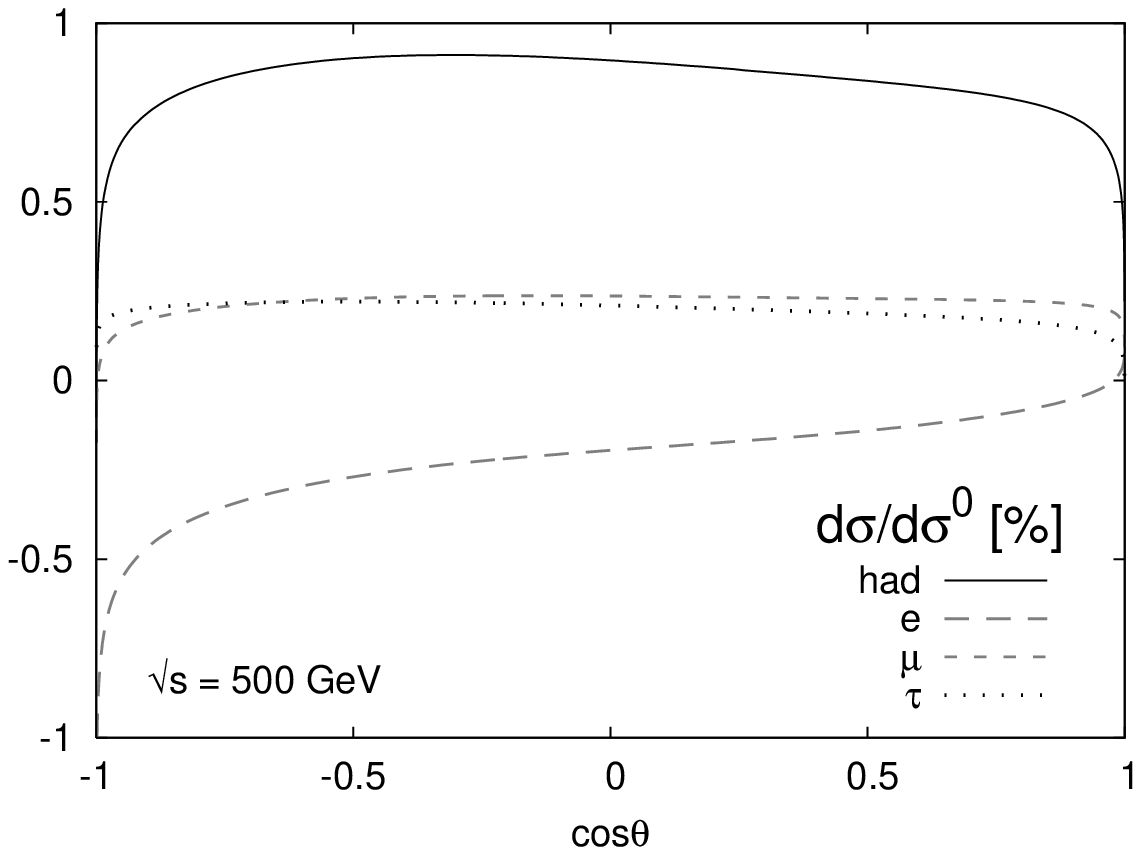, scale=0.7}
\end{tabular}
\vspace{0.1cm}
\caption{Relative corrections from hadron and lepton loops as functions of 
$\cos\theta$ for four characteristic energies.
The high energy approximation of \eqn{eq:lept-hea} is used for electron 
contribution at all energies and for muons for $\sqrt{s} \ge 10$~GeV.}
\label{fig:compare}
\vspace{0.6cm}
\end{figure}

The relative corrections from hadron and lepton ($e$, $\mu$, $\tau$) loops 
are compared in \fig{fig:compare}. 
A markedly different energy and angular dependence is observed for the four 
contributions. 
Individually and in the sum, they significantly exceed the level of one 
per mille necessary to achieve the corresponding precision of the luminosity 
measurements.
However, as discussed before, the reducible terms dominate and the 
irreducible hadronic terms are typically below one per mille.
For precise comparisons the numerical results are also listed in 
Table~\ref{tab:results} for a selected set of energies and angles.
For small angles the box contribution $d\sigma^{\!B}$ 
remains tiny, often around or below $10^{-5}$ of the Born cross section, 
and the result is dominated by the reducible correction 
$d\sigma^{\rm red}$ which is typically a factor $10$ to $100$ 
larger and is trivially obtained from existing one-loop results,
\eqn{eq:redV}-(\ref{eq:redB}).

\vspace{0.4cm}
To illustrate the relative importance of reducible and irreducible 
contributions, the results for the irreducible box $d\sigma^B$ and 
the sum $d\sigma^{B+{\rm red}}= d\sigma^B+\sigma^{\rm red}$ are listed 
in Table~\ref{tab:compare}.
The relative contribution of $d\sigma^B$ is evidently tiny. 
The results for $d\sigma^{B+{\rm red}}$ are also compared with those 
from~\cite{Actis:2007fs}. 
For the hadronic case they are in good agreement, although sometimes 
deviating in the last of the digits listed in~\cite{Actis:2007fs}. 
For the leptons perfect agreement is observed.
\clearpage

\begin{figure}[!h]
\begin{tabular}{cc}
\!\!\!\!\!
\epsfig{file=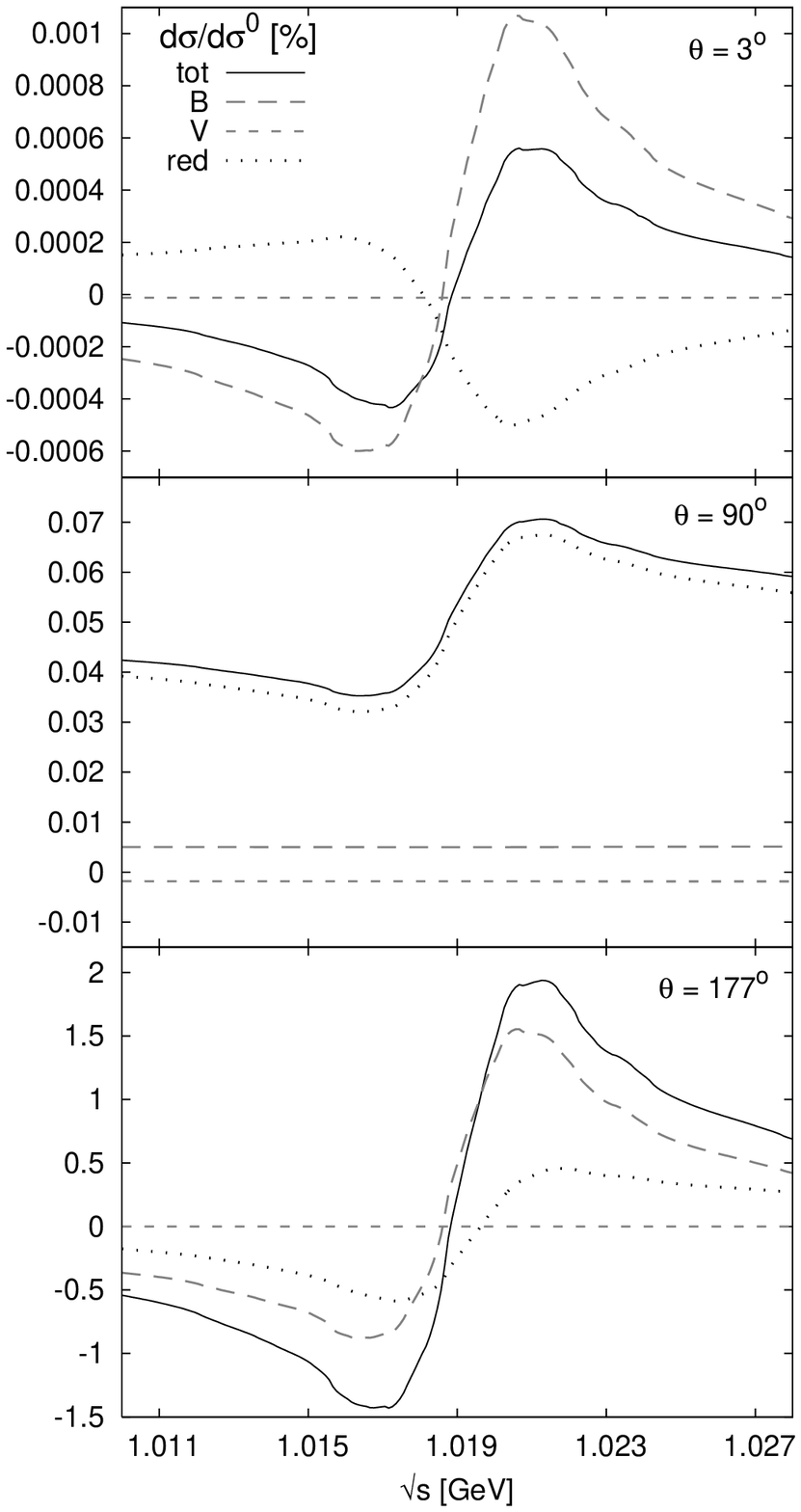, scale=0.75}
&
\!\!\!\!\!\!\!\!\!\!\!\!\!\!\!
\epsfig{file=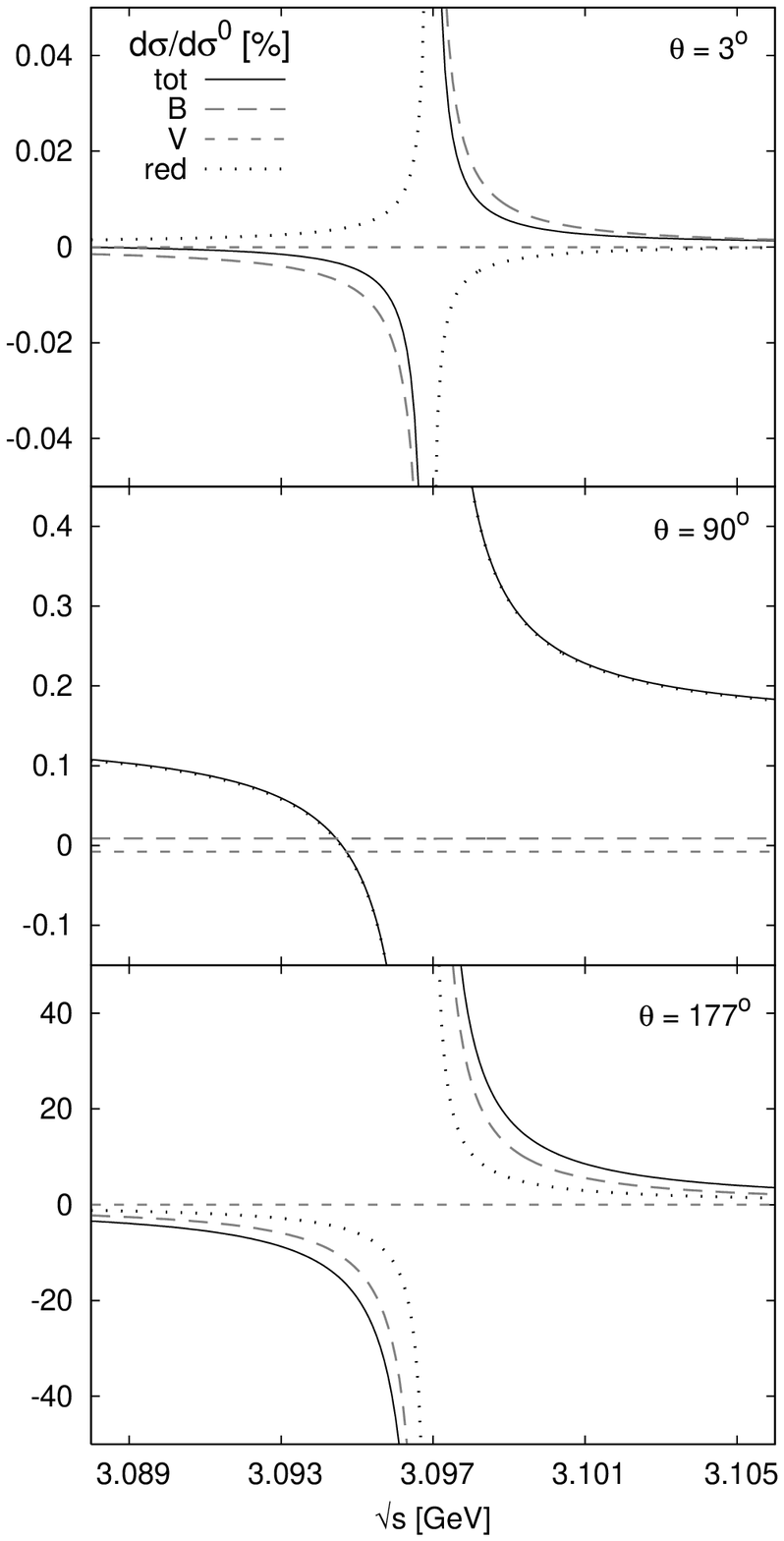, scale=0.75}
\end{tabular}
\caption{Behaviour of the hadronic corrections around the $\Phi$ and the
$J/\Psi$ resonances for three characteristic angles.}
\label{fig:res}
\end{figure}

In general, the corrections exhibit a fairly smooth energy dependence. 
However, the situation changes for energies close to narrow resonances. 
This is exemplified in \fig{fig:res} for two cases: around the $\Phi$ and 
around the $J/\Psi$ resonances for three fixed angles: 
$3^o$, $90^o$ and $177^o$.
The interference of the continuum amplitude with a Breit-Wigner enhanced 
correction is clearly visible. 
At $3^o$ ($177^o$), irreducible box and the reducible corrections are of 
comparable size and opposite (equal) sign, at $90^o$ the reducible ones 
dominate. 
The vertex corrections are always small. 
Formally for the case of the $J/\Psi$, treated as narrow resonance, 
the correction even diverges, and it is still extremely large if the 
natural width of $J/\Psi$ is introduced. In practice, however, the cross 
section has to folded with the energy spread of order MeV. 
In this case the singular amplitude with its asymmetric behaviour around 
$\sqrt{s}= M_{\rm res}$ is damped and thus remains a small correction.
From these considerations, it is clear that a precise parametrization of 
$R(s)$ is required in regions of rapidly varying cross section, if one aims 
at a precise prediction of the corrections in this region.
\clearpage
\begin{table}[!h]
\begin{center}
\begin{tabular}{c|c|c|c|c|c|c}
\hline
&\multicolumn{3}{|c|}{}&\multicolumn{3}{|c}{} 
\\[-0.4cm]
$\sqrt{s}$ & 
\multicolumn{3}{|c|}{$1$ GeV} &
\multicolumn{3}{|c}{$10$ GeV} 
\\\hline\\[-0.4cm]
$\theta$ & $3^o$ & $90^o$ & $177^o$ & $3^o$ & $90^o$ & $177^o$ 
\\\hline\hline
&&&&&&
\\[-0.35cm] 
$\!\!\!d\sigma^0\!/d\Omega\!\!\!$ &
$440994\!\cdot\!10^5$ & $46653.7$ & $20735.0$ & 
$440994\!\cdot\!10^3$ & $466.537$ & $207.350$
\\
$\!\!\![{\rm pbarn}]\!\!\!$ &&&&&&
\\\hline\hline
&&&&&&
\\[-0.35cm] 
$\!\!\!d\sigma^B\!\!/\!d\sigma^0\!\!\!$ &
$-8.626\!\cdot\!10^{-4}$ & $0.05267$ & $-1.295$ & 
$1.235\!\cdot\!10^{-4}$ & $0.1267$ & $-1.421$
\\
$[10^{-3}]$&
$-8.182\!\cdot\!10^{-4}$ & $0.05024$ & $-1.230$ & 
$-0.076(1)\!\cdot\!10^{-4}$ & $0.1168$ & $-1.532$
\\\hline
&&&&&&
\\[-0.35cm] 
$\!\!\!d\sigma^V\!\!/\!d\sigma^0\!\!\!$ &
$-1.234\!\cdot\!10^{-4}$ & $-0.01877$ & $-0.005167$ & 
$-0.004261$ & $-0.2695$ & $-0.2352$
\\
$[10^{-3}]$&
$-1.191\!\cdot\!10^{-4}$ & $-0.01796$ & $-0.004983$ & 
$-0.004084$ & $-0.2560$ & $-0.2246$
\\\hline
&&&&&&
\\[-0.35cm] 
$\!\!\!d\sigma^{\rm red}\!/\!d\sigma^0\!\!\!$ & 
$8.934\!\cdot\!10^{-4}$ & $0.4461$ & $-0.3666$ & 
$0.08860$ & $3.317$ & $2.644$
\\
$[10^{-3}]$&
$8.169\!\cdot\!10^{-4}$ & $0.4286$ & $-0.3388$ & 
$0.08529$ & $3.098$ & $2.290$
\\\hline\hline
&&&&&&
\\[-0.35cm] 
$\!\!\!d\sigma_{\!\rm had}\!/\!d\sigma^0\!\!\!$ &
$-0.9259\!\cdot\!10^{-4}$ & $0.4800$ & $-1.667$ & 
$0.08446$ & $3.175$ & $0.9880$
\\
$[10^{-3}]$&
$-1.204\!\cdot\!10^{-4}$ & $0.4609$ & $-1.575$ & 
$0.08120$ & $2.959$ & $0.5341$
\\\hline
&&&&&&
\\[-0.35cm] 
$\!\!\!d\sigma_e/\!d\sigma^0\!\!\!$ &
$0.3114$ & $0.7070$ & $-1.460$ & 
$0.6862$ & $0.4773$ & $-3.516$
\\
$[10^{-3}]$&&&&&&
\\\hline
&&&&&&
\\[-0.35cm] 
$\!\!\!d\sigma_\mu/\!d\sigma^0\!\!\!$ &
$6.623\!\cdot\!10^{-4}$ & $0.3273$ & $0.3275$ & 
$0.09040$ & $1.143$ & $0.6128$
\\
$[10^{-3}]$&&&&&&
\\\hline
&&&&&&
\\[-0.35cm] 
$\!\!\!d\sigma_\tau/\!d\sigma^0\!\!\!$ &
$4.100\!\cdot\!10^{-6}$ & $0.004869$ & $-7.525\!\cdot\!10^{-4}$ & 
$3.926\!\cdot\!10^{-4}$ & $0.2776$ & $0.3265$
\\
$[10^{-3}]$&&&&&&
\\\hline
\multicolumn{7}{c}{ }\\[0.4cm]
\hline
&\multicolumn{3}{|c|}{}&\multicolumn{3}{|c}{} 
\\[-0.4cm]
$\sqrt{s}$ & \multicolumn{3}{|c|}{$M_Z$} & \multicolumn{3}{|c}{$500$ GeV}
\\\hline\\[-0.4cm]
$\theta$   &  $3^o$ &  $90^o$ &  $177^o$ &  $3^o$  &  $90^o$  &  $177^o$
\\\hline\hline
&&&&&&
\\[-0.35cm] 
$\!\!\!d\sigma^0\!/d\Omega\!\!\!$ &
$5303480$ & $5.61067$ & $2.49363$ & $176398$ & $0.186615$ & $0.0829400$
\\
$\!\!\![{\rm pbarn}]\!\!\!$ &&&&&&
\\\hline\hline
&&&&&&
\\[-0.35cm] 
$\!\!\!d\sigma^B\!\!/\!d\sigma^0\!\!\!$ &
$0.001685$ & $0.03648$ & $-3.418$ & $0.002188$ & $-0.1682$ & $-7.017$
\\
$[10^{-3}]$&
$0.001579$ & $0.03537(1)$ & $-3.200$ & $0.002055$ & $-0.1557$ & $-6.589$
\\\hline
&&&&&&
\\[-0.35cm] 
$\!\!\!d\sigma^V\!\!/\!d\sigma^0\!\!\!$ &
$-0.08749$ & $-1.458$ & $-1.516$ & $-0.4614$ & $-3.715$ & $-3.970$
\\
$[10^{-3}]$&
$-0.08347$ & $-1.375$ & $-1.430$ & $-0.4373$ & $-3.495$ & $-3.734$
\\\hline
&&&&&&
\\[-0.35cm] 
$\!\!\!d\sigma^{\rm red}\!/\!d\sigma^0\!\!\!$ &
$1.650$ & $8.340$ & $8.001$ & $4.594$ & $13.35$ & $12.84$
\\
$[10^{-3}]$&
$1.562$ & $7.835$ & $7.526$ & $4.289$ & $12.60$ & $12.12$
\\\hline\hline
&&&&&&
\\[-0.35cm] 
$\!\!\!d\sigma_{\!\rm had}\!/\!d\sigma^0\!\!\!$ &
$1.565$ & $6.918$ & $3.066$ & $4.135$ & $9.469$ & $1.855$
\\
$[10^{-3}]$&
$1.480$ & $6.495$ & $2.895$ & $3.854$ & $8.944$ & $1.795$
\\\hline
&&&&&&
\\[-0.35cm] 
$\!\!\!d\sigma_e/\!d\sigma^0\!\!\!$ &
$0.8128$ & $-0.4993$ & $-6.769$ & $0.5561$ & $-1.954$ & $-10.33$
\\
$[10^{-3}]$&&&&&&
\\\hline
&&&&&&
\\[-0.35cm] 
$\!\!\!d\sigma_\mu/\!d\sigma^0\!\!\!$ &
$0.7078$ & $1.943$ & $0.3714$ & $1.378$ & $2.366$ & $-0.3557$
\\
$[10^{-3}]$&&&&&&
\\\hline
&&&&&&
\\[-0.35cm] 
$\!\!\!d\sigma_\tau/\!d\sigma^0\!\!\!$ &
$0.04932$ & $1.246$ & $0.9548$ & $0.4787$ & $2.100$ & $1.162$
\\
$[10^{-3}]$&&&&&&
\\\hline
\end{tabular}
\caption{Born cross section and relative corrections from hadronic 
irreducible boxes and vertices, reducible hadronic contributions and their 
sum, for selected energies and angles.
Upper/lower lines: parametrization B/HMNT.
Also shown are the corresponding contributions from electrons, muons 
and $\tau$-leptons.
}
\label{tab:results}
\end{center}
\end{table}
\clearpage
\begin{table}[!h]
\vspace{0.5cm}
\begin{center}
\begin{tabular}{c|c|l|l|l|l}
\hline
&&&&&\\[-0.4cm]
$\theta=3^o$& 
$\sqrt{s}$ & 
\multicolumn{1}{|c|}{$1$ GeV} &
\multicolumn{1}{|c|}{$10$ GeV} &
\multicolumn{1}{|c|}{$M_Z$} &
\multicolumn{1}{|c}{$500$ GeV} 
\\\hline
\hline
&&&&&\\[-0.35cm] 
& 
$\!\!\!d\sigma^B\!\!/\!d\Omega\!\!\!$ &
$-0.380382$ & $0.000544704$ & $0.0000893354$ & $0.00000386014$ 
\\\cline{2-6}
&&&&&\\[-0.35cm] 
had
& 
$\!\!\!d\sigma^{B+{\rm red}}\!\!/\!d\Omega$ &
$\;\;\;0.0136085$ & $0.391247$ & $0.0876187$ & $0.00810786$
\\
& &
$\qquad<1$ & $0.39$ & $0.0877$ & $0.0081$
\\\hline
\hline
&&&&&\\[-0.35cm] 
& 
$\!\!\!d\sigma^B\!\!/\!d\Omega\!\!\!$ &
$\;\;\;0.0419870$ & $0.00132934$ & $0.0000283258$ & $0.00000100529$
\\\cline{2-6}
&&&&&\\[-0.35cm] 
$\mu$
& 
$\!\!\!d\sigma^{B+{\rm red}}\!\!/\!d\Omega$ &
$\;\;\;0.339976$ & $0.417217$ & $0.0407916$ & $0.00287809$
\\
& &
$\qquad<1$ & $0.42$ & $0.0408$ & $0.00288$
\\\hline
\hline
&&&&&\\[-0.35cm] 
& 
$\!\!\!d\sigma^B\!\!/\!d\Omega\!\!\!$ &
$-0.000277434$ & $0.000350300$ & $0.0000118080$ & $0.000000773826$
\\\cline{2-6}
&&&&&\\[-0.35cm] 
$\tau$
& 
$\!\!\!d\sigma^{B+{\rm red}}\!\!/\!d\Omega$ &
$\;\;\;0.00227893$ & $0.00193150$ & $0.00270529$ & $0.000876352$
\\
& &
$\qquad<1$ & $\;\;<10^{-2}$ & $0.0027$ & $0.00088$
\\\hline
\multicolumn{6}{c}{ }\\[0.5cm]
\hline
&&&&&\\[-0.4cm]
$\theta=90^o$
& 
$\sqrt{s}$ & 
\multicolumn{1}{|c|}{$1$ GeV} &
\multicolumn{1}{|c|}{$10$ GeV} &
\multicolumn{1}{|c|}{$M_Z$} &
\multicolumn{1}{|c}{$500$ GeV} 
\\\hline
&&&&&\\[-0.35cm]
&
$\!\!\!d\sigma^B\!\!/\!d\Omega\!\!\!$ &
$\;\,24.5724$ & $\;\;0.591300$ & $\;\;\;0.00204702$ & $-0.000313808$
\\\cline{2-6}
&&&&&\\[-0.35cm] 
had
& 
$\!\!\!d\sigma^{B+{\rm red}}\!\!/\!d\Omega$ &
$232.674$ & $16.0671$ & $\;\;\;0.469946$ & $\;\;\;0.0246035$
\\
& &
$234$ & $16.07$ & $\;\;\;0.4701$ & $\;\;\;0.02461$
\\\hline
\hline
&&&&&\\[-0.35cm] 
& 
$\!\!\!d\sigma^B\!\!/\!d\Omega\!\!\!$ &
$\;\,12.8008$ & $\;\;0.133680$ & $-0.00115553$ & $-0.000171183$
\\\cline{2-6}
&&&&&\\[-0.35cm] 
$\mu$
& 
$\!\!\!d\sigma^{B+{\rm red}}\!\!/\!d\Omega$ &
$160.197$ & $\;\;6.08187$ & $\;\;\;0.147046$ & $\;\;\;0.00725789$
\\
& &
$160$ & $\;\;6.08$ & $\;\;\;0.1470$ & $\;\;\;0.00726$
\\\hline
\hline
&&&&&\\[-0.35cm] 
& 
$\!\!\!d\sigma^B\!\!/\!d\Omega\!\!\!$ &
$\quad\,0.465857$ & $\;\;0.0939460$ & $\;\;\;0.00188681$ & $\;\;\;0.0000195543$
\\\cline{2-6}
&&&&&\\[-0.35cm] 
$\tau$
& 
$\!\!\!d\sigma^{B+{\rm red}}\!\!/\!d\Omega$ &
$\quad\,2.38272$ & $\;\;1.33347$ & $\;\;\;0.0752669$ & $\;\;\;0.00457124$
\\
& &
$\quad\,2$ & $\;\;1.33$ & $\;\;\;0.0752$ & $\;\;\;0.00457$
\\\hline
\end{tabular}
\vspace{0.2cm}
\caption{Corrections from the irreducible boxes (first line), sum of 
box and reducible contributions (second line) and comparison 
with~\cite{Actis:2007fs} (third line), for hadrons, muons 
and $\tau$-leptons at selcted energies and angles. 
In the upper table the numbers are in units of $10^{2}$ nbarn, in the lower 
one in units of $10^{-4}$ nbarn.
}
\label{tab:compare}
\end{center}
\vspace{0.4cm}
\end{table}

The implementation of these results in a Monte Carlo generator is 
straightforward and their modular structure should lead to an efficient 
program. 
The reducible contribution $d\sigma^{\rm red}$ can be obtained from the 
one-loop corrections simply modifying the photon propagators outside the 
loop according to:
\bq
\frac{1}{q^2} \to \frac{1}{q^2} \big[ 1 + {\rm Re}\Pi(q^2) \big],
\qquad\quad
\Pi= \Pi_{\rm had} + \Pi_e + \Pi_\mu + \Pi_\tau,
\eq
and adding the terms proportional to ${\rm Im}\Pi(s)$ multiplied by the 
imaginary part of the one-loop result.
The irreducible vertex corrections $V(q^2)$ can be directly combined with 
the Born cross section. 
All these are one-dimensional functions that can be tabulated once for ever. 
The irreducible box contribution is decomposedinto terms proportional to 
$\Pi(s)$ and $\Pi(t)$ plus a remainder characterized by the functions 
$B(s,t,u)$, $B(s,u,t)$, $B(t,s,u)$ and $B(t,u,s)$. 
These are obtained through efficient and precise integration 
routines\footnote{available upon request from the authors.}.
\clearpage
\section{Conclusions}
Using published one-loop results, a compact formula has been derived which, 
in combination with dispersion relations and the by now well-measured 
R-ratio, can be used to evaluate the hadronic contributions to Bhabha 
scattering.
The same approach is applicable for leptonic contributions, in particular 
from muons and $\tau$-leptons. 
The method and result are valid in the limit $m_e^2\ll s,|t|,|u|$ for 
arbitrary $R(s)$ and arbitrary $m_{\mu,\tau}^2/s$. 
Comparing with \cite{Actis:2007fs}, our numerical results are in
perfect agreement for massive leptons, with $m_l^2$ arbitrary, while 
for hadronic contributions we observe small numerical differences.
In the high energy limit the integrals can be evaluated in analytic form 
and the results have been compared with those for lepton loops that can 
be found in the literature.
We find that overall size of the corrections, their sign and their angular 
dependence differ significantly between hadron, muon, $\tau$-lepton and 
electron contributions.
The size of the hadronic corrections varies from a fractional up to 
several permille.
However, these are dominated by the reducible ones, with the irreducible 
box and vertex terms being typically below one permille.
The modular structure of the results allows for a simple implementation 
into any Monte Carlo generator.
For such an implementation, the corrections from virtual plus soft real 
photon radiation must be complemented by hard real radiation. 
This part is evidently straightforward, since it involves tree-level 
diagrams only, with the photon propagator dressed by hadronic vacuum 
polarization.
\Acknowledgments
We would like to thank H.~Burkhardt and T.~Teubner for providing us with 
the parametrizations for the function $R(s)$ and A.~Penin and T.~Teubner 
for helpful comments.
Work supplied by BMBF contract 05HT4VKAI3.

\end{document}

%% file: macro.tex
\newcommand{\nl}{\nonumber\\}
\newcommand{\nn}{\nonumber}
\newcommand{\bq}{\begin{equation}}
\newcommand{\eq}{\end{equation}}
\newcommand{\bqa}{\arraycolsep 0.14em\begin{eqnarray}}
\newcommand{\eqa}{\end{eqnarray}}
\newcommand{\bqaa}{\begin{eqnarray*}}
\newcommand{\eqaa}{\end{eqnarray*}}
\newcommand{\ba}[1]{\begin{array}{#1}}
\newcommand{\ea}{\end{array}}
\newcommand{\bei}{\begin{itemize}}
\newcommand{\eei}{\end{itemize}}
\newcommand{\eqn}[1]{Eq.(\ref{#1})}

\newcommand{\fig}[1]{Fig.~\ref{#1}}

\newcommand{\ep}{\epsilon}
\newcommand{\li}{\mathrm{Li}}